\pdfoutput=1

\newif\iffullver
\fullvertrue

\RequirePackage[T1]{fontenc}
\RequirePackage{fix-cm}

\documentclass[smallextended,nospthms,pdftex]{svjour3}
\usepackage{microtype}
\usepackage[unicode]{hyperref}
\usepackage{amsmath, amssymb}
\usepackage{mathptmx}
\usepackage{marvosym}

\providecommand{\doi}[1]{DOI~\href{https://doi.org/#1}{#1}}

\newcommand{\secref}[1]{\autoref{sec:#1}}
\newcommand{\figref}[1]{\autoref{fig:#1}}

\usepackage{tikz}
\usepackage[nounderscore]{syntax}

\usepackage{comment}
\usepackage{xspace}

\usepackage{wrapfig}
\usepackage{listings}
\definecolor{comment}{rgb}{0, 0.5, 0}
\definecolor{primitive}{rgb}{0.5, 0, 0.5}
\definecolor{annot}{rgb}{0, 0, 0.5}
\lstdefinelanguage{michelson}
{
  language={Caml},
  basicstyle = \footnotesize\ttfamily,
  commentstyle=\color{comment},
  morecomment=[l]{//},
  morecomment=[s]{/*}{*/},
  moredelim=[s][\color{annot}]{<<}{>>},
  morekeywords = [1] {
    PAIR, UNPAIR, CAR, CDR, NIL, CONS, SWAP, DROP, DIP, DUP, PUSH, UNIT, LAMBDA, EXEC,
    ADD, PACK,
    CDDR, ASSERT, ASSERT_SOME, IFCMPEQ,
    AMOUNT, SOURCE, CONTRACT, TRANSFER_TOKENS, CHECK_SIGNATURE,
    LOOP, ITER,
  },
  keywordstyle=[1]{\color{primitive}},
  morekeywords = [2] {
    Nil, Cons, Left, Right, Some, None, True, False, Unit, Pack, Contract, SetDelegate, TransferTokens, CreateContract, Error, Overflow,
  },
  keywordstyle=[2]{\color{blue}},
}
\usepackage{mysyntax, myrules, mythm}

\newcommand{\ottnt}[1]{\mathit{#1}}
\newcommand{\ottmv}[1]{\mathit{#1}}
\newcommand{\ottkw}[1]{\mathbf{#1}}
\newcommand{\ottsym}[1]{#1}

\renewcommand{\ottkw}[1]{\mathtt{#1}\relax}
\renewcommand{\ottnt}[1]{#1}
\renewcommand{\ottmv}[1]{#1}
\renewcommand{\ottsym}[1]{#1}
\providecommand{\lBrack}{[\![}
\providecommand{\rBrack}{]\!]}

\newcommand{\HELMHOLTZ}{\textsc{Helm\-holtz}\xspace}
\newcommand{\miniMic}{{Mini-Mi\-chel\-son}\xspace}
\DeclareMathOperator{\COL}{:}

\newif\ifdraftComments
\draftCommentsfalse
\def\mkDraftFn#1#2{%
  \expandafter\def\csname #1\endcsname##1{\ifdraftComments\textcolor{#2}{[#1: ##1]}\marginpar[$\longrightarrow$]{$\longleftarrow$}\fi}%
}
\mkDraftFn{KS}{red}
\mkDraftFn{AI}{blue}
\mkDraftFn{YN}{magenta}

\begin{document}
\iffullver
% \acmConference{a}{b}{c} % workaround for acmart class bug
% \title{\HELMHOLTZ: A Verifier for Tezos Smart Contracts Based on Refinement
%   Types}
% \author{Yuki Nishida}
% \email{nishida@fos.kuis.kyoto-u.ac.jp}
% \orcid{0000-0001-5941-6770}
% \affiliation{\institution{Kyoto University}\city{Kyoto}\country{Japan}}
% \author{Hiromasa Saito}
% \email{hsaito@fos.kuis.kyoto-u.ac.jp}
% \affiliation{\institution{Kyoto University}\city{Kyoto}\country{Japan}}
% \additionalaffiliation{Current affliation: DaiLambda, Inc.}
% \author{Ran Chen}
% \email{aran@fos.kuis.kyoto-u.ac.jp}
% \affiliation{\institution{Kyoto University}\city{Kyoto}\country{Japan}}
% \additionalaffiliation{??}
% \author{Akira Kawata}
% \email{akira@fos.kuis.kyoto-u.ac.jp}
% \affiliation{\institution{Kyoto University}\city{Kyoto}\country{Japan}}
% \additionalaffiliation{Current affliation: Preferred Networks, Inc.}
% \author{Jun Furuse}
% \email{jun.furuse@dailambda.jp}
% \affiliation{\institution{DaiLambda, Inc.}\city{Kyoto}\country{Japan}}
% \author{Kohei Suenaga}
% \email{ksuenaga@fos.kuis.kyoto-u.ac.jp}
% \orcid{0000-0002-7466-8789}
% \affiliation{\institution{Kyoto University}\city{Kyoto}\country{Japan}}
% \author{Atsushi Igarashi}
% \email{igarashi@fos.kuis.kyoto-u.ac.jp}
% \orcid{0000-0002-5143-9764}
% \affiliation{\institution{Kyoto University}\city{Kyoto}\country{Japan}}
% \begin{abstract}
%   \input{sections/abstract}
% \end{abstract}
% \maketitle

\title{\HELMHOLTZ: A Verifier for Tezos Smart Contracts Based on Refinement
  Types}

\author{Yuki Nishida
  \and Hiromasa Saito
  \and Ran Chen
  \and Akira Kawata
  \and Jun Furuse
  \and Kohei Suenaga
  \and Atsushi Igarashi}

\institute{
  Yuki Nishida\textsuperscript{(\Letter)} \and Hiromasa Saito \and Ran Chen \and Akira Kawata \and
  Kohei Suenaga \and Atsushi Igarashi \at
  Kyoto University, Kyoto, Japan\\
  \email{\{nishida,hsaito,aran,akira,ksuenaga,igarashi\}@fos.kuis.kyoto-u.ac.jp}
  \and
  Jun Furuse \at
  DaiLambda, Inc., Kyoto, Japan\\
  \email{jun.furuse@dailambda.jp}}

\maketitle

\begin{abstract}
  
A \emph{smart contract} is a program executed on a blockchain, based
on which many cryptocurrencies are implemented, and is
being used for automating transactions.  Due
to the large amount of money that smart contracts deal with, there
is a surging demand for a method that can statically and formally
verify them.

This article describes our type-based static verification tool
\HELMHOLTZ{} for Michelson, which is a statically typed stack-based
language for writing smart contracts that are executed on the blockchain
platform Tezos.  \HELMHOLTZ{} is designed on top of our extension of
Michelson's type system with refinement types.  \HELMHOLTZ{} takes a
Michelson program annotated with a user-defined specification written
in the form of a refinement type as input; it then typechecks the
program against the specification based on the refinement type system,
discharging the generated verification conditions with the SMT solver
Z3.  We briefly introduce our refinement type system for the core
calculus \miniMic{} of Michelson, which incorporates the
characteristic features such as compound datatypes (e.g., lists and
pairs), higher-order functions, and invocation of another contract.
\HELMHOLTZ{} successfully verifies several practical Michelson
programs, including one that transfers money to an account and that checks a digital signature.

\KS{No mention to the full version.}
\KS{No mention to soundness proof.}
\KS{No mention to it is omitted.}
\KS{Up to 16 pages excluding bibliography}

% Our tool checks whether each contract follows a given specification.  We
% implement our tool based on a refinement type system, a part of which is
% formalized as \miniMic{} in this paper.  Our tool takes contract code and its
% specification of the form refinement types; then, automatically typechecks the
% code following the type system; and discharges generated verification conditions
% during the type checking by Z3 {SMT} solver.

% Currently our tool supports roughly 80\% of instructions of Michelson.  We show
% that our tool can verify several Tezos contracts including practical ones with a
% self benchmark.

%%% Local Variables:
%%% mode: latex
%%% TeX-master: "../paper.otex"
%%% End:

  \keywords{
    Smart contract \and
    Blockchain \and
    Formal verification \and
    Tools}
\end{abstract}

\else
\title{\HELMHOLTZ: A Verifier for Tezos Smart Contracts Based on Refinement Types}
\author{
  Yuki Nishida\inst{1}\textsuperscript{(\Letter)}\orcidID{0000-0001-5941-6770} \and
  Hiromasa Saito\inst{1} \and
  Ran Chen\inst{1} \and\\
  Akira Kawata\inst{1}\thanks{Current affiliation: Preferred Networks, Inc.} \and
  Jun Furuse\inst{2} \and\\
  Kohei Suenaga\inst{1}\orcidID{0000-0002-7466-8789} \and
  Atsushi Igarashi\inst{1}\orcidID{0000-0002-5143-9764}
}
\titlerunning{\HELMHOLTZ: Tezos Smart Contract Verifier}
\authorrunning{Y. Nishida et al.}
\institute{
  Kyoto University, Kyoto, Japan\\
  \email{\{nishida,hsaito,aran,akira,ksuenaga,igarashi\}@fos.kuis.kyoto-u.ac.jp}
  \and
  DaiLambda, Inc., Kyoto Japan\\
  \email{jun.furuse@dailambda.jp}
}
\maketitle
\begin{abstract}
  
\end{abstract}
\fi
\section{Introduction}\label{sec:introduction}

A \emph{blockchain} is a data structure to implement a
distributed ledger in a trustless yet secure way.
% an incremental data structure which
% has resistance for falsification.
The idea of blockchains is initially devised for the Bitcoin
cryptocurrency~\cite{SatoshiN08} platform.  Many cryptocurrencies are
implemented using blockchains, in which value equivalent to a significant
amount of money is exchanged.

Recently, many cryptocurrency platforms allow programs to be executed
on a blockchain.  Such programs are called \emph{smart
  contracts}~\cite{Nick1997} (or, simply \emph{contracts} in this article)
since they work as a device to enable automated execution of a
contract.
% is a program which is stored in and
% works on a blockchain.
In general, a smart contract is a program $P_a$ associated with an account $a$
on a blockchain.  When the account $a$ receives money from another account $b$
with a parameter $v$, the computation defined in $P_a$ is conducted, during
which the state of the account $a$ (e.g., the balance of the account and
values that are stored by the previous invocations of $P_a$)
which is recorded on the blockchain may be
updated.  The contract $P_a$ may execute money transactions to another account
(say $c$), which results in invocations of other contracts (say $P_c$) during or
after the computation; therefore, contract invocations may be chained.

Although smart contracts' original motivation was handling simple
transactions (e.g., money transfer) among the accounts on a
blockchain, recent contracts are being used for more complicated
purposes (e.g., establishing a fund involving multiple accounts).
Following this trend, the languages for writing smart contracts also
evolve from those that allow a contract to execute relatively
simple transactions (e.g., Script for Bitcoin) to those that allow a
program that is as complex as one written in standard programming
languages (e.g., EVM for Ethereum and Michelson~\cite{michelson}
for Tezos~\cite{tezos}).

% An early smart contract system (for instance, Script for
% the Bitcoin platform) has limited ability which is enough to realize a
% transaction system for a cryptocurrency.  Many recent smart contract
% systems, which starts from the Ethereum~\cite{} platform, provide a
% Turing-complete ability.  So smart contracts have begun to be used for
% various systems not only for transaction systems.

% which are from rather simple
% ones (e.g., just transferring money between two individuals) to more
% complicated ones (e.g., establishing a fund involving multiple
% accounts).  An early smart contract system (for instance, Script for
% the Bitcoin platform) has limited ability which is enough to realize a
% transaction system for a cryptocurrency.  Many recent smart contract
% systems, which starts from the Ethereum~\cite{} platform, provide a
% Turing-complete ability.  So smart contracts have begun to be used for
% various systems not only for transaction systems.

Due to a large amount of money they deal with, verification of smart
contracts is imperative.  \emph{Static} verification is especially
needed since a smart contract cannot be fixed once deployed on a
blockchain.  Attack on a vulnerable contract indeed happened.  For
example, the DAO attack, in which the vulnerability of a fundraising
contract was exploited, resulted in the loss of cryptocurrency
equivalent to approximately 150M USD~\cite{theDAOAttack}.

% It is no doubt that smart contracts become a target of program verification.
% Cryptocurrency is promoted its reliability which is supported by cryptography
% and novel consensus protocols.  However, there is no advantage in reliability of
% software, a part of which consists of smart contracts, which realizes a
% cryptocurrency system.  As, for example, the market capitalization of the
% cryptocurrencies goes beyond three hundred billion {US} dollars,\footnote{As of
%   October, 2020.}  smart contracts are already used in critical systems.
% Unfortunately, serious attacks~\cite{???} have happened.  A lot of
% study~\cite{???} has begun to develop a verification method for smart contracts.

% \KS{Revised up to here.}

In this article, we describe our type-based static verifier
\HELMHOLTZ{}\footnote{Hermann von Helmholtz (1821--1894), a German
  physicist and physician, was a doctoral advisor of Albert
  A. Michelson (1852--1931), whom the Michelson language is apparently
  named after.}  for smart contracts written in Michelson.  The
Michelson language is a statically and simply typed stack-based
language equipped with rich data types (e.g., lists, maps, and
higher-order functions) and primitives to manipulate them.  Although
several high-level languages that compile to Michelson are being
developed, Michelson is most widely used to write a smart contract for
Tezos as of writing.

% there are no dominant high-level language.
% It will eventually reveals that Michelson is a low-level and specific language for Tezos.    So, we choose Michelson
% as a target.

% but structured control language equipped with One promoted feature is
% the language is statically typed and purely functional, which aimed to
% make smart contracts more secure and reliable.

% we propose a verification framework for the Michelson
% language~\cite{}, the native smart contract language for the Tezos
% smart contracts platform.

% \footnote{The semantics of a transaction invocation (and
%   hence an external program invocation) depends on languages.
%   External contract invocations in Ethereum are synchronous; they
%   block until the called contract returns.  In Tezos, an external
%   transaction can be invoked only after the caller terminates as we
%   describe below.}

% A smart contract is in general a program that conducts the following
% computation when an account with which  receives certain amount of money

A Michelson program expresses the above computation in a purely
functional style, in which the Michelson program corresponding to
$P_a$ is defined as a function.  The function takes a pair of the
parameter $v$ and a value $s$ that represents the current state of the
account (called \emph{storage}) and returns a pair of a list of
\emph{operations} and the updated storage $s'$.  Here, an operation is
a Michelson value that expresses the computation (e.g., transferring
money to an account and invoking the contract associated with the
account) that is to be conducted after the current computation (i.e.,
$P_a$) terminates.
% either TransferToken(...) that expresses transferring money and
% hence may cause an invocation of another contract; CreateContract(...) that
% expresses creating a new contract from a code literal; or Delegate(...) that
% expresses delegation of certain rights to another account.
After the computation specified by $P_a$ finishes with a pair of a
storage value and an operation list, a blockchain system
% updates the contract's storage to the returned value and
invokes the computation specified in the operation list.
This purely functional
style admits static verification methods for Michelson programs
similar to those for standard functional languages.

% A contract in Tezos is an account which has the code written in the
% Michelson language, data storage for general purpose, and other data
% that usual user accounts have. A contract call happens by transferring
% a money (and arguments for the call) to the account which has
% code. Contract code calculates operations, which is operations for
% Tezos system like call another contract, update blockchain state,
% etc., and a new storage value from the given arguments and the
% original storage value. As we have introduced, smart contract itself
% is purely functional. Side-effect on the Tezos system happens by
% issued operations (and a new storage value).

As the theoretical foundation of \HELMHOLTZ{}, we design a refinement
type system for Michelson as an extension of the original simple type
system.  In contrast to standard refinement types that refine the
types of values, our type system refines the type of \emph{stacks}.
\iffullver\else
We briefly describe our type system in Section~\ref{sec:system}; a
detailed explanation is deferred to a future paper.
\fi

We show that our tool can verify several practical smart contracts.
% Since the target language is relatively new and miner one, there are
% few benchmarks which can be directly used for evaluation.  We succeed
% to verify such few benchmarks and self-made benchmark for general
% stack-based programs and smart contracts.
In addition to the contracts we wrote ourselves, we apply our tool
to the sample Michelson programs used in Mi-cho-coq~\cite{BernardoCHPT19},
a formalization of Michelson in Coq proof
assistant~\cite{coq}.  These contracts consist of practical contracts
such as one that checks a digital signature and one that transfers money.

% \KS{Describe the Mi-cho-coq benchmark.}  Our tool successfully
% verifies these contracts with a moderate amount of user annotations.

We note that \HELMHOLTZ{} currently supports approximately 80\% of the
whole instructions of the Michelson language.  Another limitation of
the current \HELMHOLTZ{} is that it can verify only a single contract,
although one often uses multiple contracts for an application, in
which a contract may call another by a money transfer operation, and
their behavior as a whole is of interest.  We are currently extending
\HELMHOLTZ{} so that it can deal with more programs.

% \KS{Mention the chaining of contracts.}  \AI{In
%   this paper, we are interested in the execution of a single contract.
%   So, although we formalize an operation, it's just another opaque
%   data structure, from which you can't extract values.  }

Our contribution is summarized as follows:
% \KS{Do we count refinement type system and soundness proof as contribution of this paper?}
% \begin{itemize}
(1) Definition of the core calculus \miniMic{} and its refinement
  type system; 
(2) Automated verification tool \HELMHOLTZ{} for Michelson contracts
  implemented based on the type system of \miniMic{}; the interface to the implementation
  can be found at \url{https://www.fos.kuis.kyoto-u.ac.jp/trylang/Helmholtz}; and
%  Prototype verification tool based on \miniMic{}.
(3) Evaluation of \HELMHOLTZ{} with various Michelson contracts, including practical ones.
% \end{itemize}
A preliminary version of this article was presented at International
Conference on Tools and Algorithms for the Construction and Analysis
of Systems (TACAS) in 2021.  We have given detailed proofs of properties of
\miniMic{} and a more detailed description
about the verifier implementation, in addition to revision of the text.

% \paragraph{Outline of the rest of paper.} \NOTE{TODO}

The rest of this article is organized as follows.  Before introducing
the technical details, we present an overview of the verifier \HELMHOLTZ{} in
\secref{overview}
 using a simple example of a Michelson
 contract.
 \iffullver
 \secref{system} introduces the core calculus
 \miniMic{} with its refinement type system and states
 soundness of the refinement type system. 
(Detailed proofs are deferred to \secref{detailedproofs}.)
 We also discuss a few extensions implemented in the verifier.
 \else
  \secref{system} introduces the core calculus
 \miniMic{}, and its refinement type system.
 \fi
\secref{implementation} describes the verifier \HELMHOLTZ{}, a
case study, and experimental results.  After discussing related work
in \secref{related}, we conclude in
\secref{conclusion}.

%%% Local Variables:
%%% mode: latex
%%% TeX-master: "../paper.otex"
%%% End:

\section{Overview of \HELMHOLTZ and Michelson}\label{sec:overview}

% \KS{Introduce the word ``contract'' somewhere in the introduction.  I'm currently using ``program'' here but this should be replaced with ``contract''.}

We give an overview of our tool \HELMHOLTZ in this section before presenting its technical details.
We also explain Michelson by example (\secref{michelsonOverview})
and user-written annotation added to a Michelson program for verification purposes (\secref{specOverview}).

% \begin{figure}
%   \centering
%   \NOTE{fill}\vrule height 4cm
%   \caption{Overview of our tool}
%   \label{fig:arch}
% \end{figure}

% In this section, we briefly introduce our tool and an underlying theory.
\subsection{\HELMHOLTZ}
\label{sec:overviewTool}

As input, \HELMHOLTZ takes a Michelson program annotated with (1) its specification expressed in a refinement type and
(2) additional user annotations such as loop invariants.
It typechecks the annotated program against the specification using our refinement type system;
the verification conditions generated during the typechecking is discharged by the SMT solver Z3~\cite{MouraB08}.
If the code successfully typechecks, then the program is guaranteed to satisfy the specification.
% Then, a post-condition of whole code can be automatically
% generated from the given pre-condition.  Finally, Z3 SMTsolver is used to check
% if the generated post-condition implies the given post-condition.

\HELMHOLTZ is implemented as a subcommand of
\texttt{tezos-client}, the client program of the Tezos blockchain.
For example, to verify \texttt{boomerang.tz} in \figref{boomerang},
% and obtain the result as
% follows.
we run \texttt{tezos-client refinement boomerang.tz}.
% \begin{lstlisting}[numbers=none, xleftmargin=0em]
% > tezos-client refinement boomerang.tz
% \end{lstlisting}
If the verification succeeds, the command outputs \texttt{VERIFIED}
to the terminal screen (with a few log messages); otherwise, it outputs \texttt{UNVERIFIED}.
% Then, our tool prints some of the outputs from Z3, followed by the verification result.
% \begin{lstlisting}
% ...
% =========== Z3 output start ===========
% unsat
% unsat
% =========== Z3 output end ===========
% =========== Z3 result start ===========
% VERIFIED
% =========== Z3 result end ===========
% \end{lstlisting}

% \lstinline$tezos-client$ is a command provided by the Tezos blockchain
% component.  We have modified the command to implement our tool so that
% \lstinline$refinement$ subcommand switches behavior to verify given annotated
% code.  You will see a lot of output as shown, but most important one is
% \lstinline$VERIFIED$ in the last, which means that the contract satisfies the
% given specification.
% The verification result output becomes to \lstinline$UNVERIFIED$ if
% the verification fails.
% You can observe that by making, for instance, the
% post-condition just \texttt{ops = [ TransferTokens Unit amount (Contract
% source) ]}.

\newcommand{\PSH}{\triangleright}
\newcommand{\PUSH}{\textcolor{comment}{$\PSH$}}
\begin{figure}[t!]
  \centering
\begin{lstlisting}[language=michelson]
parameter unit;@\label{tz:parameter}@
storage unit;@\label{tz:storage}@
<< ContractAnnot { (param, st) | True } ->@\label{tz:contractAnnotStart}@
   { (ops, st') | amount = 0 && ops = [] || @\label{tz:postcondStart}@
      amount <> 0 && (match contract_opt source with
                    | Some c -> ops = [Transfer Unit amount c]
                    | None -> False) } @\label{tz:postcondEnd}@
   & { _ | False } >> @\label{tz:contractAnnotEnd}@
code                   /* (param,st) */
 { CDR;                /* st */@\label{tz:codeStart}@
   NIL operation;      /* [] @\PUSH@ st */
   AMOUNT;             /* amount @\PUSH@ [] @\PUSH@ st */
   PUSH mutez 0;       /* 0 @\PUSH@ amount @\PUSH@ [] @\PUSH@ st */
   IFCMPEQ
    {}                 /* [] @\PUSH@ st   (amount @\textcolor{comment}{$\leq$}@ 0) */
    {                  /* [] @\PUSH@ st   (amount @\textcolor{comment}{$>$}@ 0) */
      SOURCE;          /* src @\PUSH@ [] @\PUSH@ st */
      CONTRACT unit;   /* Some (Contract src) @\PUSH@ [] @\PUSH@ st */
      ASSERT_SOME;     /* (Contract src) @\PUSH@ [] @\PUSH@ st */
      AMOUNT; UNIT;
      /* Unit @\PUSH@ amount @\PUSH@ (Contract src) @\PUSH@ [] @\PUSH@ st */
      TRANSFER_TOKENS;
      /* (Transfer Unit amount (Contract src)) @\PUSH@ [] @\PUSH@ st */
      CONS
      /* [Transfer Unit amount (Contract src)] @\PUSH@ st */
    };
    /* ops @\PUSH@ st@\textcolor{comment}{\textrm{, where \texttt{ops} is the top element at the end of each branch, namely,}}@
       @\textcolor{comment}{\textrm{\texttt{[Transfer Unit amount (Contract src)]} if \(\texttt{amount} > 0\); or \texttt{[]} otherwise}}@ */
   PAIR }              /* (ops, st) */ @\label{tz:codeEnd}@
\end{lstlisting}
  \caption{\texttt{boomerang.tz}.  The comment inside \lstinline{/* */} describes the stack at the program point.}
  \label{fig:boomerang}
\end{figure}

% \figref{boomerang} is an example of a simple contract in the Michelson language accompanied with the annotation for our tool which enclosed by \lstinline$<<$ and \lstinline$>>$ at lines 3--7.
\subsection{An Example Contract in Michelson}
\label{sec:michelsonOverview}

\figref{boomerang} shows an example of a Michelson program called
\lstinline{boomerang}.  A Michelson program is associated with an
account on the Tezos blockchain; the program is invoked by
transferring money to this account.  This artificial program in \figref{boomerang},
when it is invoked, is supposed to transfer the received money back to
the account that initiated the transaction.

A Michelson program starts with type declarations of its
\lstinline{parameter}, whose value is given by contract invocation,
and \lstinline{storage}, which is the state that the contract account stores.
Lines~\ref{tz:parameter}--\ref{tz:storage} declare that the types of both are
\lstinline{unit}, the type inhabited by the only value \lstinline{Unit}.
Lines~\ref{tz:contractAnnotStart}--\ref{tz:contractAnnotEnd} surrounded by \lstinline$<<$ and \lstinline$>>$ are a user-written annotation
used by \HELMHOLTZ for verification; we will explain this annotation later.
The \lstinline{code} section in Lines~\ref{tz:codeStart}--\ref{tz:codeEnd} is the body of this program.
% simple contract in the Michelson language accompanied with
% Lines~\ref{tz:contractAnnotStart}--\ref{tz:contractAnnotEnd} are an annotation that describes the specification of this program.

Let us take a look at the \lstinline{code} section of the program.
In the following explanation of each instruction, we describe the state of the stack after each instruction
as comments; stack elements are delimited by \(\PSH\).
% (2) the keywords \lstinline{amount} and \lstinline{source}, which we used to describe the annotation \lstinline{ContractAnnot}.
\begin{itemize}
\item Execution of a Michelson program starts with a stack with one value,
which is a pair \lstinline{(param, st)} of a parameter \lstinline$param$ and a storage value \lstinline$storage$.
\item \lstinline$CDR$ pops the pair at the top of the stack and pushes the second value of the popped pair;
  thus, after executing the instruction, the stack contains the single value \lstinline{st}.
\item \lstinline$NIL$ pushes the empty list \lstinline{[]} to the stack;
  the instruction is accompanied by the type \lstinline{operation} of the list elements for typechecking purposes.
\item \lstinline$AMOUNT$ pushes the nonnegative \lstinline{amount} of the money sent to the account to which this program is associated.
\item \lstinline$PUSH mutez 0$ pushes the value $0$.  The type \lstinline{mutez} represents a unit of money used in Tezos.
\item \lstinline$IFCMPEQ b1 b2$, if the state of the stack before executing the instruction is \lstinline{v1} $\PSH$ \lstinline{v2} $\PSH$ \lstinline{tl},
  (1) pops \lstinline{v1} and \lstinline{v2} and (2) executes the then-branch
  \lstinline{b1} (resp., the else-branch \lstinline{b2}) if \lstinline{v2} $=$ \lstinline{v1}
  (resp., \lstinline{v2} $\neq$ \lstinline{v1}).  In \lstinline{boomerang}, this instruction does nothing if \lstinline{amount} $=$ 0; otherwise, the instructions in the else-branch are executed.
\item \lstinline$SOURCE$ at the beginning of the else-branch
  pushes the address
  \lstinline{src} of the source account, which initiated the chain of contract invocations that the current contract belongs to, resulting in the stack
  \lstinline{src} $\PSH$ \lstinline{[]} $\PSH$ \lstinline{st}.  
  % \AI{Mention chains of contract invocations?}
\item \lstinline$CONTRACT$ $T$ pops an address \lstinline{addr} from the stack
  and typechecks whether the contract associated with \lstinline{addr} takes an argument of type $T$.  If the
  typechecking succeeds, then \lstinline$Some (Contract addr)$ is pushed; otherwise, \lstinline{None} is pushed.    The constructor \lstinline{Contract} creates an object that represents a typechecked contract at the given address.
  In Tezos, the source account is always a contract that takes the value \lstinline$Unit$ as a parameter;
  thus, \lstinline{Some (Contract src)} will always be pushed onto the stack.

\item \lstinline!ASSERT_SOME! pops a value \lstinline{v} from the stack and pushes \lstinline{v'} if \lstinline{v} is \lstinline{Some v'}; otherwise, it raises an exception.
\item \lstinline!UNIT! pushes the unit value \lstinline{Unit} to the stack.
\item \lstinline!TRANSFER_TOKENS!, if the stack is of the shape
  \lstinline{varg} $\PSH$ \lstinline{vamt} $\PSH$ \lstinline{vcontr} $\PSH$ \lstinline{tl}, pops \lstinline{varg},
  \lstinline{vamt}, and \lstinline{vcontr} from the stack and pushes
  \texttt{(Transfer varg vamt vcontr)} onto \lstinline{tl}.  The value
  \lstinline{Transfer varg vamt vcontr} is an \emph{operation object} expressing
  that money (of amount \lstinline{vamt}) shall be sent to the account
  \lstinline{vcontr} with the argument \lstinline{varg} after this
  program finishes without raising an exception.  Therefore, the program
  associated with \lstinline{vcontr} is invoked after this program
  finishes.  Otherwise, an operation object is an opaque tuple and
  no instruction can extract its elements.
\item \lstinline{CONS} with the stack \lstinline{v1} $\PSH$ \lstinline{v2} $\PSH$ \lstinline{tl} pops
  \lstinline{v1} and \lstinline{v2}, and pushes a cons list \lstinline{v1::v2} onto
  the stack.  (We use the list notation in OCaml here.)
\item After executing one of the branches associated with \lstinline{IFCMPEQ} in
   this program, the shape of the stack should be \lstinline{ops} $\PSH$ \lstinline{storage},
   where \lstinline{ops} is \lstinline{[]} if
  \lstinline{amount} $= 0$ or \texttt{[Transfer varg vamt vcontr]} if \lstinline{amount} $> 0$.  The instruction
  \lstinline{PAIR} pops \lstinline{ops} and \lstinline{storage}, and
  pushes \lstinline{(ops,storage)}.
\end{itemize} 
% \AI{We might want to put the shape of the stack into the code as comments.  DONE.}
A Michelson program is supposed to finish its execution with a singleton stack
whose unique element is a pair of (1) a list of operations to be executed
after the current execution of the contract finishes and (2) the new value for the storage.

Michelson is a statically typed language.  Each instruction is associated with
a typing rule that specifies the shapes of stacks before and after it by a sequence of simple types such as \lstinline{int}
and \lstinline{int list}.
For example, \lstinline$CONS$ requires the type of top element to be $T$
and that of the second to be $T$\lstinline{ list} (for any $T$); it ensures
the top element after it has type $T$\lstinline{ list}.

Other notable features of Michelson include first-class functions, hashing,
instructions related to cryptography such as signature verification,
and manipulation of a blockchain using operations.
% We will explain a part of these features in what follows.

\subsection{Specification}
\label{sec:specOverview}

A user can specify the behavior of a program by a \lstinline{ContractAnnot} annotation,
which is a part of the augmented syntax of our verification tool.
A \lstinline{ContractAnnot} annotation gives a specification of a Michelson program by the following notation inspired by the refinement types:
  \texttt{\{(param,st) | pre\} -> \{(ops,st') | post\} \& \{exc | abpost\}}
  where \lstinline{pre}, \lstinline{post}, and \lstinline{abpost} are predicates.
  % \YN{Parameter parts are actually patterns.  We use this pattern for convenience.}
This specification reads as follows:
if this program is invoked with a parameter \lstinline{param} and storage \lstinline{st}
that satisfies the property \lstinline{pre}, then
(1) if the execution of this program succeeds, then it returns a list of operations \lstinline{ops} and
new storage \lstinline{storage'} that satisfy the property \lstinline{post};
(2) if this program raises an exception with value \lstinline{exc}, then \lstinline{exc} satisfies \lstinline{abpost}.
% The contract basically does nothing but
% refunds all money sent if there is.  The specification is given as the
% annotation, and our tool verifies whether the code satisfies the specification.
The specification language, which is ML-like, is expressive enough to cover the specifications for
practical contracts, including the ones we used in the experiments in
\secref{experiments}.  In the predicates, one can use several keywords such as
\lstinline{amount} for the amount of the money sent to this program when it is
invoked and \lstinline{source} for the source account's address.

% The user-written \lstinline{ContractAnnot} annotation ,
% which is a part of the augmented syntax used by our tool,
% specifies the behavior of this program.
% % the annotation for our tool which enclosed by \lstinline$<<$ and \lstinline$>>$ at lines 3--7.

The \lstinline{ContractAnnot} annotation in \figref{boomerang}
(Lines~\ref{tz:contractAnnotStart}--\ref{tz:contractAnnotEnd}) formalizes this program's
specification as follows.  This program can take any parameter and storage
(Line~\ref{tz:contractAnnotStart}).  Successful execution of this program results in a pair
\lstinline{(ops,st')} that satisfies the condition in
Lines~\ref{tz:postcondStart}--\ref{tz:postcondEnd} that expresses (1) if $\texttt{amount}=0$, then
\lstinline{ops} is empty, that is, no operation will be issued; (2) if $\texttt{amount}>0$, then
\lstinline{ops} is a list of a single element \lstinline{Transfer Unit amount c}, where
\lstinline{c} is bound for \lstinline{Contract source}\footnote{It is one axiom of our domain
  specific theory that \lstinline{contract_opt source} always return \lstinline{Some (Contract source)}.},
which expresses transfer of money of the amount \lstinline{amount} to the account at
\lstinline{source} with the unit argument.%
\footnote{As we mentioned in \secref{introduction},
  \HELMHOLTZ{} can currently verify the behavior of a single contract,
  although there will be an invocation of the contract associated with
  \lstinline{source} after the termination of \lstinline{boomerang}.
  An operation is treated as an opaque data structure, from which one
  cannot extract values.}
In the specification language, \lstinline{source} and \lstinline{amount} are keywords that stand for
the source account and the amount of money sent to this program, respectively.
The part \lstinline!& { _ | False }! expresses that this program does not raise an exception.
This specification correctly formalizes the intended behavior of this program.

\iffullver
\section{Refinement Type System for \miniMic}\label{sec:system}

In this section, we formalize \miniMic, a core subset of Michelson with its
syntax, operational semantics, and refinement type system.  We omit many
features from the full language in favor of conciseness but includes language
constructs---such as higher-order functions and iterations---that make
verification difficult.

\subsection{Syntax}

\begin{figure}
  \centering
  \input{figures/syntax}
  \caption{Syntax of Mini-Michelson.}
  \label{fig:syntax}
\end{figure}
\figref{syntax} shows the syntax of \miniMic{}.
\emph{Values}, ranged over by \(\ottnt{V}\), consist of integers
\(\ottnt{i}\); addresses \(\ottnt{a}\); operation objects
\( \texttt{Transfer}  (  \ottnt{V} ,  \ottnt{i} ,  \ottnt{a}  ) \) to invoke a contract at
\(a\) by sending money of amount \(i\) and an argument \(V\); pairs
\(\ottsym{(}  \ottnt{V_{{\mathrm{1}}}}  \ottsym{,}  \ottnt{V_{{\mathrm{2}}}}  \ottsym{)}\) of values; the empty list \(\ottsym{[}  \ottsym{]}\); cons
\(\ottnt{V_{{\mathrm{1}}}}  ::  \ottnt{V_{{\mathrm{2}}}}\); and code \( \langle  \mathit{IS}  \rangle \) of first-class functions.\footnote{%
  Closures are not needed because functions in Michelson can access only arguments.
}
Unlike Michelson, which has primitive Boolean literals \lstinline{True} and \lstinline{False},
we use integers as a substitute for Boolean values so
that \(0\) means \lstinline{False} and the others mean \lstinline{True}.
As we have mentioned, there is no instruction to extract elements from
an operation object but the elements can be referenced in refinement types
to state what kind of operation object is constructed by a smart contract.
\emph{Simple types}, ranged over by \(\ottnt{T}\), consist of base types
(\(\ottkw{int}\), \(\ottkw{address}\), and \(\ottkw{operation}\), which are self-explanatory),
pair types \(\ottnt{T_{{\mathrm{1}}}}  \times  \ottnt{T_{{\mathrm{2}}}}\), list types \(\ottnt{T} \, \ottkw{list}\), and function types
\(\ottnt{T_{{\mathrm{1}}}}  \to  \ottnt{T_{{\mathrm{2}}}}\).
\emph{Instruction sequences}, ranged over by \(\mathit{IS}\), are a sequence
of \emph{instructions}, ranged over by \(\ottnt{I}\), enclosed by curly braces.
A \miniMic{} \emph{program} is an instruction sequence.

Instructions include those for operand stack manipulation (to $\ottkw{DROP}$,
$\ottkw{DUP}$licate, $\ottkw{SWAP}$, and $\ottkw{PUSH}$ values); $\ottkw{NOT}$ and
$\ottkw{ADD}$ for manipulating integers;
% \YN{isn't it sounds weird NOT manipulating integers...?}
$\ottkw{PAIR}$, $\ottkw{CAR}$, and $\ottkw{CDR}$ for pairs; $\ottkw{NIL}$ and
$\ottkw{CONS}$ for constructing lists; $\ottkw{LAMBDA}$ for a first-class
function; $\ottkw{EXEC}$ for calling a function; and $ \texttt{TRANSFER\symbol{95}TOKENS} $
to create an operation.  Instructions for control structures are
$\ottkw{IF}$ and $ \texttt{IF\symbol{95}CONS} $, which are for branching on integers
(whether the stack top is \lstinline{True} or not) and lists (whether the stack
top is a cons or not), respectively, and $\ottkw{LOOP}$ and $\ottkw{ITER}$,
which are for iteration on integers and lists, respectively.
$\ottkw{LAMBDA}$ pushes a function (described by its operand $\mathit{IS}$) onto the stack
and $\ottkw{EXEC}$ calls a function.  Perhaps
unfamiliar is $\ottkw{DIP} \, \mathit{IS}$, which pops and saves the stack top
somewhere else, execute $\mathit{IS}$, and then push back the saved value.

We also use a few kinds of stacks in the following definitions: operand stacks,
ranged over by \(\ottnt{S}\), type stacks, ranged over by \(\bar{T}\), and type
binding stacks, ranged over by \(\Upsilon\).  The empty stack is denoted by
\( \ddagger \) and push is by \( \triangleright \).  We often omit the empty stack and write,
for example, \(\ottnt{V_{{\mathrm{1}}}}  \triangleright  \ottnt{V_{{\mathrm{2}}}}\) for \(\ottnt{V_{{\mathrm{1}}}}  \triangleright  \ottnt{V_{{\mathrm{2}}}}  \triangleright  \ddagger\).  Intuitively,
\(\ottnt{T_{{\mathrm{1}}}}  \triangleright \, .. \, \triangleright  \ottnt{T_{\ottmv{n}}}\) and \( \ottmv{x_{{\mathrm{1}}}} \mathord:\allowbreak \ottnt{T_{{\mathrm{1}}}}   \triangleright \, .. \, \triangleright   \ottmv{x_{\ottmv{n}}} \mathord:\allowbreak \ottnt{T_{\ottmv{n}}} \) describe stacks
\(\ottnt{V_{{\mathrm{1}}}}  \triangleright \, .. \, \triangleright  \ottnt{V_{\ottmv{n}}}\) where each value \(\ottnt{V}_i\) is of type \(\ottnt{T}_i\).  We
will use variables to name stack elements in the refinement type system.

We summarize main differences from Michelson proper:
\begin{itemize}
\item Michelson has the notion of type attributes, which classify types,
  according to which generic operations such as $\ottkw{PUSH}$ can be
  applied.  For example, values of \emph{pushable} types can be put on
  the stack by $\ottkw{PUSH}$.  Since type $\ottkw{operation}$ is not
  pushable, an instruction such as
  $\ottkw{PUSH} \, \ottkw{operation} \,  \texttt{Transfer}  (  \ottnt{V} ,  \ottnt{i} ,  \ottnt{a}  ) $ is not valid in
  Michelson---all operations have to be created by designated
  instructions.  We ignore type attributes for simplicity here, but
  the implementation of \HELMHOLTZ, which calls the typechecker of
  Michelson, does not.

\item As we saw in \secref{overview}, an operation is created from an
  address in two steps via a contract value.  Since we model only one
  kind of operations, i.e., $ \texttt{Transfer}  (  \ottnt{V} ,  \ottnt{i} ,  \ottnt{a}  ) $, we simplify
  the process to let instruction $ \texttt{TRANSFER\symbol{95}TOKENS} $ directly
  creates an operation from an address in one step.  We also omit the
  typecheck of the contract associated with an address.

\item  In Michelson, each execution of a smart contract
is assigned a \emph{gas} to control how long the contract can run to
prevent contracts from running too long.

\item We do not formally model exceptions for simplicity and, thus,
  the refinement type system do not (have to) capture exceptional
  behavior.  Our verifier, however, \emph{does} handle exceptions; we
  will informally discuss how we extend the type system with
  exceptions in Section~\ref{sec:exceptions}.

\end{itemize}

\subsection{Operational Semantics}

\begin{figure}
  \centering
  \input{figures/eval}
  \caption{Operational Semantics of \miniMic{}}
  \label{fig:eval}
\end{figure}

\figref{eval} defines the operational semantics of \miniMic{}.  A
judgment of the form \(\ottnt{S}  \vdash  \ottnt{I}  \Downarrow  \ottnt{S'}\) (or \(\ottnt{S}  \vdash  \mathit{IS}  \Downarrow  \ottnt{S'}\),
resp.)  means that evaluating the instruction \(\ottnt{I}\) (or the
instruction sequence \(\mathit{IS}\), resp.) under the stack \(\ottnt{S}\)
results in the stack \(\ottnt{S'}\).  Although the defining rules are
straightforward, we will make a few remarks about them.

% we will explain a detail of rules
% only for some ones because most of ones must be straightforward, e.g.,
% \ruleref{E-Seq} says that if the first instruction of a sequence results
% \(\ottnt{S'}\) from \(\ottnt{S}\) and the rest results \(\ottnt{S''}\) from \(\ottnt{S'}\), the
% whole sequence results \(\ottnt{S''}\) from \(\ottnt{S}\); \ruleref{E-Add} says that
% \(\ottkw{ADD}\) instruction adds the top two integers of a stack; etc.  Note that

The rule \ruleref{E-Dip} means that \(\ottkw{DIP} \, \mathit{IS}\) pops and saves the stack top
somewhere else, execute $\mathit{IS}$, and then push back the saved value,
as explained above.  This instruction implicitly gives \miniMic{} (and
Michelson) a secondary stack.  \ruleref{E-Push} means that
\(\ottkw{PUSH} \, \ottnt{T} \, \ottnt{V}\) does not check if the pushed value is well formed at
run time: the check is the job of the simple type system, discussed
soon.

The rules \ruleref{E-IfT} and \ruleref{E-IfF} define the behavior of
the branching instruction \(\ottkw{IF} \, \mathit{IS}_{{\mathrm{1}}} \, \mathit{IS}_{{\mathrm{2}}}\), which executes \(\mathit{IS}_{{\mathrm{1}}}\) or
\(\mathit{IS}_{{\mathrm{2}}}\), depending on the top of the operand stack.  As we have
mentioned, nonzero integers mean \lstinline{True}.  Thus, \ruleref{E-IfT} is
used for the case in which \(\mathit{IS}_{{\mathrm{1}}}\) is executed, and otherwise,
\ruleref{E-IfF} is used.  There is another branching instruction
\(\texttt{IF\symbol{95}CONS} \, \mathit{IS}_{{\mathrm{1}}} \, \mathit{IS}_{{\mathrm{2}}}\), which executes either instruction sequence
depending on whether the list at the top of the stack is empty or not
(cf. \ruleref{E-IfConsT} and \ruleref{E-IfConsF}).

The rules \ruleref{E-LoopT} and \ruleref{E-LoopF} define the behavior of
the looping instruction \(\ottkw{LOOP} \, \mathit{IS}\).
This instruction executes \(\mathit{IS}\) repeatedly until the top of the
stack becomes \emph{false}.  \ruleref{E-LoopT} means that, if the condition is \lstinline{True}, 
\(\mathit{IS}\) is executed, and then \(\ottkw{LOOP} \, \mathit{IS}\) is executed again.
\ruleref{E-LoopF} means that, if the
condition is \emph{false}, the loop is finished after dropping
the stack top.
A similar looping instruction is \(\ottkw{ITER} \, \mathit{IS}\),
which iterates over a list (see \ruleref{E-IterNil} and \ruleref{E-IterCons}).

% Note that intuitively some code produce
% infinite loop, e.g., \(\ottkw{LOOP} \, \ottsym{\{}  \ottkw{PUSH} \, \ottkw{int} \, 1  \ottsym{\}}\) for a stack whose top is
% \emph{true}.  However our semantics cannot catch such infinite execution because
% a derivation tree must be finite.  In other words, the judgment \(\ottnt{S}  \vdash  \ottnt{I}  \Downarrow  \ottnt{S'}\)
% is also mentions that the execution of \(\ottnt{I}\) terminates.  

The rule \ruleref{E-Lambda} means that \(\ottkw{LAMBDA} \, \ottnt{T_{{\mathrm{1}}}} \, \ottnt{T_{{\mathrm{2}}}} \, \mathit{IS}\) pushes the instruction sequence
to the stack and \ruleref{E-Exec} means that \(\ottkw{EXEC}\) pops
the instruction sequence \( \langle  \mathit{IS}  \rangle \) and the stack top \(\ottnt{V}\),
saves the rest of the stack \(\ottnt{S}\) elsewhere,
runs \(\mathit{IS}\) with \(\ottnt{V}\) as the sole value in the stack,
pushes the result \(\ottnt{V'}\) back to the restored stack \(\ottnt{S}\).

The rule \ruleref{E-TransferTokens} means that
\(\texttt{TRANSFER\symbol{95}TOKENS} \, \ottnt{T}\) creates an operation object and pushes onto
the stack.  (As we have discussed, we omit a run-time check to
see if \(\ottnt{T}\) is really the argument type of the contract that the
address \(\ottnt{a}\) stores.)

\subsection{Simple Type System}

\begin{figure}
  \centering
  \input{figures/vtyping}
  \vspace{2ex}
  \input{figures/styping}
  \caption{Simple typing}
  \label{fig:styping}
\end{figure}

\miniMic (as well as Michelson) is equipped with a simple type system.  The type
judgment for instructions is written \(\bar{T}  \vdash  \ottnt{I}  \Rightarrow  \bar{T}'\), which means that
instruction \(\ottnt{I}\) transforms a stack of type \(\bar{T}\) into another stack
of type \(\bar{T}'\).  The type judgment for values is written \( \ottnt{V}  :  \ottnt{T} \),
which means that \(\ottnt{V}\) is given simple type \(\ottnt{T}\).  The typing rules, which are
shown in \figref{styping}, are fairly
straightforward.  Note that these two judgment forms depend on each other---see
\ruleref{RTV-Fun} and \ruleref{T-Push}.

\subsection{Refinement Type System}\label{sec:system/refinement}

Now we extend the simple type system to a refinement type system.  In
the refinement type system, a simple stack type \(\ottnt{T_{{\mathrm{1}}}}  \triangleright \, .. \, \triangleright  \ottnt{T_{\ottmv{n}}}\)
is augmented with a formula \(\varphi\) in an assertion language to
describe the relationship among stack elements.  More concretely, we
introduce \emph{refinement stack types}, ranged over by \(\Phi\),
of the form \(\ottsym{\{}   \ottmv{x_{{\mathrm{1}}}} \mathord:\allowbreak \ottnt{T_{{\mathrm{1}}}}   \triangleright \, ... \, \triangleright   \ottmv{x_{\ottmv{n}}} \mathord:\allowbreak \ottnt{T_{\ottmv{n}}}   \mid  \varphi  \ottsym{(}  \ottmv{x_{{\mathrm{1}}}}  \ottsym{,} \, ... \, \ottsym{,}  \ottmv{x_{\ottmv{n}}}  \ottsym{)}  \ottsym{\}}\),
which denotes a stack \(\ottnt{V_{{\mathrm{1}}}}  \triangleright \, .. \, \triangleright  \ottnt{V_{\ottmv{n}}}\) such that
\( \ottnt{V_{\ottmv{n}}}  :  \ottnt{T_{{\mathrm{1}}}} \), \ldots, \( \ottnt{V_{\ottmv{n}}}  :  \ottnt{T_{\ottmv{n}}} \) and
\(\varphi  \ottsym{(}  \ottnt{V_{{\mathrm{1}}}}  \ottsym{,} \, ... \, \ottsym{,}  \ottnt{V_{\ottmv{n}}}  \ottsym{)}\) hold, and refine the type judgment form,
accordingly.  We start with an assertion language, which is
many-sorted first-order logic and proceed to the refinement type
system.

\subsubsection{Assertion Language}

\begin{figure}
  \centering
  \input{figures/assertion}
  \caption{Syntax of Assertion Language}
  \label{fig:assertionsyntax}
\end{figure}

\begin{figure}
  \centering
  \begin{flushleft}
    \noindent {Well-Sorted Terms} \fbox{\(\Gamma  \vdash  \ottnt{t}  \ottsym{:}  \ottnt{T}\)}
  \end{flushleft}\vspace{-2ex}
  \input{figures/ttyping}
%   \caption{Well-Sorted Terms}
%   \label{fig:ttyping}
% \end{figure}
\vspace{2ex}
% \begin{figure}
%   \centering
  \begin{flushleft}
    \noindent {Well-Sorted Formulae} \fbox{\( \Gamma   \vdash   \varphi  : \mathord{*} \)}
  \end{flushleft}\vspace{-2ex}
  \input{figures/ftyping}
  \caption{Well-Sorted Terms and Formulae.}
  \label{fig:ftyping}
\end{figure}

The assertion language is many-sorted first-order logic, where sorts are simple
types.  We show the syntax of terms, ranged over by \(\ottnt{t}\), and
formulae, ranged over by \(\varphi\), in \figref{assertionsyntax}.
As usual, $\ottmv{x}$ is bound in $\exists \,  \ottmv{x} \mathord:\allowbreak \ottnt{T}   \ottsym{.}  \varphi$.
Most of them are straightforward but the formulae of the form \( \ottkw{call} ( \ottnt{t_{{\mathrm{1}}}} ,  \ottnt{t_{{\mathrm{2}}}} ) =  \ottnt{t_{{\mathrm{3}}}} \) deserves an explanation.  It means that, if instruction sequence denoted by $\ottnt{t_{{\mathrm{1}}}}$ is called with (a singleton stack that stores) a value denoted by $\ottnt{t_{{\mathrm{2}}}}$ (and terminates), it yields the value denoted by $\ottnt{t_{{\mathrm{3}}}}$.
The term constructor $ \texttt{Transfer}  (  \ottnt{t_{{\mathrm{1}}}} ,  \ottnt{t_{{\mathrm{2}}}} ,  \ottnt{t_{{\mathrm{3}}}}  ) $ allows us to refer to the elements in an operation object, which is opaque.
Conjunction $\varphi_{{\mathrm{1}}}  \wedge  \varphi$, implication $\varphi_{{\mathrm{1}}}  \implies  \varphi_{{\mathrm{2}}}$, and universal quantification $\forall \,  \ottmv{x} \mathord:\allowbreak \ottnt{T}   \ottsym{.}  \varphi$ are defined as abbreviations as usual.
We use several common abbreviations such as \(\ottnt{t_{{\mathrm{1}}}}  \neq  \ottnt{t_{{\mathrm{2}}}}\) for
\(\neg \, \ottsym{(}  \ottnt{t_{{\mathrm{1}}}}  \ottsym{=}  \ottnt{t_{{\mathrm{2}}}}  \ottsym{)}\), \(\exists \,  \ottmv{x_{{\mathrm{1}}}} \mathord:\allowbreak \ottnt{T_{{\mathrm{1}}}}   \ottsym{,} \, .. \, \ottsym{,}   \ottmv{x_{\ottmv{n}}} \mathord:\allowbreak \ottnt{T_{\ottmv{n}}}   \ottsym{.}  \varphi\) for
\(\exists  \ottmv{x_{{\mathrm{1}}}} \mathord:\allowbreak \ottnt{T_{{\mathrm{1}}}} . \dots \exists  \ottmv{x_{\ottmv{n}}} \mathord:\allowbreak \ottnt{T_{\ottmv{n}}} . \varphi\), etc.
A typing environment, ranged over by \(\Gamma\), is a sequence of type binding.
We assume all variables in \(\Gamma\) are distinct.  We abuse a comma to
concatenate typing environments, e.g., \(\Gamma_{{\mathrm{1}}}  \ottsym{,}  \Gamma_{{\mathrm{2}}}\).  We also use a type
binding stack \(\Upsilon\) as a typing environment, explicitly denoted by
\( \widehat{ \Upsilon } \), which is defined as \( \widehat{ \ddagger }  = \ottkw{empty}\) and
\( \widehat{  \ottmv{x} \mathord:\allowbreak \ottnt{T}   \triangleright  \Upsilon }  =  \ottmv{x} \mathord:\allowbreak \ottnt{T}   \ottsym{,}   \widehat{ \Upsilon } \).

Well-sorted terms and formulae are defined by the judgments \(\Gamma  \vdash  \ottnt{t}  \ottsym{:}  \ottnt{T}\)
and \( \Gamma   \vdash   \varphi  : \mathord{*} \), respectively.  The former means that the term \(\ottnt{t}\) is a
well-sorted term of the sort \(\ottnt{T}\)
under the typing environment \(\Gamma\) and the latter that the formula \(\varphi\) is well-sorted
under the typing environment \(\Gamma\), respectively.  The derivation rules for
each judgment, shown in \figref{ftyping}, are straightforward.  Note that
well-sortedness depends on the simple type system via \ruleref{WT-Val}.

Let a \emph{value assignment} \(\sigma\) be a mapping from variables to values.
We write \( \sigma  [  \ottmv{x}  \mapsto  \ottnt{V}  ] \) to denote the value assignment which maps
\(\ottmv{x}\) to \(\ottnt{V}\) and otherwise is identical to \(\sigma\). As we are
interested in well-sorted formulae, we consider a value assignment
that respects a typing environment, as follows.

\begin{prop}[type=definition]{def/typed/assignment}
  A value assignment \(\sigma\) is \emph{typed} under a typing environment \(\Gamma\),
  denoted by \(\sigma  \ottsym{:}  \Gamma\), iff \( \sigma  \ottsym{(}  \ottmv{x}  \ottsym{)}  :  \ottnt{T} \) for every \(  \ottmv{x} \mathord:\allowbreak \ottnt{T}   \in  \Gamma \).
\end{prop}

Now we define the
semantics of the well-sorted terms and formulae in a standard manner as follows.

\begin{prop}[name=Semantics of terms, type=definition]{def/sem/term}
  For a typed value assignment \(\sigma  \ottsym{:}  \Gamma\).  The semantics \( \lBrack  \ottnt{t}  \rBrack_{ \sigma : \Gamma } \) of term \(\ottnt{t}\)
    under typed value assignment \(\sigma  \ottsym{:}  \Gamma\) is defined as follows:
  \begin{align*}
     \lBrack  \ottmv{x}  \rBrack_{ \sigma : \Gamma }  = \sigma  \ottsym{(}  \ottmv{x}  \ottsym{)} & \qquad
     \lBrack  \ottnt{V}  \rBrack_{ \sigma : \Gamma }  = \ottnt{V} \\
     \lBrack   \texttt{Transfer}  (  \ottnt{t_{{\mathrm{1}}}} ,  \ottnt{t_{{\mathrm{2}}}} ,  \ottnt{t_{{\mathrm{3}}}}  )   \rBrack_{ \sigma : \Gamma }  &=  \texttt{Transfer}  (   \lBrack  \ottnt{t_{{\mathrm{1}}}}  \rBrack_{ \sigma : \Gamma }  ,   \lBrack  \ottnt{t_{{\mathrm{2}}}}  \rBrack_{ \sigma : \Gamma }  ,   \lBrack  \ottnt{t_{{\mathrm{3}}}}  \rBrack_{ \sigma : \Gamma }   )  \\
     \lBrack  \ottsym{(}  \ottnt{t_{{\mathrm{1}}}}  \ottsym{,}  \ottnt{t_{{\mathrm{2}}}}  \ottsym{)}  \rBrack_{ \sigma : \Gamma }  &= \ottsym{(}   \lBrack  \ottnt{t_{{\mathrm{1}}}}  \rBrack_{ \sigma : \Gamma }   \ottsym{,}   \lBrack  \ottnt{t_{{\mathrm{2}}}}  \rBrack_{ \sigma : \Gamma }   \ottsym{)} \\
     \lBrack  \ottnt{t_{{\mathrm{1}}}}  ::  \ottnt{t_{{\mathrm{2}}}}  \rBrack_{ \sigma : \Gamma }  &=  \lBrack  \ottnt{t_{{\mathrm{1}}}}  \rBrack_{ \sigma : \Gamma }   ::   \lBrack  \ottnt{t_{{\mathrm{2}}}}  \rBrack_{ \sigma : \Gamma }  \\
     \lBrack  \ottnt{t_{{\mathrm{1}}}}  \ottsym{+}  \ottnt{t_{{\mathrm{2}}}}  \rBrack_{ \sigma : \Gamma }  & =  \lBrack  \ottnt{t_{{\mathrm{1}}}}  \rBrack_{ \sigma : \Gamma }   \ottsym{+}   \lBrack  \ottnt{t_{{\mathrm{2}}}}  \rBrack_{ \sigma : \Gamma } .
  \end{align*}
\end{prop}

\begin{prop}[name=Semantics of formulae, type=definition]{def/sem/formula}
  For a typed value assignment \(\sigma  \ottsym{:}  \Gamma\), a \emph{valid} well-sorted formula
  \(\varphi\) under \(\Gamma\) is denoted by \(\sigma  \ottsym{:}  \Gamma  \models  \varphi\) and defined as follows.
  \begin{itemize}
  \item \(\sigma  \ottsym{:}  \Gamma  \models   \top \).
  \item \(\sigma  \ottsym{:}  \Gamma  \models  \ottnt{t_{{\mathrm{1}}}}  \ottsym{=}  \ottnt{t_{{\mathrm{2}}}}\) iff \( \lBrack  \ottnt{t_{{\mathrm{1}}}}  \rBrack_{ \sigma : \Gamma }   \ottsym{=}   \lBrack  \ottnt{t_{{\mathrm{2}}}}  \rBrack_{ \sigma : \Gamma } \).
  \item \(\sigma  \ottsym{:}  \Gamma  \models   \ottkw{call} ( \ottnt{t_{{\mathrm{1}}}} ,  \ottnt{t_{{\mathrm{2}}}} ) =  \ottnt{t_{{\mathrm{3}}}} \) iff \( \lBrack  \ottnt{t_{{\mathrm{1}}}}  \rBrack_{ \sigma : \Gamma }   \ottsym{=}   \langle  \mathit{IS}  \rangle \) and
    \( \lBrack  \ottnt{t_{{\mathrm{2}}}}  \rBrack_{ \sigma : \Gamma }   \triangleright  \ddagger  \vdash  \mathit{IS}  \Downarrow   \lBrack  \ottnt{t_{{\mathrm{3}}}}  \rBrack_{ \sigma : \Gamma }   \triangleright  \ddagger\).
  \item \(\sigma  \ottsym{:}  \Gamma  \models  \neg \, \varphi\) iff \(\sigma  \ottsym{:}  \Gamma  \not\models  \varphi\).
  \item \(\sigma  \ottsym{:}  \Gamma  \models  \varphi_{{\mathrm{1}}}  \vee  \varphi_{{\mathrm{2}}}\) iff \(\sigma  \ottsym{:}  \Gamma  \models  \varphi_{{\mathrm{1}}}\) or \(\sigma  \ottsym{:}  \Gamma  \models  \varphi_{{\mathrm{2}}}\).
  \item \(\sigma  \ottsym{:}  \Gamma  \models  \exists \,  \ottmv{x} \mathord:\allowbreak \ottnt{T}   \ottsym{.}  \varphi\) iff \( \sigma  [  \ottmv{x}  \mapsto  \ottnt{V}  ]   \ottsym{:}  \Gamma  \ottsym{,}   \ottmv{x} \mathord:\allowbreak \ottnt{T}   \models  \varphi\) for some
    \(\ottnt{V}\).
  \end{itemize}

  We write \(\Gamma  \models  \varphi\) iff \(\sigma  \ottsym{:}  \Gamma  \models  \varphi\) for any \(\sigma\).
\end{prop}

\subsubsection{Typing Rules}

\begin{figure}
  \centering
  \input{figures/typing}
  \caption{Typing rules (I)}
  \label{fig:typing}
\end{figure}

\begin{figure}
  \centering
  \input{figures/typing2}
  \caption{Typing rules (II)}
  \label{fig:typing2}
\end{figure}

The type system is defined by subtyping and typing: a subtyping judgment is of
the form \(\Gamma  \vdash  \Phi_{{\mathrm{1}}}  \ottsym{<:}  \Phi_{{\mathrm{2}}}\), which means stack type \(\Phi_{{\mathrm{1}}}\) is a
subtype of \(\Phi_{{\mathrm{2}}}\) under \(\Gamma\), and a type judgment for instructions
(resp. instruction
sequences) % \YN{Are these resp.s for readability since IS is included in I.}
is of the form \( \Gamma \vdash \Phi_{{\mathrm{1}}} \; \ottnt{I} \; \Phi_{{\mathrm{2}}} \) (resp.  \( \Gamma \vdash \Phi_{{\mathrm{1}}} \; \mathit{IS} \; \Phi_{{\mathrm{2}}} \)),
which means that if \(\ottnt{I}\) (resp. \(\mathit{IS}\)) is executed under a stack
satisfying \(\Phi_{{\mathrm{1}}}\), the resulting stack (if terminates) satisfies
\(\Phi_{{\mathrm{2}}}\).  We often call \(\Phi_{{\mathrm{1}}}\) \emph{pre-condition} and \(\Phi_{{\mathrm{2}}}\)
\emph{post-condition}, following the terminology of Hoare logic.  Note that the
scopes of variables declared in \(\Gamma\) include \(\Phi_{{\mathrm{1}}}\) and \(\Phi_{{\mathrm{2}}}\)
but those bound in \(\Phi_{{\mathrm{1}}}\) \emph{do not} include \(\Phi_{{\mathrm{2}}}\).  To express
the relationship between the initial and final stacks, we use the type
environment \(\Gamma\).  In writing down concrete specifications, it is sometimes
convenient to allow the scopes of variables bound in \(\Phi_{{\mathrm{1}}}\) to include
\(\Phi_{{\mathrm{2}}}\), but we find that it would clutter the presentation of typing
rules.

In our type system, subtyping is defined semantically as follows.
\begin{prop}[type=definition, name=Subtyping relation]{def/subty}
  A refinement stack type \(\ottsym{\{}  \Upsilon  \mid  \varphi_{{\mathrm{1}}}  \ottsym{\}}\) is called \emph{subtype} of a
  refinement stack type \(\ottsym{\{}  \Upsilon  \mid  \varphi_{{\mathrm{2}}}  \ottsym{\}}\) under a typing environment \(\Gamma\),
  denoted by \(\Gamma  \vdash  \ottsym{\{}  \Upsilon  \mid  \varphi_{{\mathrm{1}}}  \ottsym{\}}  \ottsym{<:}  \ottsym{\{}  \Upsilon  \mid  \varphi_{{\mathrm{2}}}  \ottsym{\}}\), iff
  \(\Gamma  \ottsym{,}   \widehat{ \Upsilon }   \models  \varphi_{{\mathrm{1}}}  \implies  \varphi_{{\mathrm{2}}}\).
\end{prop}

We show the typing rules in \figref{typing} and \figref{typing2}.  It is easy to
observe that the type binding stack parts in the pre- and post-conditions follows
the simple type system.  We will focus on predicate parts below.
\begin{itemize}
\item \ruleref{RT-Dip} means that \(\ottkw{DIP} \, \mathit{IS}\) is well typed if the body
  \(\mathit{IS}\) is typed under the stack type obtained by removing the top element.
  However, since a property \(\varphi\) for the initial stack relies on the
  popped value \(\ottmv{x}\), we keep the binding in the typing environment.

\item \ruleref{RT-If} means that the instruction is well typed if both branches
  have the same post-condition; the pre-conditions of the branches are
  strengthened by the assumptions that the top of the input stack is \lstinline{True}
  (\(\ottmv{x}  \neq  0\)) and \lstinline{False} (\(\ottmv{x}  \ottsym{=}  0\)).  The variable \(\ottmv{x}\) is
  existentially quantified because the top element will be removed before the
  execution of either branch.

\item \ruleref{RT-Loop} is similar to the proof rule for while-loops in Hoare
  logic.  The formula \(\varphi\) is a loop invariant.  Since the body of
  \(\ottkw{LOOP}\) is executed while the stack top is nonzero, the pre-condition for
  the body \(\mathit{IS}\) is strengthened by \(\ottmv{x}  \neq  0\), whereas the
  post-condition of \(\ottkw{LOOP} \, \mathit{IS}\) is strengthened by \(\ottmv{x}  \ottsym{=}  0\).

\item \ruleref{RT-Lambda} is for the instruction to push a first-class function
  onto the operand stack.  The premise of the rule means that the body $\mathit{IS}$
  takes a value (named \(\ottmv{y_{{\mathrm{1}}}}\)) of type \(\ottnt{T_{{\mathrm{1}}}}\) that satisfies
  \(\varphi_{{\mathrm{1}}}\) and outputs a value (named \(\ottmv{y_{{\mathrm{2}}}}\)) of type \(\ottnt{T_{{\mathrm{2}}}}\) that
  satisfies \(\varphi_{{\mathrm{2}}}\) (if it terminates).  The post-condition in the
  conclusion expresses, by using \texttt{call}, that the function \(\ottmv{x}\) has
  the property above.  The extra variable \(\ottmv{y'_{{\mathrm{1}}}}\) in the type environment of
  the premise is an alias of \(\ottmv{y_{{\mathrm{1}}}}\); being a variable declared in the type
  environment \(\ottmv{y'_{{\mathrm{1}}}}\) can appear in both \(\varphi_{{\mathrm{1}}}\) and
  \(\varphi_{{\mathrm{2}}}\)\footnote{%
    The scope of a variable in a refinement stack type is its predicate part and
    so $\ottmv{y_{{\mathrm{1}}}}$ cannot appear in the post-condition.} and can describe the
  relationship between the input and output of the function.

\item \ruleref{RT-Exec} just adds to the post-condition
  \( \ottkw{call} ( \ottmv{x_{{\mathrm{2}}}} ,  \ottmv{x_{{\mathrm{1}}}} ) =  \ottmv{x_{{\mathrm{3}}}} \), meaning the result of a call to the function
  \(\ottmv{x_{{\mathrm{2}}}}\) with \(\ottmv{x_{{\mathrm{1}}}}\) as an argument yields \(\ottmv{x_{{\mathrm{3}}}}\).  It may look
  simpler than expected; the crux here is that \(\varphi\) is expected to imply
  \(\forall \,  \ottmv{x_{{\mathrm{1}}}} \mathord:\allowbreak \ottnt{T_{{\mathrm{1}}}}   \ottsym{,}   \ottmv{x_{{\mathrm{3}}}} \mathord:\allowbreak \ottnt{T_{{\mathrm{2}}}}   \ottsym{.}  \varphi_{{\mathrm{1}}}  \wedge   \ottkw{call} ( \ottmv{x_{{\mathrm{2}}}} ,  \ottmv{x_{{\mathrm{1}}}} ) =  \ottmv{x_{{\mathrm{3}}}}   \implies  \varphi_{{\mathrm{2}}}\), where
  $\varphi_{{\mathrm{1}}}$ and $\varphi_{{\mathrm{2}}}$ represent the pre- and post-conditions,
  respectively, of function \(\ottmv{x_{{\mathrm{2}}}}\).  If \(\ottmv{x_{{\mathrm{1}}}}\) satisfies \(\varphi_{{\mathrm{1}}}\),
  then we can derive that \(\varphi_{{\mathrm{2}}}\) holds.

\item \ruleref{RT-Sub} is the rule for subsumption to strengthening the
  pre-condition and weakening the post-condition.
\end{itemize}

\subsection{Properties}\label{sec:property}

In this section, we show \emph{soundness} of our type system.  Informally, what
we show is that, for a well-typed program, if we execute it
under a stack which satisfies the pre-condition of the typing, then (if the
evaluation halts) the resulting stack satisfies the post-condition of the
typing.  We only sketch proofs with important lemmas.
The detailed proofs are found in \secref{detailedproofs}.

To state the soundness formally, we give additional definitions.

\begin{prop}[name=Free variables, type=definition]{def/fvars}
  The set of free variables in \(\varphi\) is denoted by \( \text{fvars}( \varphi ) \).
\end{prop}

% \begin{prop}[name=Domain of a typing environment, type=definition]{def/dom}
%   The \emph{domain} of typing environment \(\Gamma\), denoted by \( \text{dom}( \Gamma ) \),
%   is the set of variables in \(\Gamma\).
% \end{prop}

\begin{prop}[name=Erasure, type=definition]{def/unref}
  We define \( \lfloor  \Phi  \rfloor \), which is the simple stack type obtained by
  erasing predicates from \(\Phi\), as follows.
  \begin{gather*}
      \lfloor  \ottsym{\{}  \ddagger  \mid  \varphi  \ottsym{\}}  \rfloor  \ottsym{=} \ddagger  \qquad   \lfloor  \ottsym{\{}   \ottmv{x} \mathord:\allowbreak \ottnt{T}   \triangleright  \Upsilon  \mid  \varphi  \ottsym{\}}  \rfloor  \ottsym{=} \ottnt{T}  \triangleright   \lfloor  \ottsym{\{}  \Upsilon  \mid  \varphi  \ottsym{\}}  \rfloor  
  \end{gather*}
\end{prop}

% \begin{prop}[name=Substitution, type=definition]{def/subst}
%   We write a substitution for the free occurrence of \(\ottmv{x}\) in \(\varphi\) with
%   \(\ottnt{V}\) as \(\texttt{\textcolor{red}{<<no parses (char 2): p.***\{V/x\} >>}}\).
%   \AI{You don't really define...  Maybe remove the prop environment?}
% \end{prop}

\begin{figure}
  \centering
\begin{flushleft}
  \noindent {Stack Typing} \fbox{\( \ottnt{S}  :  \bar{T} \)}
\end{flushleft}\vspace{-2em}
  \input{figures/sttyping}
  % \caption{Stack typing}
  % \label{fig:sttyping}
  \vspace{2ex}
% \end{figure}

% \begin{figure}
  \centering
\begin{flushleft}
  \noindent {Refinement Stack Typing} \fbox{\(\sigma  \ottsym{:}  \Gamma  \models  \ottnt{S}  \ottsym{:}  \Phi\)}
\end{flushleft}\vspace{-2em}
  \input{figures/semty}
  \caption{Simple and refinement stack typing}
  \label{fig:semty}
\end{figure}

\begin{prop}[name=Stack typing, type=definition]{def/sttyping}
  \emph{Stack typing} \( \ottnt{S}  :  \bar{T} \) and \emph{refinement stack typing}
  \(\sigma  \ottsym{:}  \Gamma  \models  \ottnt{S}  \ottsym{:}  \Phi\) are defined by the
  rules in \figref{semty}.
\end{prop}

Note that the definition of refinement stack typing follows the informal
explanation of the refinement stack types in \secref{system/refinement}.

We start from soundness of simple type system as follows, that is, for a simply
well-typed instruction sequence \(\bar{T}_{{\mathrm{1}}}  \vdash  \mathit{IS}  \Rightarrow  \bar{T}_{{\mathrm{2}}}\), if evaluation starts from
correct stack \(\ottnt{S_{{\mathrm{1}}}}\), that is \( \ottnt{S_{{\mathrm{1}}}}  :  \bar{T}_{{\mathrm{1}}} \), and results in a stack
\(\ottnt{S_{{\mathrm{2}}}}\); then \(\ottnt{S_{{\mathrm{2}}}}\) respects \(\bar{T}_{{\mathrm{2}}}\), that is \( \ottnt{S_{{\mathrm{2}}}}  :  \bar{T}_{{\mathrm{2}}} \).
This lemma is not only just a desirable property but also one we use for proving
the soundness of the refinement type system in the case \(\ottkw{EXEC}\).

\begin{prop}[name=Soundness of the Simple Type System]{styping/sound'}
  If \(\bar{T}_{{\mathrm{1}}}  \vdash  \mathit{IS}  \Rightarrow  \bar{T}_{{\mathrm{2}}}\), \(\ottnt{S_{{\mathrm{1}}}}  \vdash  \mathit{IS}  \Downarrow  \ottnt{S_{{\mathrm{2}}}}\), and \( \ottnt{S_{{\mathrm{1}}}}  :  \bar{T}_{{\mathrm{1}}} \),
  then \( \ottnt{S_{{\mathrm{2}}}}  :  \bar{T}_{{\mathrm{2}}} \).
  \proof Proved with a similar statement
  \begin{quotation}
    If \(\bar{T}_{{\mathrm{1}}}  \vdash  \mathit{IS}  \Rightarrow  \bar{T}_{{\mathrm{2}}}\), \(\ottnt{S_{{\mathrm{1}}}}  \vdash  \mathit{IS}  \Downarrow  \ottnt{S_{{\mathrm{2}}}}\), and \( \ottnt{S_{{\mathrm{1}}}}  :  \bar{T}_{{\mathrm{1}}} \),
  then \( \ottnt{S_{{\mathrm{2}}}}  :  \bar{T}_{{\mathrm{2}}} \)
  \end{quotation}
 for a single instruction by simultaneous induction on \(\bar{T}_{{\mathrm{1}}}  \vdash  \mathit{IS}  \Rightarrow  \bar{T}_{{\mathrm{2}}}\) and \(\bar{T}_{{\mathrm{1}}}  \vdash  \ottnt{I}  \Rightarrow  \bar{T}_{{\mathrm{2}}}\).
\end{prop}

% \subsection{Properties for assertion language}

% We show several lemmas for the assertion language.  Actually, these are expected
% and standard ones for a first order logic language.  We put those here just for
% comprehensive proofs.

% % \input{proofs/fol}

% \subsection{Properties for subtyping}

% Now we eventually proofs the soundness of the refinement type system.  Firstly,
% we show soundness of subtyping, that is, if a stack satisfies a subtype of some
% refinement stack type, the stack satisfies the refinement stack type.

% % \input{proofs/subty}

% \subsection{Properties for semantics of refinement stack types}

% We put several lemmas for refinement stack typing.  Mainly, these are used in
% the soundness proof of the refinement type system.

% % \input{proofs/semty}

% \subsection{Properties for typing}

We state the main theorem as follows.

\begin{prop}[name=Soundness of the Refinement Type System, type=theorem]{typing/sound'}
  If \( \Gamma \vdash \Phi_{{\mathrm{1}}} \; \mathit{IS} \; \Phi_{{\mathrm{2}}} \), \(\ottnt{S_{{\mathrm{1}}}}  \vdash  \mathit{IS}  \Downarrow  \ottnt{S_{{\mathrm{2}}}}\), and \(\sigma  \ottsym{:}  \Gamma  \models  \ottnt{S_{{\mathrm{1}}}}  \ottsym{:}  \Phi_{{\mathrm{1}}}\), then
  \(\sigma  \ottsym{:}  \Gamma  \models  \ottnt{S_{{\mathrm{2}}}}  \ottsym{:}  \Phi_{{\mathrm{2}}}\).
\end{prop}

The proof is close to a proof of soundness of Hoare logic, with a few
extra complications due to the presence of first-class functions.  One
of the key lemmas is the following one, which states that a value
assignment can be represented by a logical formula or a stack element:

\begin{prop}{subty/exists=>subst'}
  The following statements are equivalent:
  \begin{statements}
  \item \(\sigma  \ottsym{:}  \Gamma  \models  \ottnt{S}  \ottsym{:}  \ottsym{\{}  \Upsilon  \mid  \exists \,  \ottmv{x} \mathord:\allowbreak \ottnt{T}   \ottsym{.}  \varphi  \wedge  \ottmv{x}  \ottsym{=}  \ottnt{V}  \ottsym{\}}\);
  \item \( \sigma  [  \ottmv{x}  \mapsto  \ottnt{V}  ]   \ottsym{:}  \Gamma  \ottsym{,}   \ottmv{x} \mathord:\allowbreak \ottnt{T}   \models  \ottnt{S}  \ottsym{:}  \ottsym{\{}  \Upsilon  \mid  \varphi  \ottsym{\}}\); and
  \item \(\sigma  \ottsym{:}  \Gamma  \models  \ottnt{V}  \triangleright  \ottnt{S}  \ottsym{:}  \ottsym{\{}   \ottmv{x} \mathord:\allowbreak \ottnt{T}   \triangleright  \Upsilon  \mid  \varphi  \ottsym{\}}\).
  \end{statements}
\end{prop}

Then, we prove a few
lemmas related to \(\ottkw{LOOP}\) (Lemma~\ref{prop:soundness/loop'}),
\(\ottkw{ITER}\) (Lemma~\ref{prop:soundness/iter'}), predicate
\texttt{call} (Lemmas~\ref{prop:soundness/lambda'} and
\ref{prop:soundness/exec'}), and subtyping
(Lemma~\ref{prop:soundness/subtyping'}).

\begin{prop}{soundness/loop'}
  Suppose $\mathit{IS}$ satisfies that
$\ottnt{S_{{\mathrm{1}}}}  \vdash  \mathit{IS}  \Downarrow  \ottnt{S_{{\mathrm{2}}}}$ and $\sigma  \ottsym{:}  \Gamma  \models  \ottnt{S_{{\mathrm{1}}}}  \ottsym{:}  \ottsym{\{}  \Upsilon  \mid  \exists \,  \ottmv{x} \mathord:\allowbreak \ottkw{int}   \ottsym{.}  \varphi  \wedge  \ottmv{x}  \neq  0  \ottsym{\}}$ imply $\sigma  \ottsym{:}  \Gamma  \models  \ottnt{S_{{\mathrm{2}}}}  \ottsym{:}  \ottsym{\{}   \ottmv{x} \mathord:\allowbreak \ottkw{int}   \triangleright  \Upsilon  \mid  \varphi  \ottsym{\}}$ for any $\ottnt{S_{{\mathrm{1}}}}$ and $\ottnt{S_{{\mathrm{2}}}}$.
  If $\ottnt{S_{{\mathrm{1}}}}  \vdash  \ottkw{LOOP} \, \mathit{IS}  \Downarrow  \ottnt{S_{{\mathrm{2}}}}$ and $\sigma  \ottsym{:}  \Gamma  \models  \ottnt{S_{{\mathrm{1}}}}  \ottsym{:}  \ottsym{\{}   \ottmv{x} \mathord:\allowbreak \ottkw{int}   \triangleright  \Upsilon  \mid  \varphi  \ottsym{\}}$, 
  then \(\sigma  \ottsym{:}  \Gamma  \models  \ottnt{S_{{\mathrm{2}}}}  \ottsym{:}  \ottsym{\{}  \Upsilon  \mid  \exists \,  \ottmv{x} \mathord:\allowbreak \ottkw{int}   \ottsym{.}  \varphi  \wedge  \ottmv{x}  \ottsym{=}  0  \ottsym{\}}\).

  \proof     By induction on the derivation of $\ottnt{S_{{\mathrm{1}}}}  \vdash  \ottkw{LOOP} \, \mathit{IS}  \Downarrow  \ottnt{S_{{\mathrm{2}}}}$.
\end{prop}

\begin{prop}{soundness/iter'}
  Suppose $ \ottmv{x_{{\mathrm{1}}}}  \notin   \text{fvars}( \varphi )  $, $ \ottmv{x_{{\mathrm{2}}}}  \notin   \text{fvars}( \varphi )  $, and that
  $\ottnt{S'_{{\mathrm{1}}}}  \vdash  \mathit{IS}  \Downarrow  \ottnt{S'_{{\mathrm{2}}}}$ and $\sigma'  \ottsym{:}  \Gamma  \ottsym{,}   \ottmv{x_{{\mathrm{2}}}} \mathord:\allowbreak \ottnt{T} \, \ottkw{list}   \models  \ottnt{S'_{{\mathrm{1}}}}  \ottsym{:}  \ottsym{\{}   \ottmv{x_{{\mathrm{1}}}} \mathord:\allowbreak \ottnt{T}   \triangleright  \Upsilon  \mid  \exists \,  \ottmv{x} \mathord:\allowbreak \ottnt{T} \, \ottkw{list}   \ottsym{.}  \varphi  \wedge  \ottmv{x_{{\mathrm{1}}}}  ::  \ottmv{x_{{\mathrm{2}}}}  \ottsym{=}  \ottmv{x}  \ottsym{\}}$  imply $\sigma'  \ottsym{:}  \Gamma  \ottsym{,}   \ottmv{x_{{\mathrm{2}}}} \mathord:\allowbreak \ottnt{T} \, \ottkw{list}   \models  \ottnt{S'_{{\mathrm{2}}}}  \ottsym{:}  \ottsym{\{}  \Upsilon  \mid  \exists \,  \ottmv{x} \mathord:\allowbreak \ottnt{T} \, \ottkw{list}   \ottsym{.}  \varphi  \wedge  \ottmv{x_{{\mathrm{2}}}}  \ottsym{=}  \ottmv{x}  \ottsym{\}}$
  for any $\ottnt{S'_{{\mathrm{1}}}}$, $\ottnt{S'_{{\mathrm{2}}}}$, and $\sigma'$.
  If $\ottnt{S_{{\mathrm{1}}}}  \vdash  \ottkw{ITER} \, \mathit{IS}  \Downarrow  \ottnt{S_{{\mathrm{2}}}}$ and $\sigma  \ottsym{:}  \Gamma  \models  \ottnt{S_{{\mathrm{1}}}}  \ottsym{:}  \ottsym{\{}   \ottmv{x} \mathord:\allowbreak \ottnt{T} \, \ottkw{list}   \triangleright  \Upsilon  \mid  \varphi  \ottsym{\}}$, 
  then \(\sigma  \ottsym{:}  \Gamma  \models  \ottnt{S_{{\mathrm{2}}}}  \ottsym{:}  \ottsym{\{}  \Upsilon  \mid  \exists \,  \ottmv{x} \mathord:\allowbreak \ottnt{T} \, \ottkw{list}   \ottsym{.}  \varphi  \wedge  \ottmv{x}  \ottsym{=}  \ottsym{[}  \ottsym{]}  \ottsym{\}}\).

  \proof By induction on the derivation of  $\ottnt{S_{{\mathrm{1}}}}  \vdash  \ottkw{ITER} \, \mathit{IS}  \Downarrow  \ottnt{S_{{\mathrm{2}}}}$.
\end{prop}

\begin{prop}{soundness/lambda'}
  If $\ottmv{y_{{\mathrm{1}}}}  \neq  \ottmv{y_{{\mathrm{2}}}}$, $  \ottmv{y'_{{\mathrm{1}}}} \mathord:\allowbreak \ottnt{T_{{\mathrm{1}}}}   \ottsym{,}   \ottmv{y_{{\mathrm{1}}}} \mathord:\allowbreak \ottnt{T_{{\mathrm{1}}}}    \vdash   \varphi_{{\mathrm{1}}}  : \mathord{*} $,  $  \ottmv{y'_{{\mathrm{1}}}} \mathord:\allowbreak \ottnt{T_{{\mathrm{1}}}}   \ottsym{,}   \ottmv{y_{{\mathrm{2}}}} \mathord:\allowbreak \ottnt{T_{{\mathrm{2}}}}    \vdash   \varphi_{{\mathrm{2}}}  : \mathord{*} $,
    $  \langle  \mathit{IS}  \rangle   :  \ottnt{T_{{\mathrm{1}}}}  \to  \ottnt{T_{{\mathrm{2}}}} $, and \[
      \begin{split}
        \text{for any } \ottnt{V_{{\mathrm{1}}}}, \ottnt{V_{{\mathrm{2}}}}, \sigma,
        \text{ if } & \ottnt{V_{{\mathrm{1}}}}  \triangleright  \ddagger  \vdash  \mathit{IS}  \Downarrow  \ottnt{V_{{\mathrm{2}}}}  \triangleright  \ddagger \text{ and } \\ 
         &\sigma  \ottsym{:}   \ottmv{y'_{{\mathrm{1}}}} \mathord:\allowbreak \ottnt{T_{{\mathrm{1}}}}   \models  \ottnt{V_{{\mathrm{1}}}}  \triangleright  \ddagger  \ottsym{:}  \ottsym{\{}   \ottmv{y_{{\mathrm{1}}}} \mathord:\allowbreak \ottnt{T_{{\mathrm{1}}}}   \triangleright  \ddagger  \mid  \ottmv{y'_{{\mathrm{1}}}}  \ottsym{=}  \ottmv{y_{{\mathrm{1}}}}  \wedge  \varphi_{{\mathrm{1}}}  \ottsym{\}} \\
        \text{ then } & \sigma  \ottsym{:}   \ottmv{y'_{{\mathrm{1}}}} \mathord:\allowbreak \ottnt{T_{{\mathrm{1}}}}   \models  \ottnt{V_{{\mathrm{2}}}}  \triangleright  \ddagger  \ottsym{:}  \ottsym{\{}   \ottmv{y_{{\mathrm{2}}}} \mathord:\allowbreak \ottnt{T_{{\mathrm{2}}}}   \triangleright  \ddagger  \mid  \varphi_{{\mathrm{2}}}  \ottsym{\}},
      \end{split}
    \]
  then
  \(\Gamma  \models  \forall \,  \ottmv{y'_{{\mathrm{1}}}} \mathord:\allowbreak \ottnt{T_{{\mathrm{1}}}}   \ottsym{,}   \ottmv{y_{{\mathrm{1}}}} \mathord:\allowbreak \ottnt{T_{{\mathrm{1}}}}   \ottsym{,}   \ottmv{y_{{\mathrm{2}}}} \mathord:\allowbreak \ottnt{T_{{\mathrm{2}}}}   \ottsym{.}  \ottmv{y'_{{\mathrm{1}}}}  \ottsym{=}  \ottmv{y_{{\mathrm{1}}}}  \wedge  \varphi_{{\mathrm{1}}}  \wedge   \ottkw{call} (  \langle  \mathit{IS}  \rangle  ,  \ottmv{y'_{{\mathrm{1}}}} ) =  \ottmv{y_{{\mathrm{2}}}}   \implies  \varphi_{{\mathrm{2}}}\) for any \(\Gamma\).

  \proof By the definition of the semantics of $\mathtt{call}$.
\end{prop}

\begin{prop}{soundness/exec'}
  If $\ottnt{V_{{\mathrm{1}}}}  \triangleright  \ddagger  \vdash  \mathit{IS}  \Downarrow  \ottnt{V_{{\mathrm{2}}}}  \triangleright  \ddagger$,
    $ \ottnt{V_{{\mathrm{1}}}}  :  \ottnt{T_{{\mathrm{1}}}} $,
    $ \ottnt{V_{{\mathrm{2}}}}  :  \ottnt{T_{{\mathrm{2}}}} $, and
    $  \langle  \mathit{IS}  \rangle   :  \ottnt{T_{{\mathrm{1}}}}  \to  \ottnt{T_{{\mathrm{2}}}} $, then
  \(\Gamma  \models   \ottkw{call} (  \langle  \mathit{IS}  \rangle  ,  \ottnt{V_{{\mathrm{1}}}} ) =  \ottnt{V_{{\mathrm{2}}}} \) for any \(\Gamma\).

  \proof By the definition of the semantics of $\mathtt{call}$.
\end{prop}

\begin{prop}{soundness/subtyping'}
  If $\Gamma  \vdash  \Phi_{{\mathrm{1}}}  \ottsym{<:}  \Phi_{{\mathrm{2}}}$ and $\sigma  \ottsym{:}  \Gamma  \models  \ottnt{S}  \ottsym{:}  \Phi_{{\mathrm{1}}}$,
  then \(\sigma  \ottsym{:}  \Gamma  \models  \ottnt{S}  \ottsym{:}  \Phi_{{\mathrm{2}}}\).

  \proof Straightforward from \propref{def/subty}.
\end{prop}

\paragraph{Proof of Theorem~\ref{prop:typing/sound'}.}
It is proved together with a similar statement
\begin{quotation}
  If \( \Gamma \vdash \Phi_{{\mathrm{1}}} \; \ottnt{I} \; \Phi_{{\mathrm{2}}} \), \(\ottnt{S_{{\mathrm{1}}}}  \vdash  \ottnt{I}  \Downarrow  \ottnt{S_{{\mathrm{2}}}}\), and \(\sigma  \ottsym{:}  \Gamma  \models  \ottnt{S_{{\mathrm{1}}}}  \ottsym{:}  \Phi_{{\mathrm{1}}}\), then
  \(\sigma  \ottsym{:}  \Gamma  \models  \ottnt{S_{{\mathrm{2}}}}  \ottsym{:}  \Phi_{{\mathrm{2}}}\).
\end{quotation}
for a single instruction by simultaneous induction on \( \Gamma \vdash \Phi_{{\mathrm{1}}} \; \mathit{IS} \; \Phi_{{\mathrm{2}}} \) and \( \Gamma \vdash \Phi_{{\mathrm{1}}} \; \ottnt{I} \; \Phi_{{\mathrm{2}}} \)
with case analysis on the last typing rule used.
We show a few representative cases.

\def\hyp#1{\relax}
\def\hypref#1{\relax}
\begin{description}
\item[Case \ruleref{RT-Dip}:]
We have
\( \ottnt{I} \ottsym{=} \ottkw{DIP} \, \mathit{IS} \) and \( \Phi_{{\mathrm{1}}} \ottsym{=} \ottsym{\{}   \ottmv{x} \mathord:\allowbreak \ottnt{T}   \triangleright  \Upsilon  \mid  \varphi  \ottsym{\}} \) and
\( \Phi_{{\mathrm{2}}} \ottsym{=} \ottsym{\{}   \ottmv{x} \mathord:\allowbreak \ottnt{T}   \triangleright  \Upsilon'  \mid  \varphi'  \ottsym{\}} \) and \( \Gamma  \ottsym{,}   \ottmv{x} \mathord:\allowbreak \ottnt{T}  \vdash \ottsym{\{}  \Upsilon  \mid  \varphi  \ottsym{\}} \; \mathit{IS} \; \ottsym{\{}  \Upsilon'  \mid  \varphi'  \ottsym{\}} \)
    for some \(\mathit{IS}\), \(\ottmv{x}\), \(\ottnt{T}\), \(\Upsilon\), \(\Upsilon'\),
    \(\varphi\), and \(\varphi'\).
    By \ruleref{E-Dip}, we have
      \( \ottnt{S_{{\mathrm{1}}}} \ottsym{=} \ottnt{V}  \triangleright  \ottnt{S'_{{\mathrm{1}}}} \) and
      \( \ottnt{S_{{\mathrm{2}}}} \ottsym{=} \ottnt{V}  \triangleright  \ottnt{S'_{{\mathrm{2}}}} \) and 
      \(\ottnt{S'_{{\mathrm{1}}}}  \vdash  \mathit{IS}  \Downarrow  \ottnt{S'_{{\mathrm{2}}}}\) \hyp{7}
    for some \(\ottnt{V}\), \(\ottnt{S'_{{\mathrm{1}}}}\), and \(\ottnt{S'_{{\mathrm{2}}}}\).
    By \propref{subty/exists=>subst'}, we have
\(
       \sigma  [  \ottmv{x}  \mapsto  \ottnt{V}  ]   \ottsym{:}  \Gamma  \ottsym{,}   \ottmv{x} \mathord:\allowbreak \ottnt{T}   \models  \ottnt{S'_{{\mathrm{1}}}}  \ottsym{:}  \ottsym{\{}  \Upsilon  \mid  \varphi  \ottsym{\}}
\).
    By applying IH, we have
\(
       \sigma  [  \ottmv{x}  \mapsto  \ottnt{V}  ]   \ottsym{:}  \Gamma  \ottsym{,}   \ottmv{x} \mathord:\allowbreak \ottnt{T}   \models  \ottnt{S'_{{\mathrm{2}}}}  \ottsym{:}  \ottsym{\{}  \Upsilon'  \mid  \varphi'  \ottsym{\}}
\)
from which \(\sigma  \ottsym{:}  \Gamma  \models  \ottnt{V}  \triangleright  \ottnt{S'_{{\mathrm{2}}}}  \ottsym{:}  \ottsym{\{}   \ottmv{x} \mathord:\allowbreak \ottnt{T}   \triangleright  \Upsilon'  \mid  \varphi'  \ottsym{\}}\) follows.
\item[Case \ruleref{RT-Loop}:] We have
\(      \ottnt{I} \ottsym{=} \ottkw{LOOP} \, \mathit{IS} \) and
\(       \Phi_{{\mathrm{1}}} \ottsym{=} \ottsym{\{}   \ottmv{x} \mathord:\allowbreak \ottkw{int}   \triangleright  \Upsilon  \mid  \varphi  \ottsym{\}}  \) and
\(       \Phi_{{\mathrm{2}}} \ottsym{=} \ottsym{\{}  \Upsilon  \mid  \exists \,  \ottmv{x} \mathord:\allowbreak \ottkw{int}   \ottsym{.}  \varphi  \wedge  \ottmv{x}  \ottsym{=}  0  \ottsym{\}}  \) and
\(       \Gamma \vdash \ottsym{\{}  \Upsilon  \mid  \exists \,  \ottmv{x} \mathord:\allowbreak \ottkw{int}   \ottsym{.}  \varphi  \wedge  \ottmv{x}  \neq  0  \ottsym{\}} \; \mathit{IS} \; \ottsym{\{}   \ottmv{x} \mathord:\allowbreak \ottkw{int}   \triangleright  \Upsilon  \mid  \varphi  \ottsym{\}}  \) and
\(       \ottnt{S_{{\mathrm{1}}}} \ottsym{=} \ottnt{i}  \triangleright  \ottnt{S}  \)
    for some \(\mathit{IS}\), \(\ottmv{x}\), \(\Upsilon\), \(\ottnt{S}\), and
    \(\varphi\).  By IH, we have that,
\(
      \text{for any } \ottnt{S'_{{\mathrm{1}}}}, \ottnt{S'_{{\mathrm{2}}}}, \text{ if } \ottnt{S'_{{\mathrm{1}}}}  \vdash  \mathit{IS}  \Downarrow  \ottnt{S'_{{\mathrm{2}}}} \text{
        and } \sigma  \ottsym{:}  \Gamma  \models  \ottnt{S'_{{\mathrm{1}}}}  \ottsym{:}  \ottsym{\{}  \Upsilon  \mid  \exists \,  \ottmv{x} \mathord:\allowbreak \ottkw{int}   \ottsym{.}  \varphi  \wedge  \ottmv{x}  \neq  0  \ottsym{\}}, \text{ then
      } \sigma  \ottsym{:}  \Gamma  \models  \ottnt{S'_{{\mathrm{2}}}}  \ottsym{:}  \ottsym{\{}   \ottmv{x} \mathord:\allowbreak \ottkw{int}   \triangleright  \Upsilon  \mid  \varphi  \ottsym{\}}.\)
    The goal easily follows from \propref{soundness/loop'}.

  \item[Case \ruleref{RT-Lambda}:] We have
\(       \ottnt{I} \ottsym{=} \ottkw{LAMBDA} \, \ottnt{T_{{\mathrm{1}}}} \, \ottnt{T_{{\mathrm{2}}}} \, \mathit{IS}  \) and
\(       \Phi_{{\mathrm{1}}} \ottsym{=} \ottsym{\{}  \Upsilon  \mid  \varphi  \ottsym{\}}  \) and 
\(       \Phi_{{\mathrm{2}}} \ottsym{=} \ottsym{\{}   \ottmv{x} \mathord:\allowbreak \ottnt{T_{{\mathrm{1}}}}  \to  \ottnt{T_{{\mathrm{2}}}}   \triangleright  \Upsilon  \mid  \varphi  \wedge  \forall \,  \ottmv{y'_{{\mathrm{1}}}} \mathord:\allowbreak \ottnt{T_{{\mathrm{1}}}}   \ottsym{,}   \ottmv{y_{{\mathrm{1}}}} \mathord:\allowbreak \ottnt{T_{{\mathrm{1}}}}   \ottsym{,}   \ottmv{y_{{\mathrm{2}}}} \mathord:\allowbreak \ottnt{T_{{\mathrm{2}}}}   \ottsym{.}  \ottmv{y'_{{\mathrm{1}}}}  \ottsym{=}  \ottmv{y_{{\mathrm{1}}}}  \wedge  \varphi_{{\mathrm{1}}}  \wedge   \ottkw{call} ( \ottmv{x} ,  \ottmv{y'_{{\mathrm{1}}}} ) =  \ottmv{y_{{\mathrm{2}}}}   \implies  \varphi_{{\mathrm{2}}}  \ottsym{\}}  \) and
\(        \ottmv{y'_{{\mathrm{1}}}} \mathord:\allowbreak \ottnt{T_{{\mathrm{1}}}}  \vdash \ottsym{\{}   \ottmv{y_{{\mathrm{1}}}} \mathord:\allowbreak \ottnt{T_{{\mathrm{1}}}}   \triangleright  \ddagger  \mid  \ottmv{y'_{{\mathrm{1}}}}  \ottsym{=}  \ottmv{y_{{\mathrm{1}}}}  \wedge  \varphi_{{\mathrm{1}}}  \ottsym{\}} \; \mathit{IS} \; \ottsym{\{}   \ottmv{y_{{\mathrm{2}}}} \mathord:\allowbreak \ottnt{T_{{\mathrm{2}}}}   \triangleright  \ddagger  \mid  \varphi_{{\mathrm{2}}}  \ottsym{\}}  \) and
\(       \ottmv{x}  \notin    \text{dom}( \Gamma  \ottsym{,}   \widehat{ \Upsilon }  )   \cup  \ottsym{\{}  \ottmv{y_{{\mathrm{1}}}}  \ottsym{,}  \ottmv{y'_{{\mathrm{1}}}}  \ottsym{,}  \ottmv{y_{{\mathrm{2}}}}  \ottsym{\}}   \) and
\(      \ottmv{y_{{\mathrm{1}}}}  \neq  \ottmv{y'_{{\mathrm{1}}}} \) and
\(        \ottmv{y'_{{\mathrm{1}}}} \mathord:\allowbreak \ottnt{T_{{\mathrm{1}}}}   \ottsym{,}   \ottmv{y_{{\mathrm{1}}}} \mathord:\allowbreak \ottnt{T_{{\mathrm{1}}}}    \vdash   \varphi_{{\mathrm{1}}}  : \mathord{*}  \) and
    for some \(\mathit{IS}\), \(\ottmv{x}\), \(\ottmv{y_{{\mathrm{1}}}}\), \(\ottmv{y'_{{\mathrm{1}}}}\), \(\ottmv{y_{{\mathrm{2}}}}\),
    \(\ottnt{T_{{\mathrm{1}}}}\), \(\ottnt{T_{{\mathrm{2}}}}\), \(\Upsilon\), \(\varphi\), \(\varphi_{{\mathrm{1}}}\), and \(\varphi_{{\mathrm{2}}}\).
    By \ruleref{E-Lambda}, we also have \(       \ottnt{S_{{\mathrm{2}}}} \ottsym{=}  \langle  \mathit{IS}  \rangle   \triangleright  \ottnt{S_{{\mathrm{1}}}}  \).
    By IH, we have that,
\(
      \text{for any } \ottnt{V_{{\mathrm{1}}}}, \ottnt{V_{{\mathrm{2}}}}, \sigma, \text{ if } \ottnt{V_{{\mathrm{1}}}}  \triangleright  \ddagger  \vdash  \mathit{IS}  \Downarrow  \ottnt{V_{{\mathrm{2}}}}  \triangleright  \ddagger \text{ and } \sigma  \ottsym{:}   \ottmv{y'_{{\mathrm{1}}}} \mathord:\allowbreak \ottnt{T_{{\mathrm{1}}}}   \models  \ottnt{V_{{\mathrm{1}}}}  \triangleright  \ddagger  \ottsym{:}  \ottsym{\{}   \ottmv{y_{{\mathrm{1}}}} \mathord:\allowbreak \ottnt{T_{{\mathrm{1}}}}   \triangleright  \ddagger  \mid  \ottmv{y'_{{\mathrm{1}}}}  \ottsym{=}  \ottmv{y_{{\mathrm{1}}}}  \wedge  \varphi_{{\mathrm{1}}}  \ottsym{\}},
      \text{ then } \sigma  \ottsym{:}   \ottmv{y'_{{\mathrm{1}}}} \mathord:\allowbreak \ottnt{T_{{\mathrm{1}}}}   \models  \ottnt{V_{{\mathrm{2}}}}  \triangleright  \ddagger  \ottsym{:}  \ottsym{\{}   \ottmv{y_{{\mathrm{2}}}} \mathord:\allowbreak \ottnt{T_{{\mathrm{2}}}}   \triangleright  \ddagger  \mid  \varphi_{{\mathrm{2}}}  \ottsym{\}}.\)
    By \propref{soundness/lambda'}, 
\(
      \Gamma  \ottsym{,}   \widehat{ \Upsilon }   \models  \forall \,  \ottmv{y'_{{\mathrm{1}}}} \mathord:\allowbreak \ottnt{T_{{\mathrm{1}}}}   \ottsym{,}   \ottmv{y_{{\mathrm{1}}}} \mathord:\allowbreak \ottnt{T_{{\mathrm{1}}}}   \ottsym{,}   \ottmv{y_{{\mathrm{2}}}} \mathord:\allowbreak \ottnt{T_{{\mathrm{2}}}}   \ottsym{.}  \ottmv{y'_{{\mathrm{1}}}}  \ottsym{=}  \ottmv{y_{{\mathrm{1}}}}  \wedge  \varphi_{{\mathrm{1}}}  \wedge   \ottkw{call} (  \langle  \mathit{IS}  \rangle  ,  \ottmv{y'_{{\mathrm{1}}}} ) =  \ottmv{y_{{\mathrm{2}}}}   \implies  \varphi_{{\mathrm{2}}}.
\)
    Then, it is easy to show
\(
      \sigma  \ottsym{:}  \Gamma  \models  \ottnt{S_{{\mathrm{1}}}}  \ottsym{:}  \ottsym{\{}  \Upsilon  \mid  \varphi  \wedge  \forall \,  \ottmv{y'_{{\mathrm{1}}}} \mathord:\allowbreak \ottnt{T_{{\mathrm{1}}}}   \ottsym{,}   \ottmv{y_{{\mathrm{1}}}} \mathord:\allowbreak \ottnt{T_{{\mathrm{1}}}}   \ottsym{,}   \ottmv{y_{{\mathrm{2}}}} \mathord:\allowbreak \ottnt{T_{{\mathrm{2}}}}   \ottsym{.}  \ottmv{y'_{{\mathrm{1}}}}  \ottsym{=}  \ottmv{y_{{\mathrm{1}}}}  \wedge  \varphi_{{\mathrm{1}}}  \wedge   \ottkw{call} (  \langle  \mathit{IS}  \rangle  ,  \ottmv{y'_{{\mathrm{1}}}} ) =  \ottmv{y_{{\mathrm{2}}}}   \implies  \varphi_{{\mathrm{2}}}  \ottsym{\}} \hyp{7}.
\)
    Then, we have
    \begin{multline*}
      \sigma  \ottsym{:}  \Gamma  \models  \ottnt{S_{{\mathrm{1}}}}  \ottsym{:}  \ottsym{\{}  \Upsilon  \mid  \exists \,  \ottmv{x} \mathord:\allowbreak \ottnt{T_{{\mathrm{1}}}}  \to  \ottnt{T_{{\mathrm{2}}}}   \ottsym{.}  \ottsym{(}   \varphi  \wedge {} \\  \forall \,  \ottmv{y'_{{\mathrm{1}}}} \mathord:\allowbreak \ottnt{T_{{\mathrm{1}}}}   \ottsym{,}   \ottmv{y_{{\mathrm{1}}}} \mathord:\allowbreak \ottnt{T_{{\mathrm{1}}}}   \ottsym{,}   \ottmv{y_{{\mathrm{2}}}} \mathord:\allowbreak \ottnt{T_{{\mathrm{2}}}}   \ottsym{.}  \ottmv{y'_{{\mathrm{1}}}}  \ottsym{=}  \ottmv{y_{{\mathrm{1}}}}   \wedge  \varphi_{{\mathrm{1}}}  \wedge   \ottkw{call} ( \ottmv{x} ,  \ottmv{y'_{{\mathrm{1}}}} ) =  \ottmv{y_{{\mathrm{2}}}}   \implies  \varphi_{{\mathrm{2}}}  \ottsym{)}  \wedge  \ottmv{x}  \ottsym{=}   \langle  \mathit{IS}  \rangle   \ottsym{\}}.
    \end{multline*}
    Therefore, by \propref{subty/exists=>subst'}, we have
    \begin{multline*}
      \sigma  \ottsym{:}  \Gamma  \models   \langle  \mathit{IS}  \rangle   \triangleright  \ottnt{S_{{\mathrm{1}}}}  \ottsym{:}  \ottsym{\{}   \ottmv{x} \mathord:\allowbreak \ottnt{T_{{\mathrm{1}}}}  \to  \ottnt{T_{{\mathrm{2}}}}   \triangleright  \Upsilon  \mid   \varphi  \wedge {} \\  \forall \,  \ottmv{y'_{{\mathrm{1}}}} \mathord:\allowbreak \ottnt{T_{{\mathrm{1}}}}   \ottsym{,}   \ottmv{y_{{\mathrm{1}}}} \mathord:\allowbreak \ottnt{T_{{\mathrm{1}}}}   \ottsym{,}   \ottmv{y_{{\mathrm{2}}}} \mathord:\allowbreak \ottnt{T_{{\mathrm{2}}}}   \ottsym{.}  \ottmv{y'_{{\mathrm{1}}}}  \ottsym{=}  \ottmv{y_{{\mathrm{1}}}}   \wedge  \varphi_{{\mathrm{1}}}  \wedge   \ottkw{call} ( \ottmv{x} ,  \ottmv{y'_{{\mathrm{1}}}} ) =  \ottmv{y_{{\mathrm{2}}}}   \implies  \varphi_{{\mathrm{2}}}  \ottsym{\}}
    \end{multline*}
    as required.

  \item[Case \ruleref{RT-Exec}:]
    We have
    \(  \ottnt{I} \ottsym{=} \ottkw{EXEC}  \) and,
        for some \(\ottmv{x_{{\mathrm{1}}}}\), \(\ottmv{x_{{\mathrm{2}}}}\), \(\ottmv{x_{{\mathrm{3}}}}\), \(\ottnt{T_{{\mathrm{1}}}}\),
    \(\ottnt{T_{{\mathrm{2}}}}\), \(\Upsilon\), and \(\varphi\),
\(       \Phi_{{\mathrm{1}}} \ottsym{=} \ottsym{\{}   \ottmv{x_{{\mathrm{1}}}} \mathord:\allowbreak \ottnt{T_{{\mathrm{1}}}}   \triangleright   \ottmv{x_{{\mathrm{2}}}} \mathord:\allowbreak \ottnt{T_{{\mathrm{1}}}}  \to  \ottnt{T_{{\mathrm{2}}}}   \triangleright  \Upsilon  \mid  \varphi  \ottsym{\}}  \) and
\(       \Phi_{{\mathrm{2}}} \ottsym{=} \ottsym{\{}   \ottmv{x_{{\mathrm{3}}}} \mathord:\allowbreak \ottnt{T_{{\mathrm{2}}}}   \triangleright  \Upsilon  \mid  \exists \,  \ottmv{x_{{\mathrm{1}}}} \mathord:\allowbreak \ottnt{T_{{\mathrm{1}}}}   \ottsym{,}   \ottmv{x_{{\mathrm{2}}}} \mathord:\allowbreak \ottnt{T_{{\mathrm{1}}}}  \to  \ottnt{T_{{\mathrm{2}}}}   \ottsym{.}  \varphi  \wedge   \ottkw{call} ( \ottmv{x_{{\mathrm{2}}}} ,  \ottmv{x_{{\mathrm{1}}}} ) =  \ottmv{x_{{\mathrm{3}}}}   \ottsym{\}}  \) and
\(       \ottmv{x_{{\mathrm{3}}}}  \notin   \text{dom}( \Gamma  \ottsym{,}   \widehat{  \ottmv{x_{{\mathrm{1}}}} \mathord:\allowbreak \ottnt{T_{{\mathrm{1}}}}   \triangleright   \ottmv{x_{{\mathrm{2}}}} \mathord:\allowbreak \ottnt{T_{{\mathrm{1}}}}  \to  \ottnt{T_{{\mathrm{2}}}}   \triangleright  \Upsilon }  )   \).  By \ruleref{E-Exec}, we have
\(       \ottnt{S_{{\mathrm{1}}}} \ottsym{=} \ottnt{V_{{\mathrm{1}}}}  \triangleright   \langle  \mathit{IS}  \rangle   \triangleright  \ottnt{S}  \) and
\(       \ottnt{S_{{\mathrm{2}}}} \ottsym{=} \ottnt{V_{{\mathrm{2}}}}  \triangleright  \ottnt{S}  \) and
\(      \ottnt{V_{{\mathrm{1}}}}  \triangleright  \ddagger  \vdash  \mathit{IS}  \Downarrow  \ottnt{V_{{\mathrm{2}}}}  \triangleright  \ddagger\)
for some \(\ottnt{V_{{\mathrm{1}}}}\), \(\ottnt{V_{{\mathrm{2}}}}\), \(\mathit{IS}\), and \(\ottnt{S}\).
By the assumption \(       \sigma  \ottsym{:}  \Gamma  \models  \ottnt{S_{{\mathrm{1}}}}  \ottsym{:}  \Phi_{{\mathrm{1}}} \), we have
\(
      \sigma  \ottsym{:}  \Gamma  \models  \ottnt{S}  \ottsym{:}  \ottsym{\{}  \Upsilon  \mid  \exists \,  \ottmv{x_{{\mathrm{2}}}} \mathord:\allowbreak \ottnt{T_{{\mathrm{1}}}}  \to  \ottnt{T_{{\mathrm{2}}}}   \ottsym{.}  \ottsym{(}  \exists \,  \ottmv{x_{{\mathrm{1}}}} \mathord:\allowbreak \ottnt{T_{{\mathrm{1}}}}   \ottsym{.}  \varphi  \wedge  \ottmv{x_{{\mathrm{1}}}}  \ottsym{=}  \ottnt{V_{{\mathrm{1}}}}  \ottsym{)}  \wedge  \ottmv{x_{{\mathrm{2}}}}  \ottsym{=}   \langle  \mathit{IS}  \rangle   \ottsym{\}}.
\)
    By \propref{soundness/exec'}, we have
\(
      \Gamma  \ottsym{,}   \widehat{ \Upsilon }   \models   \ottkw{call} (  \langle  \mathit{IS}  \rangle  ,  \ottnt{V_{{\mathrm{1}}}} ) =  \ottnt{V_{{\mathrm{2}}}}  \hyp{10}.
\)
    Therefore, we have
\(
      \sigma  \ottsym{:}  \Gamma  \models  \ottnt{S}  \ottsym{:}  \ottsym{\{}  \Upsilon  \mid  \ottsym{(}  \exists \,  \ottmv{x_{{\mathrm{2}}}} \mathord:\allowbreak \ottnt{T_{{\mathrm{1}}}}  \to  \ottnt{T_{{\mathrm{2}}}}   \ottsym{.}  \ottsym{(}  \exists \,  \ottmv{x_{{\mathrm{1}}}} \mathord:\allowbreak \ottnt{T_{{\mathrm{1}}}}   \ottsym{.}  \varphi  \wedge  \ottmv{x_{{\mathrm{1}}}}  \ottsym{=}  \ottnt{V_{{\mathrm{1}}}}  \ottsym{)}  \wedge  \ottmv{x_{{\mathrm{2}}}}  \ottsym{=}   \langle  \mathit{IS}  \rangle   \ottsym{)}  \wedge   \ottkw{call} (  \langle  \mathit{IS}  \rangle  ,  \ottnt{V_{{\mathrm{1}}}} ) =  \ottnt{V_{{\mathrm{2}}}}   \ottsym{\}}.
\)
    Finally, we have
    $\sigma  \ottsym{:}  \Gamma  \models  \ottnt{S}  \ottsym{:}  \ottsym{\{}  \Upsilon  \mid  \exists \,  \ottmv{x_{{\mathrm{3}}}} \mathord:\allowbreak \ottnt{T_{{\mathrm{2}}}}   \ottsym{.}  \ottsym{(}  \exists \,  \ottmv{x_{{\mathrm{1}}}} \mathord:\allowbreak \ottnt{T_{{\mathrm{1}}}}   \ottsym{,}   \ottmv{x_{{\mathrm{2}}}} \mathord:\allowbreak \ottnt{T_{{\mathrm{1}}}}  \to  \ottnt{T_{{\mathrm{2}}}}   \ottsym{.}  \varphi  \wedge   \ottkw{call} ( \ottmv{x_{{\mathrm{2}}}} ,  \ottmv{x_{{\mathrm{1}}}} ) =  \ottmv{x_{{\mathrm{3}}}}   \ottsym{)}  \wedge  \ottmv{x_{{\mathrm{3}}}}  \ottsym{=}  \ottnt{V_{{\mathrm{2}}}}  \ottsym{\}}$ and, thus,
    $\sigma  \ottsym{:}  \Gamma  \models  \ottnt{V_{{\mathrm{2}}}}  \triangleright  \ottnt{S}  \ottsym{:}  \ottsym{\{}   \ottmv{x_{{\mathrm{3}}}} \mathord:\allowbreak \ottnt{T_{{\mathrm{2}}}}   \triangleright  \Upsilon  \mid  \ottsym{(}  \exists \,  \ottmv{x_{{\mathrm{1}}}} \mathord:\allowbreak \ottnt{T_{{\mathrm{1}}}}   \ottsym{,}   \ottmv{x_{{\mathrm{2}}}} \mathord:\allowbreak \ottnt{T_{{\mathrm{1}}}}  \to  \ottnt{T_{{\mathrm{2}}}}   \ottsym{.}  \varphi  \wedge   \ottkw{call} ( \ottmv{x_{{\mathrm{2}}}} ,  \ottmv{x_{{\mathrm{1}}}} ) =  \ottmv{x_{{\mathrm{3}}}}   \ottsym{)}  \ottsym{\}}$ as required.

  \item[Case \ruleref{RT-Sub}:] We have
\(
      \Gamma  \vdash  \Phi_{{\mathrm{1}}}  \ottsym{<:}  \Phi'_{{\mathrm{1}}} \) and
\(      \Gamma  \vdash  \Phi'_{{\mathrm{2}}}  \ottsym{<:}  \Phi_{{\mathrm{2}}} \) and
\(       \Gamma \vdash \Phi'_{{\mathrm{1}}} \; \ottnt{I} \; \Phi'_{{\mathrm{2}}}  \)
    for some \(\Phi'_{{\mathrm{1}}}\) and \(\Phi'_{{\mathrm{2}}}\).
    By \propref{soundness/subtyping'}, we have
\(
      \sigma  \ottsym{:}  \Gamma  \models  \ottnt{S_{{\mathrm{1}}}}  \ottsym{:}  \Phi'_{{\mathrm{1}}}.
\)
    By IH, we have
\(
      \sigma  \ottsym{:}  \Gamma  \models  \ottnt{S_{{\mathrm{2}}}}  \ottsym{:}  \Phi'_{{\mathrm{2}}}.
\)
    Then, the goal follows from \propref{soundness/subtyping'}.
    \qed
\end{description}

% \input{proofs/typing}

%%% Local Variables:
%%% mode: latex
%%% TeX-master: "../paper.otex"
%%% End:

\subsection{Extension with Exceptions}

\label{sec:exceptions}

The type system implemented in \HELMHOLTZ{} is extended to handle
instruction \texttt{FAILWITH}, which immediately aborts the execution,
discarding all the stack elements but the top element.  The type
judgment form is extended to
\[
   \Gamma \vdash \Phi_{{\mathrm{1}}} \; \mathit{IS} \; \Phi_{{\mathrm{2}}}  \mathrel{\& } \Phi_{{\mathrm{3}}}, 
\]
which means that, if \(\mathit{IS}\) is executed under a stack
satisfying \(\Phi_{{\mathrm{1}}}\), then the resulting stack satisfies
\(\Phi_{{\mathrm{2}}}\) (if normally terminates),
or \(\Phi_{{\mathrm{3}}}\) (if aborted by \texttt{FAILWITH}).
The typing rule for instruction \texttt{FAILWITH},
which raises an exception with the value at the stack top,
is given as follows:
\begin{gather*}
  \Gamma \vdash \ottsym{\{}   \ottmv{x} \mathord:\allowbreak \ottnt{T}   \triangleright  \Upsilon  \mid  \varphi  \ottsym{\}}\; \texttt{FAILWITH} \;
  \ottsym{\{}  \Upsilon  \mid   \bot   \ottsym{\}} \mathrel{\&} \ottsym{\{} \, \ottkw{err} \, \mid  \exists \,  \ottmv{x} \mathord:\allowbreak \ottnt{T}   \ottsym{,}   \widehat{ \Upsilon }   \ottsym{.}  \varphi  \wedge  \ottmv{x}  \ottsym{=}  \ottkw{err}  \ottsym{\}}.
\end{gather*}
The rule expresses that, if \texttt{FAILWITH} is executed under a
non-empty stack that satisfies $\varphi$, then the program point just
after the instruction is not reachable (hence, $\ottsym{\{}  \Upsilon  \mid   \bot   \ottsym{\}}$).
The refinement $\exists \,  \ottmv{x} \mathord:\allowbreak \ottnt{T}   \ottsym{,}   \widehat{ \Upsilon }   \ottsym{.}  \varphi  \wedge  \ottmv{x}  \ottsym{=}  \ottkw{err}$ for the
exception case states that $\varphi$ in the pre-condition with the top
element $\ottmv{x}$ is equal to the raised value $\ottkw{err}$; since $\ottmv{x}$
is not in the scope in the exception refinement, $\ottmv{x}$ is bound by
an existential quantifier.  Most of the other typing rules can be
extended with the ``\&'' part easily.

For \ruleref{RT-Lambda} and \ruleref{RT-Exec}, we first extend the the
assertion language with a new predicate \( \texttt{call\symbol{95}err}( \ottnt{t_{{\mathrm{1}}}} ,  \ottnt{t_{{\mathrm{2}}}} ) =  \ottnt{t_{{\mathrm{3}}}} \)
meaning the call of \(\ottnt{t_{{\mathrm{1}}}}\) with \(\ottnt{t_{{\mathrm{2}}}}\) aborts with the value
\(\ottnt{t_{{\mathrm{3}}}}\).  (The semantics of \texttt{call} is unchanged.)  Using the new predicate,
\ruleref{RT-Lambda} and \ruleref{RT-Exec} are modified as in \figref{newrules}.
\begin{figure}
   \centering
\begin{rules}
  \infun{ \setlength\lineskip{2pt}%
    \vbox{%
      \hbox{\(( \ottmv{x}  \notin    \text{dom}( \Gamma  \ottsym{,}   \widehat{ \Upsilon }  )   \cup  \ottsym{\{}  \ottmv{y_{{\mathrm{1}}}}  \ottsym{,}  \ottmv{y'_{{\mathrm{1}}}}  \ottsym{,}  \ottmv{y_{{\mathrm{2}}}}  \ottsym{\}}  )\) \quad
        \((\ottmv{y_{{\mathrm{1}}}}  \neq  \ottmv{y_{{\mathrm{2}}}})\) \quad \(  \ottmv{y'_{{\mathrm{1}}}} \mathord:\allowbreak \ottnt{T_{{\mathrm{1}}}}   \ottsym{,}   \ottmv{y_{{\mathrm{1}}}} \mathord:\allowbreak \ottnt{T_{{\mathrm{1}}}}    \vdash   \varphi_{{\mathrm{1}}}  : \mathord{*} \)}%
      \hbox{\(  \ottmv{y'_{{\mathrm{1}}}} \mathord:\allowbreak \ottnt{T_{{\mathrm{1}}}}  \vdash \ottsym{\{}   \ottmv{y_{{\mathrm{1}}}} \mathord:\allowbreak \ottnt{T_{{\mathrm{1}}}}   \triangleright  \ddagger  \mid  \ottmv{y'_{{\mathrm{1}}}}  \ottsym{=}  \ottmv{y_{{\mathrm{1}}}}  \wedge  \varphi_{{\mathrm{1}}}  \ottsym{\}} \; \mathit{IS} \; \ottsym{\{}   \ottmv{y_{{\mathrm{2}}}} \mathord:\allowbreak \ottnt{T_{{\mathrm{2}}}}   \triangleright  \ddagger  \mid  \varphi_{{\mathrm{2}}}  \ottsym{\}}  \mathrel{\&} \ottsym{\{} \, \ottkw{err} \, \mid  \varphi_{{\mathrm{3}}}  \ottsym{\}}\)}%
    }
  }{
    \parbox{\textwidth-\ScoreOverhang-\ScoreOverhang-\labelSpacing-1.5pt}{%
      \( \Gamma \vdash \ottsym{\{}  \Upsilon  \mid  \varphi  \ottsym{\}} \; \ottkw{LAMBDA} \, \ottnt{T_{{\mathrm{1}}}} \, \ottnt{T_{{\mathrm{2}}}} \, \mathit{IS} \; \ottsym{\{}   \ottmv{x} \mathord:\allowbreak \ottnt{T_{{\mathrm{1}}}}  \to  \ottnt{T_{{\mathrm{2}}}}   \triangleright  \Upsilon  \mid  \varphi  \wedge  \forall \,  \ottmv{y'_{{\mathrm{1}}}} \mathord:\allowbreak \ottnt{T_{{\mathrm{1}}}}   \ottsym{,}   \ottmv{y_{{\mathrm{1}}}} \mathord:\allowbreak \ottnt{T_{{\mathrm{1}}}}   \ottsym{,}   \ottmv{y_{{\mathrm{2}}}} \mathord:\allowbreak \ottnt{T_{{\mathrm{2}}}}   \ottsym{.}  \ottmv{y'_{{\mathrm{1}}}}  \ottsym{=}  \ottmv{y_{{\mathrm{1}}}}  \wedge  \varphi_{{\mathrm{1}}}  \implies  \ottsym{(}   \ottkw{call} ( \ottmv{x} ,  \ottmv{y'_{{\mathrm{1}}}} ) =  \ottmv{y_{{\mathrm{2}}}}   \implies  \varphi_{{\mathrm{2}}}  \ottsym{)}  \wedge  \ottsym{(}   \texttt{call\symbol{95}err}( \ottmv{x} ,  \ottmv{y'_{{\mathrm{1}}}} ) =  \ottkw{err}   \implies  \varphi_{{\mathrm{3}}}  \ottsym{)}  \ottsym{\}}  \mathrel{\&} \ottsym{\{} \, \ottkw{err} \, \mid   \bot   \ottsym{\}}\)}
  }
  \infun{
    \(( \ottmv{x_{{\mathrm{3}}}}  \notin   \text{dom}( \Gamma  \ottsym{,}   \widehat{  \ottmv{x_{{\mathrm{1}}}} \mathord:\allowbreak \ottnt{T_{{\mathrm{1}}}}   \triangleright   \ottmv{x_{{\mathrm{2}}}} \mathord:\allowbreak \ottnt{T_{{\mathrm{1}}}}  \to  \ottnt{T_{{\mathrm{2}}}}   \triangleright  \Upsilon }  )  )\)
  }{
    \parbox{\textwidth-\ScoreOverhang-\ScoreOverhang-\labelSpacing-1.5pt}{%
    \( \Gamma \vdash \ottsym{\{}   \ottmv{x_{{\mathrm{1}}}} \mathord:\allowbreak \ottnt{T_{{\mathrm{1}}}}   \triangleright   \ottmv{x_{{\mathrm{2}}}} \mathord:\allowbreak \ottnt{T_{{\mathrm{1}}}}  \to  \ottnt{T_{{\mathrm{2}}}}   \triangleright  \Upsilon  \mid  \varphi  \ottsym{\}} \; \ottkw{EXEC} \; \ottsym{\{}   \ottmv{x_{{\mathrm{3}}}} \mathord:\allowbreak \ottnt{T_{{\mathrm{2}}}}   \triangleright  \Upsilon  \mid  \exists \,  \ottmv{x_{{\mathrm{1}}}} \mathord:\allowbreak \ottnt{T_{{\mathrm{1}}}}   \ottsym{,}   \ottmv{x_{{\mathrm{2}}}} \mathord:\allowbreak \ottnt{T_{{\mathrm{1}}}}  \to  \ottnt{T_{{\mathrm{2}}}}   \ottsym{.}  \varphi  \wedge   \ottkw{call} ( \ottmv{x_{{\mathrm{2}}}} ,  \ottmv{x_{{\mathrm{1}}}} ) =  \ottmv{x_{{\mathrm{3}}}}   \ottsym{\}} 
    \mathrel{\&} \ottsym{\{} \, \ottkw{err} \, \mid  \exists \,  \ottmv{x_{{\mathrm{1}}}} \mathord:\allowbreak \ottnt{T_{{\mathrm{1}}}}   \ottsym{,}   \ottmv{x_{{\mathrm{2}}}} \mathord:\allowbreak \ottnt{T_{{\mathrm{1}}}}  \to  \ottnt{T_{{\mathrm{2}}}}   \ottsym{,}   \widehat{ \Upsilon }   \ottsym{.}  \varphi  \wedge   \texttt{call\symbol{95}err}( \ottmv{x_{{\mathrm{2}}}} ,  \ottmv{x_{{\mathrm{1}}}} ) =  \ottkw{err}   \ottsym{\}}\)}
  }
\end{rules}
  \caption{Modified Typing Rules for \lstinline{LAMBDA} and \lstinline{EXEC}.}
  \label{fig:newrules}
\end{figure}

\AI{Maybe some explanation how we conduct run-time typechecking?
I don't know how \HELMHOLTZ deals with statically unknown addresses.}

\fi
\section{Tool Implementation}\label{sec:implementation}
% \begin{figure}
%   \centering
%   \includegraphics[width=\linewidth]{figures/tacas2021.pdf}
%   \caption{Architecture of \HELMHOLTZ}
%   \label{fig:arch}
% \end{figure}

% We implemented the verifier \HELMHOLTZ{} based on the formalization of
% Section~\ref{sec:system}.  Figure~\ref{fig:arch} presents its
% architecture.  As is mentioned already, \HELMHOLTZ{} is implemented as
% a part of the client program \lstinline{tezos-client} for the Tezos
% blockchain.  It takes an annotated Michelson contract as input,
% conducts simple typechecking of the contract by using a functionality
% of the original \lstinline{tezos-client}, and then verifies against
% the annotated specification using Z3 for discharging verification
% conditions.  \AI{Not much new information...}

In this section, we present $\HELMHOLTZ$, the verification tool based
on the refinement type system.  We first discuss how Michelson code
can be annotated.  Then, we give an overview of the verification
algorithm, which reduces the verification problem to SMT solving, and
discuss how Michelson-specific features are encoded.  Finally, we show
a case study of contract verification and present verification
experiments.

% as .  Of course, the implementation has a lot of
% extensions because we aim for a fully-fledged verification tool.  In what
% follows, we show an overview architecture of the implementation, some
% extensions, and example programs we have verified.

% The largest difference from the theory is a support for exceptions.  Our tool
% can also verify a \emph{error-condition} which is a condition when a contract
% fails.

% \subsection{Extended Type System}\label{sec:implementation/extension}

\subsection{Annotations}\label{sec:implementation/usage}

% In the annotation, there are three
% S-expressions which describes post-condition, pre-condition, and extra logical
% variables declarations.  A pre- (and post-) condition S-expression consists of
% logical variable name referring an initial (and final) stack, a type of
% corresponding stack, and logical formula describing a condition.  As we have
% explained in \secref{introduction}, contract code itself is given a pair of an
% input parameter and a storage value and results a pair of an operation list and
% an updated storage value.  So, the type of initial and final stack of the
% add-arg contract is specified as \lstinline$(Pair Int Int)$ and
% \lstinline$(Pair (List Operation) Int)$, respectively.  The specification for
% the pre-condition is specified as \lstinline$true$, which means add-arg contract
% can call under any condition.  On the other hand, one for the post-condition is
% specified as \lstinline$(= (second ret) (+ (first arg) (second arg)))$, which
% means the updated storage value represented as \lstinline$(second ret)$ is equal
% to the sum of the original storage value represented as \lstinline$(first arg)$
% and a given integer represented as \lstinline$(second arg)$---and actually this
% describes a specification of add-arg contract.

% \begin{figure}
%   \centering
%   \input{figures/tsyntax}
%   \caption{Syntax of annotations}
%   \label{fig:annot}
% \end{figure}

\HELMHOLTZ{} supports several forms of annotations (surrounded by
\lstinline$<<$ and \lstinline$>>$ in the source code), other than
\lstinline{ContractAnnot} explained in \secref{overview}.
As we have already mentioned, ML-like
notations can be used to describe formulae, which have to be quantifier free
(mainly because state-of-the-art SMT solvers do not handle
quentifiers very well).
% We think most of them are
% familiar to readers but we will explain constructs special to
% \HELMHOLTZ{} as they appear.
We explain several constructs for an annotation in the following.

% The following BNF defines
% the syntax of annotations,
% which are enclosed by 
% %
% % \input{figures/tsyntax}
% %
% Metavariables $p$ and $e$ represent
% ML-like patterns and expressions; the full syntax will be found in the
% accompanied artifact.  We can write a Boolean expression in the
% predicate part of a refinement stack type.

\begin{figure}[t]
  \centering
\begin{lstlisting}[language=michelson]
parameter unit ;
storage int;
<< ContractAnnot { _ | True } -> { _ | True }
                 & { _ | False } >>
code { DROP;
       << LambdaAnnot { p | p = (3, 1) } -> { x | x = 4 } @\label{tz:lambda}@
                      & { _ | False }
          (a:int, b:int) >>
       LAMBDA (pair int int) int
              { << Assume { p | p = (a, b) } >> @\label{tz:lambda_assume}@
                UNPAIR; ADD
                << Assert { p | p = a + b } >> @\label{tz:lambda_assert}@
              };
       PUSH int 1; PUSH int 3; PAIR; EXEC;
       << Assert { x | x = 4 } >>
       NIL operation; PAIR
     }
\end{lstlisting}
\begin{lstlisting}[language=michelson]
parameter (list int);
storage int;
<< Measure len : list int -> int
     where [] = 0 | h :: t = (1 + len t) >>
<< ContractAnnot
   { (p, _) | True } -> { (_, ret) | len p = ret }
   & { _ | False } >>
code { CAR; PUSH int 0; SWAP;
       << LoopInv { l : n | len l + n = len p } >>
       ITER { DROP; PUSH int 1; ADD };
       NIL operation;
       PAIR
     }
\end{lstlisting}
  \caption{\texttt{lambda.tz}, which uses higher-order functions, and \texttt{length.tz}, which uses a measure function in the contract annotation.}
  \label{fig:lambda_and_length}
\end{figure}

%\begin{itemize}
%\item

\texttt{Assert} $\Phi$ and \texttt{Assume} $\Phi$ can appear before or
after an instruction.  The former asserts that the stack at the annotated
program location satisfies the type $\Phi$; the assertion is verified by
\HELMHOLTZ{}.  If there is an annotation \texttt{Assume} $\Phi$, \HELMHOLTZ{}
assumes that the stack satisfies the type $\Phi$ at the annotated program
location.  A user can give a hint to \HELMHOLTZ{} by using \texttt{Assume}
$\Phi$.  The user has to make sure that it is correct; if an \texttt{Assume}
annotation is incorrect, the verification result may also be incorrect.
\AI{What are typical uses of assume?}

  % annotation forces our tool to check
  % whether a given condition holds in arbitrary code point.  So it does
  % not affect essential verification process but will useful for
  % experiments and debugging.

% \item

  % annotation is used to strengthen a
  % automatically generated condition in arbitrary code point.  Our tool
  % blindly trusts assuming conditions, that is, if we give an invalid
  % assumption, the result of verification becomes meaningless.  The
  % primary intended usage of this annotation is to refer stack values
  % in some code point with extra logical variables like
  % \texttt{lambda.tz} does.

% \item
An annotation \texttt{LoopInv} $\Phi$ may appear before a loop instruction (e.g., \lstinline{LOOP} and \lstinline{ITER}).
It asserts that $\Phi$ is a loop invariant of the loop instruction.
In the current implementation, annotating a loop invariant using \texttt{LoopInv} $\Phi$ is mandatory for a loop instruction.
\HELMHOLTZ{} checks that $\Phi$ is indeed a loop invariant and uses it to verify the rest of the program.

% \item
\begin{sloppypar}
In the current implementation, a \lstinline{LAMBDA} instruction, which pushes a function on the top of the stack, must be accompanied by a \lstinline{LambdaAnnot} annotation.
\lstinline{LambdaAnnot} comes with a specification of the pushed function written in the same way as \lstinline{ContractAnnot}.
Concretely, the specification of the form \(\Phi_\text{pre} \rightarrow \Phi_\text{post} \mathrel{\&} \Phi_\text{abpost} (\ottmv{x_{{\mathrm{1}}}}:\ottnt{T_{{\mathrm{1}}}},\dots,\ottmv{x_{\ottmv{n}}}:\ottnt{T_{\ottmv{n}}})\) specifies the precondition $\Phi_\text{pre}$, the (normal) postcondition $\Phi_\text{post}$, and the (abnormal) postcondition $\Phi_\text{abpost}$ as refinement stack types.
The binding $(\ottmv{x_{{\mathrm{1}}}}:\ottnt{T_{{\mathrm{1}}}},\dots,\ottmv{x_{\ottmv{n}}}:\ottnt{T_{\ottmv{n}}})$ introduces the ghost variables that can be used in the annotations in the body of the annotated \lstinline{LAMBDA} instruction;\footnote{%
  \lstinline{ContractAnnot} also allows declarations of ghost variable used in
  the \lstinline{code} section.}; one can omit if it is empty.  \AI{Can't we reference ghost variables in the
  specification?}
\end{sloppypar}

The first contract in \figref{lambda_and_length}, which
pushes a function that takes a pair of integers and returns the sum of them,
exemplifies \lstinline{LambdaAnnot}.
The annotated type of the function (Line~\ref{tz:lambda}) expresses that it returns $4$ if it is fed with
a pair $(3,1)$.
The ghost variables $a$ and $b$ are used in the annotations \lstinline{Assume} (Line~\ref{tz:lambda_assume}) and \lstinline{Assert} (Line~\ref{tz:lambda_assert}) in the body to denote the first and the second arguments of the pair passed to this function.

To describe a property for recursive data structures, \HELMHOLTZ{} supports \emph{measure functions} introduced by Kawaguchi et al.~\cite{KawaguchiRJ09} and also supported in Liquid Haskell~\cite{VazouSJVJ2014}.
A measure function is a (recursive) function over a recursive data structure that can be used in assertions. 
The annotation \lstinline{Measure} $\ottmv{x} : \ottnt{T_{{\mathrm{1}}}} \rightarrow \ottnt{T_{{\mathrm{2}}}}\ \mathtt{where}\ p_1 = e_1 \mid \dots \mid p_n = e_n$ defines a measure function $x$ over the type $\ottnt{T_{{\mathrm{1}}}}$.
The measure function $\ottmv{x}$ takes a value of type $\ottnt{T_{{\mathrm{1}}}}$, destructs it by the pattern matching, and returns a value of type $\ottnt{T_{{\mathrm{2}}}}$.
Metavariables $p$ and $e$ represent ML-like patterns and expressions.
The second contract in \figref{lambda_and_length}, which computes the length of the list passed as a parameter, exemplifies the usage of the \lstinline{Measure} annotation.
This contract defines a measure function \lstinline{len} that takes a list of integers and returns its type; it is used in \lstinline{ContractAnnot} and \lstinline{LoopInv}.
  % The current implementation supports limited forms of patterns, depending on $\ottnt{T_{{\mathrm{1}}}}$.
  % \AI{Do we need the last sentence?}
  \AI{Do we support measure functions for datatypes other than lists?}

\subsection{Overview of the Verification Algorithm}
\label{sec:overviewVC}

\HELMHOLTZ{} takes an annotated Michelson program and conducts typechecking based on the refinement type system in Section~\ref{sec:system/refinement}.
The typechecking procedure (1) computes the verification conditions (VCs) for the program to be well-typed and (2) discharges it using an SMT solver.
The latter is standard: We decide the validity of the generated VCs using an SMT solver (Z3 in the current implementation.)
We explain the VC-generation step in detail.

% Helmholtz verifies code following a refinement type system.  As Michelson does typecheck following
% the typing rules, we have extended the typing rules with refinements, first-order logic predicates.
% Following the refinement type system, given contract annotations, Helmholtz does:
% \begin{itemize}
% \item calculating the post-condition of program code under the given pre-condition;
% \item trying to fill the gap between the calculated condition and given post-condition; and
% \item showing the result whether the trial succeeds or not.
% \end{itemize}
% Now let's see each step in more detail.

For an annotated contract, \HELMHOLTZ{} conducts forward reasoning starting from the precondition and generates VCs if necessary.
During the forward reasoning, \HELMHOLTZ{} keeps track of the $\Gamma$-and-$\Upsilon$ part of the type judgment.

The typing rules are designed so that they enable the forward reasoning if a program is simply typed.
For example, consider the rule \ruleref{RT-Add} in Figure~\ref{fig:typing}.
This rule can be read as a rule to compute a postcondition $\exists \,  \ottmv{x_{{\mathrm{1}}}} \mathord:\allowbreak \ottkw{int}   \ottsym{,}   \ottmv{x_{{\mathrm{2}}}} \mathord:\allowbreak \ottkw{int}   \ottsym{.}  \varphi  \wedge  \ottmv{x_{{\mathrm{1}}}}  \ottsym{+}  \ottmv{x_{{\mathrm{2}}}}  \ottsym{=}  \ottmv{x_{{\mathrm{3}}}}$ from a precondition $\varphi$ if the first two elements in $\Upsilon$ are $\ottmv{x_{{\mathrm{1}}}}$ and $\ottmv{x_{{\mathrm{2}}}}$.
The other rules can also be read as postcondition-generation rules in the same way.

% \begin{gather*}
% \Gamma \vdash \{ x_1:\mathtt{int} \rhd x_2:\mathtt{int} \rhd \Upsilon \mid \phi \} \; \mathtt{ADD}
% \; \{ x_3:\mathtt{int} \rhd \Upsilon \mid \exists x_1 x_2. \phi \wedge x_1 + x_2 = x_3 \}
% \end{gather*}

% The point of our type system is it is designed so that, given a pre-condition, we can syntactically
% calculate the post-condition of contract code.  For instance, given
% $\{x_1 \rhd x_2 \rhd [] \mid x_1 = 1 \wedge x_2 = 2\}$, which means the top of stack is 1 and the
% second from the top is 2, as a pre-condition of \texttt{ADD}, we can have the post-condition
% $\{x_3 \rhd [] \mid \exists x_1 x_2. x_1 = 1 \wedge x_2 = 2 \wedge x_1 + x_2 = x_3\}$, which means
% the top of stack is 3, following by the typing rule.  In this manner, Helmholtz calculate the
% post-condition of contract code from a given pre-condition.

% \subsubsection{Generating Verification Condition (VC)}

There are three places where \HELMHOLTZ{} generates a verification condition.
\begin{itemize}
\item At the end of the program: \HELMHOLTZ{} generates a condition
  that ensures that the computed postcondition of the entire program
  implies the postcondition annotated to the program.
\item Before and after instruction \texttt{LAMBDA}: \HELMHOLTZ{}
  generates conditions that ensure that the pre- and post- conditions
  of the instruction \texttt{LAMBDA} is as annotated in \texttt{LambdaAnnot}.
\item At a loop instruction: \HELMHOLTZ{} generates verification
  conditions that ensure the condition annotated by \texttt{LoopInv}
  is indeed a loop invariant of this instruction.
  % A loop invariant of loop instructions,
  % which is also given as a LoopInv annotation.  In this case, the
  % calculated post-condition of instruction just before a loop-like
  % instruction and the calculated post-condition of the body of the
  % loop-like instruction differ from the given ones.
\end{itemize}
A VC generated by \HELMHOLTZ{} at these places is of the form $\forall \vec{x}\COL\vec{T}. \varphi_1 \Rightarrow \varphi_2$, where $\vec{x}\COL\vec{T}$ is a sequence of bindings.

To discharge each VC, as many verification condition discharging procedures do, \HELMHOLTZ{} checks whether its negation, $\exists \vec{x}\COL\vec{T}. \varphi_1 \land \neg\varphi_2$, is satisfiable; if it is unsatisfiable, then the original VC is successfully discharged.
We remark that our type system is designed so that $\varphi_1$ and $\varphi_2$ are quantifier free for a program that does not use \lstinline{LAMBDA} and \lstinline{EXEC}.
Indeed, $\varphi_2$ comes only from the annotations, which are quantifier-free.
$\varphi_1$ comes from the postcondition-computation procedure, which is of the form $\exists \vec{x'}\COL\vec{T'}. \varphi_1'$ for quantifier-free $\varphi_1'$ for the instructions other than \lstinline{LAMBDA} and \lstinline{EXEC}; the formula $\exists \vec{x}\COL\vec{T}. \varphi_1 \land \neg\varphi_2$ is equivalent to $\exists \vec{x}\COL\vec{T},\vec{x'}\COL\vec{T'}. \varphi_1' \land \neg\varphi_2$.
%
% The exception is the rules for \lstinline{LAMBDA} and \lstinline{EXEC}, in which case \HELMHOLTZ{} ... \KS{Fill here.}

% \subsubsection{Discharging VC}

% In the theoretical domain, the typing rule used to fill the gap is RT-Sub.  To apply the rule,
% Helmholtz checks if the subtyping relation holds.  By definition of the relation, for example,
% Helmholtz needs to know the validity of

% \begin{gather*}
% \forall \Gamma x_3 . (\exists x_1 x_2. x_1 = 1 \wedge x_2 = 2 \wedge x_1 + x_2 = x_3) \Rightarrow
% (x_3 = 3)
% \end{gather*}

% in the case mentioned in the first item of the last itemization.  To do so, Helmholtz uses the Z3
% SMT-solver to check satisfiability of the following formula, called \emph{verification condition}.

% \begin{gather*}
% \neg (\forall \Gamma x_3 . (\exists x_1 x_2. x_1 = 1 \wedge x_2 = 2 \wedge x_1 + x_2 = x_3)
% \Rightarrow (x_3 = 3))
% \end{gather*}

% The verification condition is the negation of the formula that we want to know its validity.  So, if
% Z3 reports the verification condition is satisfiable, which means there is a counter-example; the
% original formula is invalid.  Otherwise, if Z3 reports unsatisfiable, which means there is no
% counter-example, the original formula is valid.

% % This step is rather simple.  Helmholtz reads a response of Z3.  Then, if all verification conditions
% % are unsatisfiable, Helmholtz reports VERIFIED message, or otherwise, UNVERIFIED message.

\subsection{Encoding Micheslon-Specific Features}

% The refinement type system uses a Michelson specialized first-order theory to describe the predicate
% formulae.  For example, the theory contains all Michelson types as sorts and various functions and
% predicates simulating Michelson instructions.  Z3 is implemented rather rich theory, but of course,
% our Tezos specific theory is out of the scope of the native theory in Z3.  Here we explain how our
% theory is implemented on Z3.  Note that the method explained here is a well-known method among logic
% textbooks.

Since our assertion language includes several specific features that originates from Michelson and our assertion language, \HELMHOLTZ{} needs to encode them in discharging VCs so that Z3 can handle them.
We explain how this encoding is conducted.

\subsubsection{Michelson-Specific Functions and Predicates}

We use encode several Michelson-specific values using uninterpreted functions.
For example, \HELMHOLTZ{} assumes the following typing rule for instruction \texttt{SHA256}, which
converts the top element to its SHA256 hash.
\begin{gather*}
\Gamma \vdash \{ x:\mathtt{bytes} \rhd \Upsilon \mid \phi \} \; \mathtt{SHA256} \; \{
y:\mathtt{bytes} \rhd \Upsilon \mid \exists x. \phi \wedge y = \text{sha256}(x) \}
\end{gather*}
In the post condition, we use an uninterpreted function sha256 to express the SHA256 hash of $x$.
% For example, the typing rule for the following is the typing rule for \texttt{SHA256}.
% where \text{sha256} is a logical function respecting \texttt{SHA256} behavior, that is, it (ideally)
% returns the sha256 hash of a given argument.  To represent the function in Z3, we declare the
% function symbol sha256 called an \emph{uninterpreted function} and give axioms for the functions.
% Concretely, we give the following code for Z3 input.
In Z3, this uninterpreted function is declared as follows.
\begin{lstlisting}[language=michelson]
(declare-fun sha256 (String) String)
(assert (forall ((x String)) (= (str.len (sha256 x)) 32)))
\end{lstlisting}
The first line declares the signature of sha256.
The second line is the axiom for sha256 that the length of a hash is always 32.

% In the first line, the function symbol is declared and, in the second line, we give the axiom that
% says the length of the hash is 32.  Note that \texttt{bytes} sort is encoded as \texttt{String} of Z3 sorts, and
% \texttt{str.len} is a Z3 provided logic function that returns the length of a given string.

Notice that, we cannot assert that a hash is equal to a specific constant since sha256 is an uninterpreted function.
Therefore, \HELMHOLTZ{} cannot prove:
\begin{gather*}\small
  \Gamma \vdash \{x:\mathtt{bytes} \mid \ottmv{x}  \ottsym{=}  0 \} \; \mathtt{SHA256} \; \{y:\mathtt{bytes} \rhd \Upsilon \mid \exists x. \phi \wedge y =\\
  \text{``6e340b9cffb37a989ca544e6bb780a2c78901d3fb33738768511a30617afa01d''} \}.
\end{gather*}

% Many readers have noticed that the axiom given is insufficient since it says nothing about how the
% hash is calculated.  Actually, most logical functions in Helmholtz which are encoded as
% uninterpreted functions are underspecified because precise axiomatization requires quite complicated
% formulae, which lead Z3 execution to diverge.  As a drawback, for instance, the formula
% \texttt{sha256 0x0 = 0x6e34 0b9c ffb3 7a98 9ca5 44e6 bb78 0a2c 7890 1d3f b337 3876 8511 a306 17af
%   a01d} cannot be validate by Helmholtz, despite the fact that the SHA-256 hash value of 0x0 is
% actually the value in the right side of the equality.  However, we observe that such a lightweight
% encoding well works because the fact that a function is applied to some terms itself is more
% important than what the result of the application is in many cases.  Moreover, we could manually
% supply such insufficient facts as the pre-condition of a contract if we really need it.

\subsubsection{Implementation of Measure Functions}

A measure function is encoded as an uninterpreted function accompanied with axioms that specify the behavior of the function, which is defined by the \texttt{Measure} annotation.
% Thanks to the restricted form of measure annotations, Z3 input can be easily obtained.
For example, for the following form of a measure function definition for lists:
\begin{lstlisting}[language=michelson]
<< Measure f : list T -> T' where [] = e1 | h :: t = e2 >>
\end{lstlisting}
Theoretically, one could insert the following declarations and assertions when it generates an input to Z3 to encode this definition:
% It is easily imagined that the following Z3 input can be derived from the annotation.
\begin{lstlisting}[language=Lisp]
(declare-fun f ((List T)) T')
(assert (= (f []) e1))
(assert (forall ((h T) (t (List T)))
   (= (f (cons h t)) e2)))
\end{lstlisting}
% \KS{the following two paragaraphs are yet to be rewritten.}
However, Z3 tends to timeout if we naively insert the above axioms to Z3 input which contains a universal quantifier in the encoded definition.
%
% Another source of quantifiers is the axioms of the theory.  As we have already seen, various axioms
% involve the universal quantifier.  Naively put such axiom in Z3 input, Z3 soon diverges.  Especially,
% verifying contract code expected UNVERIFIED tends to diverge because Z3 needs to find a
% counter-example for the given specification, which involves finding a model of uninterpreted
% functions.  Such a task is sometimes too hard if the functions are constrained by universally
% quantified axioms.
%

To address this problem, \HELMHOLTZ{} rewrites VCs so that heuristically instantiated conditions on a measure function are available where necessary.
Consider the above definition of \lstinline{f} as an example.
Suppose \HELMHOLTZ{} obtains a VC of the form $\exists \vec{x}\COL\vec{T}. \varphi_1 \land \neg\varphi_2$ mentioned in Section~\ref{sec:overviewVC} and $\varphi_1$ and $\varphi_2$ contains \texttt{(cons $e_{h,i}$ $e_{t,i}$)} for $i \in \{1,\dots,N\}$.
Then, \HELMHOLTZ{} constructs a formula $\varphi_{\mathit{meas}} := \bigwedge_{i \in \{1,\dots,N\}} \texttt{(= (f (cons}\ e_{h,i}\ e_{t,i}\texttt{)) e2)}$ and rewrites the VC to $\exists \vec{x}\COL\vec{T}. \varphi_{\mathit{meas}} \land \varphi_1 \land \neg\varphi_2$.

% inserts the axioms heuristically instantiated with terms that appear in VC to Z3 input instead of the universally quantified ones.
% %
% For example, for the above measure function \lstinline{f}, \HELMHOLTZ{} insert \lstinline{(assert (= (f (cons eh et)) e2))} for each \texttt{(cons eh et)} that appears in VC to Z3 input.

% difficulty, we give instantiated axioms for Z3 input instead of original universally
% quantified ones.  Of course, instantiated ones are less precise than the original ones.  For
% example, consider the axioms for a list measure.  We naively put the first axiom, which is the case
% for the nil list, because it is not quantified.  Regarding the second axiom, which is the case for a
% cons list, we do not give the axiom to Z3 input.  Instead, if we find a term of the form
% \texttt{(cons $e_h$ $e_t$)} in formulae dealt with, we give the axiom obtained by replacing \texttt{h}
% and \texttt{t} with \texttt{$e_h$} and \texttt{$e_t$}, respectively, in \texttt{(= (f (cons h t)) e2)},
% which is one instance of the second axiom.  What terms are chosen for instantiation is determined by
% heuristics.  More instances would lead Helmholtz to verify more contracts but take more time.

We remark that, in LiquidHaskell, measure functions are treated as a part of the type
system~\cite{KawaguchiRJ09}: the asserted axioms are systematically
(instantiated and) embedded into the typing rules.
In \HELMHOLTZ{}, measure functions are treated as an ingredient that is orthogonal to the type system; the type system is oblivious of measure functions until its definition is inserted to Z3 input.

% We find that measures can be better handled at logic level than a type system because a type system does not need to concern measures, which now become usual logical functions, and logic can reason about properties of measures independently from a type system.

% Note that the original
% idea is inherited as the axiom instantiation described in \secref{}.
% So our treatment can be seen as an extension of the original work.

\subsubsection{Overloaded Functions}

Due to the polymorphically-typed instructions in Michelson, our assertion language incorporates polymorphic uninterpreted functions.
For example, Michelson has an instruction \lstinline{PACK}, which pops a value (of any type) from the stack, serializes it to a binary representation, and pushes the serialized value.
\HELMHOLTZ{} typechecks this instruction based on the following rule.
\begin{gather*}
  \Gamma \vdash \{x:T \rhd \Upsilon \mid \varphi \} \; \mathtt{PACK} \; \{y:\mathtt{bytes} \rhd \Upsilon \mid \exists x:T. \varphi \land y = \mathrm{pack}(x) \}
\end{gather*}
The term $\mathrm{pack}(x)$ in the postcondition represents a serialized value created from $x$.
Since $x$ may be of any simple type $T$, $\mathrm{pack}$ must be polymorphic.

Having a polymorphic uninterpreted function in assertions is tricky because Z3 does not support a polymorphic value.
\HELMHOLTZ{} encodes polymorphic uninterpreted functions to a monomorphic function whose name is generated by mangling the name of its instantiated parameter type.
For example, the above $\mathrm{pack}(x)$ is encoded as a Z3 function $\mathtt{pack!int}(x)$ whose type is $\ottkw{int} \rightarrow \mathtt{bytes}$.
Although there are infinitely many types, the number of the encoded functions is finite since only finitely many types appear in a single contract.

% Corresponding to the fact that Michelson has overloaded instructions, there are overloaded logical
% functions like \texttt{pack}.  Z3 has no facility to declare an overloaded function, but an obvious idea to
% deal with such functions is to declare functions for every sort, e.g., \texttt{pack!int}, \texttt{pack!string},
% etc.  One worry is that there are infinite functions for an overloaded function since there are
% infinite sorts because of compound data types, but declaring finite ones in Z3 input suffices since
% only finite sorts occur in each contract code.  (Of course, a set of declarations varies according
% to the contract code.)

\subsubsection{Michelson-Specific Types}

In encoding a VC as Z3 constraints, \HELMHOLTZ{} maps types in Michelson into sorts in Z3, e.g., the Michelson type \texttt{nat} for nonnegative integers to the Z3 sort \texttt{Int}.
A naive mapping from Michelson types to Z3 sorts is problematic; for example, $\forall x:\mathtt{nat}. x \ge 0$ in \HELMHOLTZ{} is valid, but a naively encoded formula \texttt{(forall ((x Int)) (>= x 0))} is invalid in Z3.
This naive encoding ignores that a value of sort \texttt{nat} is nonnegative.

To address the problem, we adapt the method encoding a many-sorted logic formula into a single-sorted logic formula~\cite{Enderton2001}.
Concretely, we define a \emph{sort predicate} $P_T(x)$ for each sort $T$, which characterizes the property of the sort $T$.
For example, $P_\mathtt{nat}(x) := x \ge 0$.
We also define sort predicates for compound data types.
% Due to the compound data types, the definition of the sort predicates are more complicated
% than the first thought, but honestly speaking, those can be defined in first-order predicate logic and,
% by similar reason to the overloaded functions, finite sort predicates suffices for each verification.

Using the sort predicates, we can encode a VC into a Z3 constraint as follows: $\forall x:T. \phi$ is encoded into $\forall x:[\![ T ]\!]. P_T(x) \Rightarrow \phi$ and $\exists x:T. \phi$ is encoded into $\exists x:[\![ T ]\!]. P_T(x) \wedge \phi$, where $[\![ T ]\!]$ denotes the target sort of $T$ (e.g., $[\![\mathtt{nat} ]\!] = \mathtt{Int}$).
Furthermore, we also add axioms about co-domain of uninterpreted functions as $\forall \vec{x_i}:\vec{T_i}. P_T(f(\vec{x_i}))$ for the function $f$ of $\vec{T_i} \rightarrow T$.

\subsection{Case Study: Contract with Signature Verification}
\label{sec:casestudy}

\KS{kokomade mita.}

\begin{figure}[t]
  \centering
\begin{lstlisting}[language=michelson]
parameter (pair signature string);
storage (pair address key);
<< ContractAnnot
   { ((sign, data), (addr, pubkey)) |
       match contract_opt addr with
       | Some (Contract<string> _) -> True | _ -> False } ->
   { (ops, new_store) | (addr, pubkey) = new_store &&
             sig pubkey sign (pack data) &&
             match contract_opt addr with
             | Some c -> ops = [ TransferTokens data 1 c ]
             | None -> False }
   & { _ | not (sig pubkey sign (pack data)) } >>
code  { DUP; DUP; DUP;
        DIP { CAR; UNPAIR; DIP { PACK } }; CDDR;
        CHECK_SIGNATURE; ASSERT; @\label{checksig:assert}@

        UNPAIR; CDR; SWAP; CAR;
        CONTRACT string; ASSERT_SOME; SWAP;
        PUSH mutez 1; SWAP;
        TRANSFER_TOKENS;

        NIL operation; SWAP;
        CONS; DIP { CDR };
        PAIR
      }
\end{lstlisting}
  \caption{\texttt{checksig.tz}, which involves signature verification.}
  \label{fig:checksig}
\end{figure}

\figref{checksig} presents the code of the contract
\lstinline{checksig.tz}, which verifies that a sender indeed signed
certain data using her private key.  This contract uses instruction
\lstinline{CHECK_SIGNATURE}, which is supposed to be executed under a
stack of the form \lstinline{key} $\PSH$ \lstinline{sig} $\PSH$
\lstinline{bytes} $\PSH$ \lstinline{tl}, where \lstinline{key} is a
public key, \lstinline{sig} is a signature, and \lstinline{bytes} is
some data.  \lstinline{CHECK_SIGNATURE} pops these three values from
the stack and pushes \lstinline{True} if \lstinline{sig} is the valid
signature for \lstinline{bytes} with the private key corresponding to
\lstinline{key}.  The instruction \lstinline{ASSERT} after
\lstinline{CHECK_SIGNATURE} checks if the signature checking has
succeeded; it aborts the execution of the contract if the stack top is
\lstinline{False}; otherwise, it pops the stack top (\lstinline{True})
and proceeds the next instruction.

The intended behavior of \texttt{checksig.tz} is as follows.  It stores
a pair of an address \lstinline{addr}, which is the address of a contract
that takes a \lstinline{string} parameter, and a public key
\lstinline{pubkey} in its storage.  It takes a pair \texttt{(sign,data)} of type
\texttt{(pair signature string)} as a parameter; here,
\lstinline{signature} is the primitive Michelson type for signatures.
This contract terminates without exception if
\lstinline{sign} is created from the serialized (packed) representation of
\lstinline{data} and signed by the private key corresponding to
\lstinline{pubkey}.  In a normal termination, this contract transfers $1$ \lstinline{mutez} to
the contract with address \lstinline{addr}.
If this signature verification fails, then an exception is raised.

This behavior is expressed as a specification in the
\lstinline{ContractAnnot} annotation in \lstinline{checksig.tz} as follows.
\begin{itemize}
\item The refinement of its pre-condition part expresses that the address stored in the first
  element \lstinline{addr} of the storage is an address of a contract that takes a value of type
  \lstinline{string} as a parameter.  This is expressed by the pattern-matching on
  \lstinline{contract_opt addr}, which checks if there is an intended parameter type of contract
  stored at the address \lstinline{addr} and returns the contract (wrapped by \lstinline{Some}) if there is.  The intended parameter
  type is given by the pattern expression \lstinline{Contract<string> _}, which matches a contract
  that takes a \lstinline{string}.
  % \lstinline{Contract} is a primitive constructor
  % that can be used in an annotation of \HELMHOLTZ{}.  \AI{\lstinline{Contract} has already appeared (but has not been explained well).}
  
  % pattern-matches \lstinline{Contract store.first}.  \lstinline{Contract store.first} represents the
  % contract corresponding to the address \lstinline{store.first} stored
  % at the first element of the storage \lstinline{store};
  % \lstinline{Pack param.second} is the packed representation of the
  % second element \lstinline{param.second} of the parameter.  The
  % constructors \lstinline{Contract} and \lstinline{Pack} are
  % primitives that can be used in annotations.  The current
  % implementation requires a pattern expression matching to a contract
  % needs to be annotated with its parameter type like
  % \lstinline$Contract<string>$ in \texttt{checksig.tz}.
  
% \item The refinement predicate of its precondition part
%   pattern-matches a pair \lstinline{(Contract store.first, Pack param.second)}.  \lstinline{Contract store.first} represents the
%   contract corresponding to the address \lstinline{store.first} stored
%   at the first element of the storage \lstinline{store};
%   \lstinline{Pack param.second} is the packed representation of the
%   second element \lstinline{param.second} of the parameter.  The
%   constructors \lstinline{Contract} and \lstinline{Pack} are
%   primitives that can be used in annotations.  The current
%   implementation requires a pattern expression matching to a contract
%   needs to be annotated with its parameter type like
%   \lstinline$Contract<string>$ in \texttt{checksig.tz}.
\item The refinement of the post-condition forces the following three conditions: (1) the store is
  not updated by this contract (\texttt{(addr, pubkey) = new\symbol{95}store}); (2) \texttt{sign} is
  the signature created from the packed bytes \texttt{pack data} of the string in the second element
  of the parameter and signed by the private key corresponding to the second element \texttt{pubkey}
  of the store (\texttt{sig pubkey sign (pack data)}); and (3) the operations \texttt{ops} returned
  by this contract is \texttt{[ Transfer data 1 c ]}, which represents an operation of transferring
  $1$ \texttt{mutez} to the contract \texttt{c}, which is bound to \texttt{Contract addr}, with
  the parameter \texttt{data}.  The predicate \texttt{sig} and the function \texttt{pack} are
  primitives of the assertion language of \HELMHOLTZ{}.
  % proposition  in
  % the body of the pattern-matching expression, which uses a primitive
  % predicate \lstinline{sig} of \HELMHOLTZ{}, holds if and only if
  % \lstinline{param.first} is created from \lstinline{byt} and signed
  % by the private key corresponding to \lstinline{store.second}.

\item
    The refinement in the exception part expresses that if an
  exception is raised, then the signature verification should have
  failed (\texttt{not (sig pubkey sign (pack data))}).
\end{itemize}

\HELMHOLTZ{} successfully verifies \texttt{checksig.tz}
without any additional annotation in the \lstinline{code} section.
If we change the instruction \lstinline{ASSERT} in Line~\ref{checksig:assert}
to \lstinline{DROP} to let the contract drop the result of
the signature verification (hence, an exception is not raised
even if the signature verification fails), the verification fails as intended.

\subsection{Experiments}
\label{sec:experiments}

\KS{Examples (including annotations) can be found at the web interface.}
\AI{Done.}

We applied \HELMHOLTZ{} to various contracts; \autoref{tab:result} is an
excerpt of the result, in which we show (1) the number of the instructions in
each contract (column \#instr.) and (2) time (ms) spent to verify each contract.
The experiments are conducted on MacOS Big Sur 11.4 with Quad-Core Intel
Core i7 (2.3 GHz), 32 GB RAM.  We used Z3 version 4.8.10.  The contracts
\texttt{boomerang.tz}, \texttt{deposit.tz}, \texttt{manager.tz},
\texttt{vote.tz}, and \texttt{reservoir.tz} are taken from the benchmark of
Mi-cho-coq~\cite{BernardoCHPT19}.  \texttt{checksig.tz}, discussed above, is derived from \texttt{weather\symbol{95}insurance.tz} of the official Tezos
test
suite.\footnote{\url{https://gitlab.com/tezos/tezos/-/tree/ee2f75bb941522acbcf6d5065a9f3b2/tests_python/contracts/mini_scenarios}}
\texttt{vote\symbol{95}for\symbol{95}delegate.tz} and \texttt{xcat.tz} are taken from the official
test suite; \texttt{xcat.tz} is simplified from the original.  \texttt{tzip.tz} is taken from Tezos
Improvement proposals.\footnote{\url{https://gitlab.com/tezos/tzip/-/blob/b73c7cd5df8e045bbf7ad9ac20a45fb3cb862c87/proposals/tzip-7/tzip-7.md}}
% experiments is a part of the original \AI{?}  code.
\texttt{triangular\symbol{95}num.tz} is a simple test case that we made as an example of using \lstinline{LOOP}.
% \KS{Explain from where they are
%   taken for the other contracts.}.
The source code of these contracts
can be found at the Web interface
of \HELMHOLTZ{}.
% We explain each contract below.
Each contract is supposed to work as follows.
% The following is the
% explanation of each contract.
\begin{itemize}
\item \texttt{boomerang.tz}: Transfers the received amount of money to
  the source account.
\item \texttt{deposit.tz}: Transfers money to the sender if the address of the sender is identical
  to that is stored in the storage.
\item \texttt{manager.tz}: Calls the passed function if the address of
  the caller matches the address stored in the storage.
\item \texttt{vote.tz}: Accepts a vote to a candidate if the voter transfers enough voting fee, and stores the tally.
  % Once a voter sends her choice from a candidate list with a voting fee.  The tally is stored in the contract.
  % Store votes.
\item \texttt{tzip.tz}: One of the components implementing Tezos smart contracts API.  We verify one
  entrypoint of the contract.
\item \texttt{checksig.tz}: The one explained in \secref{casestudy}.
\item \texttt{vote\symbol{95}for\symbol{95}delegate.tz}: Delegates
  one's ballot in voting by stakeholders, which is one of the
  fundamental features of Tezos, to another using a primitive
  operation of Tezos.
\item \texttt{xcat.tz}: Transfers all stored money to
  one of the two accounts specified beforehand if
  called with the correct password.  The account that gets money is decided
  based on whether the contract is called before or after a deadline.
\item \texttt{reservoir.tz}: Sends a certain amount of money to either
  a contract or another depending on whether the contract is executed
  before or after the deadline.
\item \texttt{triangular\symbol{95}num.tz}: Calculates the sum from $1$ to $n$, which is the passed parameter.
\end{itemize}
In the experiments, we verified that each contract indeed works according to the intention explained above.
\texttt{triangular\symbol{95}num.tz} was the only contract
that required a manual annotation for verification in the \lstinline{code} section;
we needed to specify a loop invariant in this contract.
% \AI{I guess you mean this by ``All ... not''.}
% \AI{I suspect none of them uses iteration or higher-order functions.  So, probably it's not surprising.}

\begin{table}
  \caption{Benchmark result}
  \label{tab:result}
  \centering
  \begin{tabular}{|>{\ttfamily}lrr||>{\ttfamily}lrr|}
    \hline
    \rmfamily Filename & \#instr. & time (ms) &  \rmfamily Filename & \#instr. & time (ms) \\ \hline
    boomerang.tz & 17 & 435  &  checksig\symbol{95}unverified.tz & 30 & 462 \\
    deposit.tz & 24 & 451    &  vote\symbol{95}for\symbol{95}delegate.tz & 78 & 608 \\
    manager.tz & 24 & 449    &  xcat.tz & 52 & 513 \\
    vote.tz & 24 & 450       &  reservoir.tz & 45 & 482 \\
    tzip.tz & 537 & 15578    &  triangular\symbol{95}num.tz & 16 & 517 \\
    checksig.tz & 32 & 468 &&&\\
    \hline
  \end{tabular}
\end{table}

Although the numbers of instructions in these contracts are not large,
they capture essential features of smart contracts; every contract
except \texttt{triangular\symbol{95}num.tz}
executes transactions; \texttt{deposit.tz} and \texttt{manager.tz}
check the identity of the caller; and \texttt{checksig.tz} conducts
signature verification.  The time spent on verification is small.

%%% Local Variables:
%%% mode: latex
%%% TeX-master: "../paper.otex"
%%% End:

\section{Related Work}\label{sec:related}

% Smart contracts are growing topic nowadays.  As mentioned in
% \secref{introduction}, after serious attack brought a notorious impact on
% cryptocurrency community, many studies have been proposed to verify correctness
% of smart contracts.

There are several publications on the formalization of programming languages for
writing smart contracts.  Hirai~\cite{Hirai2017} formalizes EVM, a low-level
smart contract language of Ethereum and its implementation, using
Lem~\cite{Owens2011}, a language to specify semantic definitions; definitions
written in Lem can be compiled into definitions in Coq, HOL4, and Isabelle/HOL.
Based on the generated definition, he verifies several properties of Ethereum
smart contracts using Isabelle/HOL.  Bernardo et al.~\cite{BernardoCHPT19}
implemented Mi-Cho-Coq, a formalization of the semantics of Michelson using the
Coq proof assistant.  They also verified several Michelson contracts.  Compared
to their approach, we aim to develop an automated verification tool for smart
contracts.  Park et al.~\cite{Park2018-jz} developed a formal verification tool
for EVM by using the K-framework~\cite{Rosu2010-hu}, which can be used to
derive a symbolic model checker from a formally specified language semantics (in
this case, formalized EVM semantics~\cite{Hildenbrandt2018-wu}), and
successfully applied the derived model checker to a few EVM contracts.  It would
be interesting to formalize the semantics of Michelson in the K-framework to
compare \HELMHOLTZ with the derived model checker.
\AI{K-Michelson?}

The DAO attack~\cite{theDAOAttack}, mentioned in \secref{introduction}, is
one of the notorious attacks on a smart contract.  It exploits a
vulnerability of a smart contract that is related to a callback.
Grossman et al.~\cite{Grossman2017} proposed a type-based technique to
verify that execution of a smart contract that may contain
callbacks is equivalent to another execution without any callback.
This property, called \emph{effectively callback freedom}, can be seen
as one of the criteria for execution of a smart contract not to be
vulnerable to the DAO-like attack.  Their type system focuses on
verifying the ECF property of the \emph{execution} of a smart contract,
whereas ours concerns the verification of generic functional properties
of a smart contract.

Benton proposes a program logic for a minimal stack-based programming
language~\cite{Benton2005-ir}.  His program logic can give an assertion
to a stack as our stack refinement types do.  However, his language
does not support first-class functions nor instructions for dealing
with smart contracts (e.g., signature verification).

Our type system is an extension of the Michelson type system with
refinement types, which have been successfully applied to various
programming languages~\cite{%
  RondonKJ08,UnnoK09,KawaguchiRJ09,KobayashiSU11,Terauchi10,ZhuJ13,VazouSJVJ2014,Xi2007,XiH2001}.
%
% are more sophisticated type systems than traditional simply typed
% systems.  The key component of refinement type systems is
% \emph{refinement types} of the form \(\ottsym{\{}   \ottmv{x} \mathord:\allowbreak \ottnt{T}   \mid  \varphi  \ottsym{\}}\).
% % , where \(\ottmv{x}\) is bound in \emph{refinement predicate}
% % \(\varphi\).  A Refinement type \(\ottsym{\{}   \ottmv{x} \mathord:\allowbreak \ottnt{T}   \mid  \varphi  \ottsym{\}}\) denotes the subset values of the
% % \emph{underlying type} \(\ottnt{T}\) which satisfy \(\varphi\).  So, for instance,
% % \(\texttt{\textcolor{red}{<<no parses (char 9): \{x:int\mbox{$\mid$}x>***0\} >>}}\) denotes positive numbers.  Using refinement types, we can
% % express more detailed properties by types and deal with the properties by a type
% % system.
% %
% A refinement type system is mainly motivated by program verification and
% typically developed on functional programming languages like
% \(\lambda\)-calculus and more practically ML-like language~\cite{mumon} and
% Haskell-like language~\cite{VazouSJVJ2014}.  For low-level languages, just a few
% work~\cite{XiH2001} is proposed for a register-base machine.  So, our work is
% the first attempt to apply the methodology on a stack-based language.
DTAL~\cite{XiH2001} is a notable example of an application of refinement
types to an assembly language, a low-level language like Michelson.  A
DTAL program defines a computation using registers; we are not aware
of refinement types for stack-based languages like Michelson.

% Someone might notice that our type system looks like the Floyd-Hoare
% logic~\cite{Floyd1967,Hoare1969}.  For stack-based languages, some programming
% logic has been developed~\cite{Benton2005,Quigley2003}.  However, those are a
% bit complicated because of uncontrolled structures of there aimed languages (JVM
% and CLI) for.  Fortunately, since our target language is structured programming
% one, reasoning system becomes rather straightforward like the original
% Floyd-Hoare logic.

We notice the resemblance between our type system and a program logic
for PCF proposed by Honda and Yoshida~\cite{HondaY04},
although the targets of verification are different.  % \AI{Removed
  % ``(i.e., a functional program in theirs; a stack-based program in
  % ours)'' because, I think, Michelson is functional enough, although the use of higher-order functions may be rare.}
Their logic supports a judgment of the form $A \vdash e \COL_u B$, where $e$
is a PCF program, $A$ is a pre-condition assertion, $B$ is a
post-condition assertion, and $u$ represents the value that $e$
evaluates to and can be used in $B$, which resembles our type
judgment in the formalization in \secref{system}.  Their
assertion language also incorporates a term expression $f \bullet x$,
which expresses the value resulting from the application of $f$ to
$x$; this expression resembles the formula $ \ottkw{call} ( \ottnt{t_{{\mathrm{1}}}} ,  \ottnt{t_{{\mathrm{2}}}} ) =  \ottnt{t_{{\mathrm{3}}}} $
used in a refinement predicate.  We have not noticed an automated verifier
implemented based on their logic.  Further comparison
is interesting future work.

% \YN{Liquid Haskell~\cite{VazouSJVJ2014}, K-framework, F*, spec\#, why3,Albert~\cite{BernardoCPT2020}}

%%% Local Variables:
%%% mode: latex
%%% TeX-master: "../paper.otex"
%%% End:

\section{Conclusion}\label{sec:conclusion}

We described our automated verification tool \HELMHOLTZ{} for the
smart contract language Michelson based on the refinement
type system for \miniMic{}.  \HELMHOLTZ{} verifies whether a Michelson
program follows a specification given in the form of a refinement
type.  We also demonstrated that \HELMHOLTZ{} successfully verifies
various practical Michelson contracts.

% The conditions are written in a first-order logic which contains a
% lot of smart contract related constants.  Verification process is
% standard one, trying to typecheck given code against a specified
% type, namely specification, and discharge verification conditions
% obtained during the typecheck by a SMT-solver (Z3).

Currently, \HELMHOLTZ{} supports approximately 80\% of the whole
instructions of the Michelson language.  The definition of a measure
function is limited in the sense that, for example, it can define only
a function with one argument.  We are currently extending \HELMHOLTZ{}
so that it can deal with more programs.
\AI{Does LiquidHaskell support multi-argument measure functions?}

% We verify several Michelson code from a kind of test cases to
% practical and pre-existing contract code.

% \paragraph{Future Work.}

\HELMHOLTZ{} currently verifies the behavior of a single contract, although a blockchain application often consists of multiple contracts
in which contract calls are chained.  To verify such an application as
a whole, we plan to extend \HELMHOLTZ{} so that it can verify an
inter-contract behavior compositionally by combining the verification
results of each contract.

% a blockchain platform, a transaction is the meaningful
% (recorded on the blockchain) unit.  So, it is quite interest and
% important to verify correctness of behavior involving multiple
% contract invocations.  We have been investigating a verification
% technique to verify such correctness just by examining refinement
% types of each contract which can be obtained by \HELMHOLTZ.

% We show that \HELMHOLTZ can verifies most existing benchmarks but one
% exception (multisig) in mi-cho-coq project.  This is because our
% reasoning mechanism is still weak for handling inductive properties.
% Measures can handle such properties but can define limited functions.
% There is an existing work~\cite{Vazou2017} to use more general
% user-defined functions in an assertion language.  We could adopt their
% contribution, but it seems not trivial since their technique relies on
% what the target language is a functional programming language.

%%% Local Variables:
%%% mode: latex
%%% TeX-master: "../paper.otex"
%%% End:

% \KS{Akira's current affiliation.}

\begin{acknowledgements}
  This work has been partially supported by a research grant from
  Tezos Foundations.  Igarashi's first encounter with formal proofs
  was when he visited Prof. Hagiya's group as an intern and played
  Boomborg PC.
\end{acknowledgements}

\iffullver
\bibliographystyle{spmpsci}
\else
\bibliographystyle{splncs04}
\fi
\bibliography{reference}

\iffullver
\appendix

% \section{Proofs}\label{sec:detailedproofs}
\materialtrue
\section{Proof of Soundness}\label{sec:detailedproofs}

\emph{This Appendix is only for reviewing.  If this is accepted, we
  will remove it, upload a preprint version with the proof to arXiv.
  The article will refer to the preprint version.}

We prove our main theorem (Theorem~\ref{prop:typing/sound}), whose
first item is Theorem~\ref{prop:typing/sound'}.  The proof is close to
a proof of soundness of Hoare logic, with a few extra complications
due to the presence of first-class functions.  We first prove a few
lemmas related to \(\ottkw{LOOP}\) (Lemma~\ref{prop:soundness/loop}),
\(\ottkw{ITER}\) (Lemma~\ref{prop:soundness/iter}), predicate
\texttt{call} (Lemmas~\ref{prop:soundness/lambda} and
\ref{prop:soundness/exec}), and subtyping
(Lemma~\ref{prop:soundness/subtyping}).  We often use
Theorem~\ref{prop:styping/sound'} implicitly.

\begin{prop}[number=11]{soundness/loop} Suppose $\mathit{IS}$ satisfies
  \begin{gather}
    \ottnt{S_{{\mathrm{1}}}}  \vdash  \mathit{IS}  \Downarrow  \ottnt{S_{{\mathrm{2}}}} \text{ and }
    \sigma  \ottsym{:}  \Gamma  \models  \ottnt{S_{{\mathrm{1}}}}  \ottsym{:}  \ottsym{\{}  \Upsilon  \mid  \exists \,  \ottmv{x} \mathord:\allowbreak \ottkw{int}   \ottsym{.}  \varphi  \wedge  \ottmv{x}  \neq  0  \ottsym{\}} \text{ imply } \sigma  \ottsym{:}  \Gamma  \models  \ottnt{S_{{\mathrm{2}}}}  \ottsym{:}  \ottsym{\{}   \ottmv{x} \mathord:\allowbreak \ottkw{int}   \triangleright  \Upsilon  \mid  \varphi  \ottsym{\}} \hyp{OIH}
  \end{gather}
  for any $\ottnt{S_{{\mathrm{1}}}}$ and $\ottnt{S_{{\mathrm{2}}}}$.
  If
  \begin{gather}
    \ottnt{S_{{\mathrm{1}}}}  \vdash  \ottkw{LOOP} \, \mathit{IS}  \Downarrow  \ottnt{S_{{\mathrm{2}}}} \hyp{eval}\\
    \sigma  \ottsym{:}  \Gamma  \models  \ottnt{S_{{\mathrm{1}}}}  \ottsym{:}  \ottsym{\{}   \ottmv{x} \mathord:\allowbreak \ottkw{int}   \triangleright  \Upsilon  \mid  \varphi  \ottsym{\}} \hyp{semty}
  \end{gather}
  then \(\sigma  \ottsym{:}  \Gamma  \models  \ottnt{S_{{\mathrm{2}}}}  \ottsym{:}  \ottsym{\{}  \Upsilon  \mid  \exists \,  \ottmv{x} \mathord:\allowbreak \ottkw{int}   \ottsym{.}  \varphi  \wedge  \ottmv{x}  \ottsym{=}  0  \ottsym{\}}\).

  \proof By induction on the derivation of \hypref{eval}.
  We conduct the last rule that derives \hypref{eval}, which is either \ruleref{E-LoopT} or \ruleref{E-LoopF}.
  \begin{match}
  \item[\ruleref{E-LoopT}] We have
    \begin{gather}
       \ottnt{S_{{\mathrm{1}}}} \ottsym{=} \ottnt{i}  \triangleright  \ottnt{S}  \hyp{h1}\\
      \ottnt{i}  \neq  0 \hyp{h2}\\
      \ottnt{S}  \vdash  \mathit{IS}  \Downarrow  \ottnt{S'} \hyp{h3}\\
      \ottnt{S'}  \vdash  \ottkw{LOOP} \, \mathit{IS}  \Downarrow  \ottnt{S_{{\mathrm{2}}}}. \hyp{h4}
    \end{gather}
    From \hypref{h1}, \hypref{h2}, and \hypref{semty}, we have
    \begin{gather}
      % \texttt{\textcolor{red}{<<no parses (char 8): G\mbox{$\mid$}= i : ***int >>}} \hyp{h5}\\
      \sigma  \ottsym{:}  \Gamma  \models  \ottnt{S}  \ottsym{:}  \ottsym{\{}  \Upsilon  \mid  \exists \,  \ottmv{x} \mathord:\allowbreak \ottkw{int}   \ottsym{.}  \varphi  \wedge  \ottmv{x}  \neq  0  \ottsym{\}} \hyp{h6}
    \end{gather}
    % Then, we have \(\sigma  \ottsym{:}  \Gamma  \models  \ottnt{S}  \ottsym{:}  \ottsym{\{}  \Upsilon  \mid  \exists \,  \ottmv{x} \mathord:\allowbreak \ottkw{int}   \ottsym{.}  \varphi  \wedge  \ottmv{x}  \neq  0  \ottsym{\}}\) by  \propref{semty/strengthen(ift)}, \hypref{h2}, \hypref{h6}.
    From \hypref{OIH}, we have \(\sigma  \ottsym{:}  \Gamma  \models  \ottnt{S'}  \ottsym{:}  \ottsym{\{}   \ottmv{x} \mathord:\allowbreak \ottkw{int}   \triangleright  \Upsilon  \mid  \varphi  \ottsym{\}}\).
    Then, the goal from \hypref{h4}, \hypref{h6}, and IH.
    % Applying \hypref{OIH} to \hypref{h3}, \(\sigma  \ottsym{:}  \Gamma  \models  \ottnt{S'}  \ottsym{:}  \ottsym{\{}   \ottmv{x} \mathord:\allowbreak \ottkw{int}   \triangleright  \Upsilon  \mid  \varphi  \ottsym{\}}\).  Now the
    % goal follows by IH.

  \item[\ruleref{E-LoopF}] We have $ \ottnt{S_{{\mathrm{1}}}} \ottsym{=} 0  \triangleright  \ottnt{S_{{\mathrm{2}}}} $.  The goal
    \(\sigma  \ottsym{:}  \Gamma  \models  \ottnt{S_{{\mathrm{2}}}}  \ottsym{:}  \ottsym{\{}  \Upsilon  \mid  \exists \,  \ottmv{x} \mathord:\allowbreak \ottkw{int}   \ottsym{.}  \varphi  \wedge  \ottmv{x}  \ottsym{=}  0  \ottsym{\}}\) follows from
    \propref{subty/exists=>subst'} and \hypref{semty}.
  \end{match}
\end{prop}

\begin{prop}[number=12]{soundness/iter}
  Suppose % \AI{Do we need $ \ottmv{x_{{\mathrm{1}}}}  \notin   \text{fvars}( \varphi )  $, too?}
  \begin{gather}
     \ottmv{x_{{\mathrm{1}}}}  \notin   \text{fvars}( \varphi )   \hyp{fv1} \\
     \ottmv{x_{{\mathrm{2}}}}  \notin   \text{fvars}( \varphi )   \hyp{fv2}
    %  \Gamma  \ottsym{,}   \ottmv{x_{{\mathrm{2}}}} \mathord:\allowbreak \ottnt{T} \, \ottkw{list}  \vdash \ottsym{\{}   \ottmv{x_{{\mathrm{1}}}} \mathord:\allowbreak \ottnt{T}   \triangleright  \Upsilon  \mid  \exists \,  \ottmv{x} \mathord:\allowbreak \ottnt{T} \, \ottkw{list}   \ottsym{.}  \varphi  \wedge  \ottmv{x_{{\mathrm{1}}}}  ::  \ottmv{x_{{\mathrm{2}}}}  \ottsym{=}  \ottmv{x}  \ottsym{\}} \; \mathit{IS} \; \ottsym{\{}  \Upsilon  \mid  \exists \,  \ottmv{x} \mathord:\allowbreak \ottnt{T} \, \ottkw{list}   \ottsym{.}  \varphi  \wedge  \ottmv{x_{{\mathrm{2}}}}  \ottsym{=}  \ottmv{x}  \ottsym{\}}  \hyp{type}
    % \Gamma, \Phi_{{\mathrm{1}}}, \Phi_{{\mathrm{2}}}, \ottnt{S_{{\mathrm{1}}}}, \ottnt{S_{{\mathrm{2}}}}, \text{ if }  \Gamma \vdash \Phi_{{\mathrm{1}}} \; \mathit{IS} \; \Phi_{{\mathrm{2}}} , \ottnt{S_{{\mathrm{1}}}}  \vdash  \mathit{IS}  \Downarrow  \ottnt{S_{{\mathrm{2}}}}, \text{ and } \sigma  \ottsym{:}  \Gamma  \models  \ottnt{S_{{\mathrm{1}}}}  \ottsym{:}  \Phi_{{\mathrm{1}}}, \text{ then } \sigma  \ottsym{:}  \Gamma  \models  \ottnt{S_{{\mathrm{2}}}}  \ottsym{:}  \Phi_{{\mathrm{2}}}. \hyp{OIH}
  \end{gather}
  Suppose also that
  \begin{gather*}
    \ottnt{S'_{{\mathrm{1}}}}  \vdash  \mathit{IS}  \Downarrow  \ottnt{S'_{{\mathrm{2}}}}  \text{ and } \sigma'  \ottsym{:}  \Gamma  \ottsym{,}   \ottmv{x_{{\mathrm{2}}}} \mathord:\allowbreak \ottnt{T} \, \ottkw{list}   \models  \ottnt{S'_{{\mathrm{1}}}}  \ottsym{:}  \ottsym{\{}   \ottmv{x_{{\mathrm{1}}}} \mathord:\allowbreak \ottnt{T}   \triangleright  \Upsilon  \mid  \exists \,  \ottmv{x} \mathord:\allowbreak \ottnt{T} \, \ottkw{list}   \ottsym{.}  \varphi  \wedge  \ottmv{x_{{\mathrm{1}}}}  ::  \ottmv{x_{{\mathrm{2}}}}  \ottsym{=}  \ottmv{x}  \ottsym{\}} \text{ imply }\\
    \sigma'  \ottsym{:}  \Gamma  \ottsym{,}   \ottmv{x_{{\mathrm{2}}}} \mathord:\allowbreak \ottnt{T} \, \ottkw{list}   \models  \ottnt{S'_{{\mathrm{2}}}}  \ottsym{:}  \ottsym{\{}  \Upsilon  \mid  \exists \,  \ottmv{x} \mathord:\allowbreak \ottnt{T} \, \ottkw{list}   \ottsym{.}  \varphi  \wedge  \ottmv{x_{{\mathrm{2}}}}  \ottsym{=}  \ottmv{x}  \ottsym{\}} \hyp{OIH}
  \end{gather*}
  for any $\ottnt{S'_{{\mathrm{1}}}}$, $\ottnt{S'_{{\mathrm{2}}}}$, and $\sigma'$.
  If
  \begin{gather}
    \ottnt{S_{{\mathrm{1}}}}  \vdash  \ottkw{ITER} \, \mathit{IS}  \Downarrow  \ottnt{S_{{\mathrm{2}}}} \hyp{eval}\\
    \sigma  \ottsym{:}  \Gamma  \models  \ottnt{S_{{\mathrm{1}}}}  \ottsym{:}  \ottsym{\{}   \ottmv{x} \mathord:\allowbreak \ottnt{T} \, \ottkw{list}   \triangleright  \Upsilon  \mid  \varphi  \ottsym{\}} \hyp{semty}
  \end{gather}
  then \(\sigma  \ottsym{:}  \Gamma  \models  \ottnt{S_{{\mathrm{2}}}}  \ottsym{:}  \ottsym{\{}  \Upsilon  \mid  \exists \,  \ottmv{x} \mathord:\allowbreak \ottnt{T} \, \ottkw{list}   \ottsym{.}  \varphi  \wedge  \ottmv{x}  \ottsym{=}  \ottsym{[}  \ottsym{]}  \ottsym{\}}\).

  \proof By induction on the derivation of
  \hypref{eval}.  We conduct case analysis on the last rule that
  derives \hypref{eval}, which is either \ruleref{E-IterNil} or
  \ruleref{E-IterCons}.
  \begin{match}
  \item[\ruleref{E-IterNil}] We have
    \begin{gather}
       \ottnt{S_{{\mathrm{1}}}} \ottsym{=} \ottsym{[}  \ottsym{]}  \triangleright  \ottnt{S_{{\mathrm{2}}}} . \hyp{h1}
    \end{gather}
    The goal follows from \propref{subty/exists=>subst'} and \hypref{semty}.

  \item[\ruleref{E-IterCons}] We have
    \begin{gather}
       \ottnt{S_{{\mathrm{1}}}} \ottsym{=} \ottnt{V_{{\mathrm{1}}}}  ::  \ottnt{V_{{\mathrm{2}}}}  \triangleright  \ottnt{S}  \hyp{h1}\\
      \ottnt{V_{{\mathrm{1}}}}  \triangleright  \ottnt{S}  \vdash  \mathit{IS}  \Downarrow  \ottnt{S'} \hyp{h2}\\
      \ottnt{V_{{\mathrm{2}}}}  \triangleright  \ottnt{S'}  \vdash  \ottkw{ITER} \, \mathit{IS}  \Downarrow  \ottnt{S_{{\mathrm{2}}}}. \hyp{h3}
    \end{gather}
    Therefore, from \hypref{h1}, \hypref{semty}, and \propref{subty/exists=>subst'}, we have
    \begin{gather}
      % \texttt{\textcolor{red}{<<no parses (char 14): \mbox{$\mid$}= V1 :: V2 : ***T list >>}} \hyp{h4}\\
      \sigma  \ottsym{:}  \Gamma  \models  \ottnt{S}  \ottsym{:}  \ottsym{\{}  \Upsilon  \mid  \exists \,  \ottmv{x} \mathord:\allowbreak \ottnt{T} \, \ottkw{list}   \ottsym{.}  \varphi  \wedge  \ottmv{x}  \ottsym{=}  \ottnt{V_{{\mathrm{1}}}}  ::  \ottnt{V_{{\mathrm{2}}}}  \ottsym{\}}. \hyp{h5}
    \end{gather}
    Therefore, we have $\sigma  \ottsym{:}  \Gamma  \models  \ottnt{S}  \ottsym{:}  \ottsym{\{}  \Upsilon  \mid  \exists \,  \ottmv{x} \mathord:\allowbreak \ottnt{T} \, \ottkw{list}   \ottsym{,}   \ottmv{x_{{\mathrm{1}}}} \mathord:\allowbreak \ottnt{T}   \ottsym{,}   \ottmv{x_{{\mathrm{2}}}} \mathord:\allowbreak \ottnt{T} \, \ottkw{list}   \ottsym{.}  \varphi  \wedge  \ottmv{x}  \ottsym{=}  \ottmv{x_{{\mathrm{1}}}}  ::  \ottmv{x_{{\mathrm{2}}}}  \wedge  \ottmv{x_{{\mathrm{1}}}}  \ottsym{=}  \ottnt{V_{{\mathrm{1}}}}  \wedge  \ottmv{x_{{\mathrm{2}}}}  \ottsym{=}  \ottnt{V_{{\mathrm{2}}}}  \ottsym{\}}$
    and hence
    \begin{gather}
       \sigma  [  \ottmv{x_{{\mathrm{2}}}}  \mapsto  \ottnt{V_{{\mathrm{2}}}}  ]   \ottsym{:}  \Gamma  \ottsym{,}   \ottmv{x_{{\mathrm{2}}}} \mathord:\allowbreak \ottnt{T} \, \ottkw{list}   \models  \ottnt{V_{{\mathrm{1}}}}  \triangleright  \ottnt{S}  \ottsym{:}  \ottsym{\{}   \ottmv{x_{{\mathrm{1}}}} \mathord:\allowbreak \ottnt{T}   \triangleright  \Upsilon  \mid  \exists \,  \ottmv{x} \mathord:\allowbreak \ottnt{T} \, \ottkw{list}   \ottsym{.}  \varphi  \wedge  \ottmv{x}  \ottsym{=}  \ottmv{x_{{\mathrm{1}}}}  ::  \ottmv{x_{{\mathrm{2}}}}  \ottsym{\}} \hyp{h6}
      % \sigma  \ottsym{:}  \Gamma  \models  \ottnt{V_{{\mathrm{1}}}}  \triangleright  \ottnt{S}  \ottsym{:}  \ottsym{\{}   \ottmv{x_{{\mathrm{1}}}} \mathord:\allowbreak \ottnt{T}   \triangleright  \Upsilon  \mid  \exists \,  \ottmv{x} \mathord:\allowbreak \ottnt{T} \, \ottkw{list}   \ottsym{,}   \ottmv{x_{{\mathrm{2}}}} \mathord:\allowbreak \ottnt{T} \, \ottkw{list}   \ottsym{.}  \varphi  \wedge  \ottmv{x}  \ottsym{=}  \ottnt{V_{{\mathrm{1}}}}  ::  \ottmv{x_{{\mathrm{2}}}}  \wedge  \ottmv{x_{{\mathrm{2}}}}  \ottsym{=}  \ottnt{V_{{\mathrm{2}}}}  \ottsym{\}}
    \end{gather}
    % Applying \propref{vtyping/inv(cons)} to \hypref{h4}, we also have
    % \begin{gather}
    %   \texttt{\textcolor{red}{<<no parses (char 8): \mbox{$\mid$}= V1 : ***T >>}} \hyp{h6}\\
    %   \texttt{\textcolor{red}{<<no parses (char 8): \mbox{$\mid$}= V2 : ***T list >>}} \hyp{h7}.
    % \end{gather}
    From \hypref{OIH} and \hypref{h2}, we have
    \begin{gather}
       \sigma  [  \ottmv{x_{{\mathrm{2}}}}  \mapsto  \ottnt{V_{{\mathrm{2}}}}  ]   \ottsym{:}  \Gamma  \ottsym{,}   \ottmv{x_{{\mathrm{2}}}} \mathord:\allowbreak \ottnt{T} \, \ottkw{list}   \models  \ottnt{S'}  \ottsym{:}  \ottsym{\{}  \Upsilon  \mid  \exists \,  \ottmv{x} \mathord:\allowbreak \ottnt{T} \, \ottkw{list}   \ottsym{.}  \varphi  \wedge  \ottmv{x}  \ottsym{=}  \ottmv{x_{{\mathrm{2}}}}  \ottsym{\}} \hyp{h7}.
    \end{gather}
    From \propref{subty/exists=>subst'}, we have
    \begin{gather}
      \sigma  \ottsym{:}  \Gamma  \models  \ottnt{V_{{\mathrm{2}}}}  \triangleright  \ottnt{S'}  \ottsym{:}  \ottsym{\{}   \ottmv{x_{{\mathrm{2}}}} \mathord:\allowbreak \ottnt{T} \, \ottkw{list}   \triangleright  \Upsilon  \mid  \exists \,  \ottmv{x} \mathord:\allowbreak \ottnt{T} \, \ottkw{list}   \ottsym{.}  \varphi  \wedge  \ottmv{x}  \ottsym{=}  \ottmv{x_{{\mathrm{2}}}}  \ottsym{\}} \hyp{h8}.
    \end{gather}
    Knowing \hypref{fv2}, we have
    \begin{gather}
      \sigma  \ottsym{:}  \Gamma  \models  \ottnt{V_{{\mathrm{2}}}}  \triangleright  \ottnt{S'}  \ottsym{:}  \ottsym{\{}   \ottmv{x} \mathord:\allowbreak \ottnt{T} \, \ottkw{list}   \triangleright  \Upsilon  \mid  \varphi  \ottsym{\}} \hyp{9}
    \end{gather}
    from \hypref{h8}.
    Now the goal follows from IH, \hypref{h2}, and \hypref{9}.

  \end{match}
\end{prop}

\begin{prop}[number=13]{soundness/lambda}
  If
  \begin{gather}
    \ottmv{y_{{\mathrm{1}}}}  \neq  \ottmv{y_{{\mathrm{2}}}} \hyp{neq}\\
      \ottmv{y'_{{\mathrm{1}}}} \mathord:\allowbreak \ottnt{T_{{\mathrm{1}}}}   \ottsym{,}   \ottmv{y_{{\mathrm{1}}}} \mathord:\allowbreak \ottnt{T_{{\mathrm{1}}}}    \vdash   \varphi_{{\mathrm{1}}}  : \mathord{*}  \hyp{wf1}\\
      \ottmv{y'_{{\mathrm{1}}}} \mathord:\allowbreak \ottnt{T_{{\mathrm{1}}}}   \ottsym{,}   \ottmv{y_{{\mathrm{2}}}} \mathord:\allowbreak \ottnt{T_{{\mathrm{2}}}}    \vdash   \varphi_{{\mathrm{2}}}  : \mathord{*}  \hyp{wf2}\\
      \langle  \mathit{IS}  \rangle   :  \ottnt{T_{{\mathrm{1}}}}  \to  \ottnt{T_{{\mathrm{2}}}}  \hyp{typ}\\
    \left(
      \begin{split}
        \text{For any } & \ottnt{V_{{\mathrm{1}}}}, \ottnt{V_{{\mathrm{2}}}}, \sigma, \\
        \text{ if } & \ottnt{V_{{\mathrm{1}}}}  \triangleright  \ddagger  \vdash  \mathit{IS}  \Downarrow  \ottnt{V_{{\mathrm{2}}}}  \triangleright  \ddagger \text{ and } \\
         &\sigma  \ottsym{:}   \ottmv{y'_{{\mathrm{1}}}} \mathord:\allowbreak \ottnt{T_{{\mathrm{1}}}}   \models  \ottnt{V_{{\mathrm{1}}}}  \triangleright  \ddagger  \ottsym{:}  \ottsym{\{}   \ottmv{y_{{\mathrm{1}}}} \mathord:\allowbreak \ottnt{T_{{\mathrm{1}}}}   \triangleright  \ddagger  \mid  \ottmv{y'_{{\mathrm{1}}}}  \ottsym{=}  \ottmv{y_{{\mathrm{1}}}}  \wedge  \varphi_{{\mathrm{1}}}  \ottsym{\}} \\
        \text{ then } & \sigma  \ottsym{:}   \ottmv{y'_{{\mathrm{1}}}} \mathord:\allowbreak \ottnt{T_{{\mathrm{1}}}}   \models  \ottnt{V_{{\mathrm{2}}}}  \triangleright  \ddagger  \ottsym{:}  \ottsym{\{}   \ottmv{y_{{\mathrm{2}}}} \mathord:\allowbreak \ottnt{T_{{\mathrm{2}}}}   \triangleright  \ddagger  \mid  \varphi_{{\mathrm{2}}}  \ottsym{\}}
      \end{split}
    \right) \hyp{OIH}
  \end{gather}
  then
  \(\Gamma  \models  \forall \,  \ottmv{y'_{{\mathrm{1}}}} \mathord:\allowbreak \ottnt{T_{{\mathrm{1}}}}   \ottsym{,}   \ottmv{y_{{\mathrm{1}}}} \mathord:\allowbreak \ottnt{T_{{\mathrm{1}}}}   \ottsym{,}   \ottmv{y_{{\mathrm{2}}}} \mathord:\allowbreak \ottnt{T_{{\mathrm{2}}}}   \ottsym{.}  \ottmv{y'_{{\mathrm{1}}}}  \ottsym{=}  \ottmv{y_{{\mathrm{1}}}}  \wedge  \varphi_{{\mathrm{1}}}  \wedge   \ottkw{call} (  \langle  \mathit{IS}  \rangle  ,  \ottmv{y'_{{\mathrm{1}}}} ) =  \ottmv{y_{{\mathrm{2}}}}   \implies  \varphi_{{\mathrm{2}}}\) for any \(\Gamma\).

  \proof By the definition of the semantics of $\mathtt{call}$.
\end{prop}

\begin{prop}[number=14]{soundness/exec}
  If
  \begin{gather}
    \ottnt{V_{{\mathrm{1}}}}  \triangleright  \ddagger  \vdash  \mathit{IS}  \Downarrow  \ottnt{V_{{\mathrm{2}}}}  \triangleright  \ddagger \hyp{eval}\\
     \ottnt{V_{{\mathrm{1}}}}  :  \ottnt{T_{{\mathrm{1}}}}  \hyp{typ1}\\
     \ottnt{V_{{\mathrm{2}}}}  :  \ottnt{T_{{\mathrm{2}}}}  \hyp{typ2}\\
      \langle  \mathit{IS}  \rangle   :  \ottnt{T_{{\mathrm{1}}}}  \to  \ottnt{T_{{\mathrm{2}}}}  \hyp{3}
  \end{gather}
  then
  \(\Gamma  \models   \ottkw{call} (  \langle  \mathit{IS}  \rangle  ,  \ottnt{V_{{\mathrm{1}}}} ) =  \ottnt{V_{{\mathrm{2}}}} \) for any \(\Gamma\).

  \proof By the definition of the semantics of $\mathtt{call}$.
\end{prop}

\begin{prop}[number=15]{soundness/subtyping}
  If
  \begin{gather}
    \Gamma  \vdash  \Phi_{{\mathrm{1}}}  \ottsym{<:}  \Phi_{{\mathrm{2}}} \hyp{subty}\\
    \sigma  \ottsym{:}  \Gamma  \models  \ottnt{S}  \ottsym{:}  \Phi_{{\mathrm{1}}} \hyp{semty}
  \end{gather}
  then \(\sigma  \ottsym{:}  \Gamma  \models  \ottnt{S}  \ottsym{:}  \Phi_{{\mathrm{2}}}\).

  \proof Straightforward from \propref{def/subty}.

  % By \propref{def/subty} of \hypref{subty}, we have
  % \begin{gather}
  %    \Phi_{{\mathrm{1}}} \ottsym{=} \ottsym{\{}  \Upsilon  \mid  \varphi_{{\mathrm{1}}}  \ottsym{\}}  \hyp{1}\\
  %    \Phi_{{\mathrm{2}}} \ottsym{=} \ottsym{\{}  \Upsilon  \mid  \varphi_{{\mathrm{2}}}  \ottsym{\}}  \hyp{2}\\
  %   \Gamma  \ottsym{,}   \widehat{ \Upsilon }   \models  \varphi_{{\mathrm{1}}}  \implies  \varphi_{{\mathrm{2}}} \hyp{3}
  % \end{gather}
  % for some \(\Upsilon\), \(\varphi_{{\mathrm{1}}}\), and \(\varphi_{{\mathrm{2}}}\).
  % So, the goal follows by \propref{semty/logic/imply}, \hypref{semty}, and \hypref{3}.
\end{prop}

\begin{prop}[name=Soundness, type=theorem, number=16]{typing/sound}
  The following two statements hold.
  \begin{statements}
  \item If
    \begin{gather}
       \Gamma \vdash \Phi_{{\mathrm{1}}} \; \mathit{IS} \; \Phi_{{\mathrm{2}}}  \hyp{type1}\\
      \ottnt{S_{{\mathrm{1}}}}  \vdash  \mathit{IS}  \Downarrow  \ottnt{S_{{\mathrm{2}}}} \hyp{eval1}\\
      \sigma  \ottsym{:}  \Gamma  \models  \ottnt{S_{{\mathrm{1}}}}  \ottsym{:}  \Phi_{{\mathrm{1}}} \hyp{semty1}
    \end{gather}
    then \(\sigma  \ottsym{:}  \Gamma  \models  \ottnt{S_{{\mathrm{2}}}}  \ottsym{:}  \Phi_{{\mathrm{2}}}\).
  \item If
    \begin{gather}
       \Gamma \vdash \Phi_{{\mathrm{1}}} \; \ottnt{I} \; \Phi_{{\mathrm{2}}}  \hyp{type2}\\
      \ottnt{S_{{\mathrm{1}}}}  \vdash  \ottnt{I}  \Downarrow  \ottnt{S_{{\mathrm{2}}}} \hyp{eval2}\\
      \sigma  \ottsym{:}  \Gamma  \models  \ottnt{S_{{\mathrm{1}}}}  \ottsym{:}  \Phi_{{\mathrm{1}}} \hyp{semty2}
    \end{gather}
    then \(\sigma  \ottsym{:}  \Gamma  \models  \ottnt{S_{{\mathrm{2}}}}  \ottsym{:}  \Phi_{{\mathrm{2}}}\).
  \end{statements}

  \proof The proof is done by mutual induction on the given derivation of
  \hypref{type1} and \hypref{type2}.
  \begin{match}
  \item[\ruleref{RT-Nop}] We have $ \mathit{IS} \ottsym{=} \ottsym{\{}  \ottsym{\}} $ and $ \Phi_{{\mathrm{1}}} \ottsym{=} \Phi_{{\mathrm{2}}} $.
    % \begin{gather}
    %    \mathit{IS} \ottsym{=} \ottsym{\{}  \ottsym{\}}  \hyp{1}\\
    %    \Phi_{{\mathrm{1}}} \ottsym{=} \Phi_{{\mathrm{2}}}  \hyp{2}.
    % \end{gather}
    % By \propref{eval/inv(nop)}, \hypref{eval1}, and \hypref{1}, we have
    % \begin{gather}
    %    \ottnt{S_{{\mathrm{1}}}} \ottsym{=} \ottnt{S_{{\mathrm{2}}}} . \hyp{3}
    % \end{gather}
    % Now we can see the goal is \hypref{semty1}.
    The last rule that is used to derive \hypref{eval1} is \ruleref{E-Nop}.
    Therefore, we have $ \ottnt{S_{{\mathrm{1}}}} \ottsym{=} \ottnt{S_{{\mathrm{2}}}} $, which is followed by \hypref{semty1}.
  \item[\ruleref{RT-Seq}] We have
    \begin{gather}
       \mathit{IS} \ottsym{=} \ottsym{\{}  \ottnt{I'}  \ottsym{;}  \mathit{IS}'  \ottsym{\}}  \hyp{1}\\
       \Gamma \vdash \Phi_{{\mathrm{1}}} \; \ottnt{I'} \; \Phi'  \hyp{2}\\
       \Gamma \vdash \Phi' \; \mathit{IS}' \; \Phi_{{\mathrm{2}}}  \hyp{3}
    \end{gather}
    for some \(\ottnt{I'}\), \(\mathit{IS}'\), and \(\Phi'\).
    The last rule that derives \hypref{eval1} is \ruleref{E-Seq}.
    Therefore, we have

    % By
    % \propref{eval/inv(seq)}, \hypref{eval1}, and \hypref{1}, we have
    \begin{gather}
      \ottnt{S_{{\mathrm{1}}}}  \vdash  \ottnt{I'}  \Downarrow  \ottnt{S'} \hyp{4}\\
      \ottnt{S'}  \vdash  \mathit{IS}'  \Downarrow  \ottnt{S_{{\mathrm{2}}}} \hyp{5}
    \end{gather}
    for some \(\ottnt{S'}\).  Now we have
    \begin{gather}
      \sigma  \ottsym{:}  \Gamma  \models  \ottnt{S'}  \ottsym{:}  \Phi' \hyp{6}
    \end{gather}
    by applying IH to \hypref{2}, \hypref{4} and \hypref{semty1}.
    Then, the goal follows by applying IH to \hypref{3}, \hypref{5}
    and \hypref{6}.
  \item[\ruleref{RT-Dip}] We have
    \begin{gather}
       \ottnt{I} \ottsym{=} \ottkw{DIP} \, \mathit{IS}  \hyp{1}\\
       \Phi_{{\mathrm{1}}} \ottsym{=} \ottsym{\{}   \ottmv{x} \mathord:\allowbreak \ottnt{T}   \triangleright  \Upsilon  \mid  \varphi  \ottsym{\}}  \hyp{2}\\
       \Phi_{{\mathrm{2}}} \ottsym{=} \ottsym{\{}   \ottmv{x} \mathord:\allowbreak \ottnt{T}   \triangleright  \Upsilon'  \mid  \varphi'  \ottsym{\}}  \hyp{3}\\
       \Gamma  \ottsym{,}   \ottmv{x} \mathord:\allowbreak \ottnt{T}  \vdash \ottsym{\{}  \Upsilon  \mid  \varphi  \ottsym{\}} \; \mathit{IS} \; \ottsym{\{}  \Upsilon'  \mid  \varphi'  \ottsym{\}}  \hyp{4}
    \end{gather}
    for some \(\mathit{IS}\), \(\ottmv{x}\), \(\ottnt{T}\), \(\Upsilon\), \(\Upsilon'\),
    \(\varphi\), and \(\varphi'\).
    The last rule that derives \hypref{eval2} is \ruleref{E-Dip}.
    Therefore, we have
    \begin{gather}
       \ottnt{S_{{\mathrm{1}}}} \ottsym{=} \ottnt{V}  \triangleright  \ottnt{S'_{{\mathrm{1}}}}  \hyp{5}\\
       \ottnt{S_{{\mathrm{2}}}} \ottsym{=} \ottnt{V}  \triangleright  \ottnt{S'_{{\mathrm{2}}}}  \hyp{6}\\
      \ottnt{S'_{{\mathrm{1}}}}  \vdash  \mathit{IS}  \Downarrow  \ottnt{S'_{{\mathrm{2}}}} \hyp{7}
    \end{gather}
    for some \(\ottnt{V}\), \(\ottnt{S'_{{\mathrm{1}}}}\), and \(\ottnt{S'_{{\mathrm{2}}}}\).
    % By \propref{semty/inv(push)}, \hypref{semty2}, \hypref{2}, and \hypref{5}, we have
    By \propref{subty/exists=>subst'}, we have
    \begin{gather}
       \sigma  [  \ottmv{x}  \mapsto  \ottnt{V}  ]   \ottsym{:}  \Gamma  \ottsym{,}   \ottmv{x} \mathord:\allowbreak \ottnt{T}   \models  \ottnt{S'_{{\mathrm{1}}}}  \ottsym{:}  \ottsym{\{}  \Upsilon  \mid  \varphi  \ottsym{\}} \hyp{11}
    \end{gather}
    from \hypref{semty2}, \hypref{2}, and \hypref{5}.
    % So by applying IH of \hypref{4} to \hypref{7} and \hypref{11}, we have
    By applying IH to \hypref{4}, \hypref{7}, and \hypref{11}, we have
    \begin{gather}
       \sigma  [  \ottmv{x}  \mapsto  \ottnt{V}  ]   \ottsym{:}  \Gamma  \ottsym{,}   \ottmv{x} \mathord:\allowbreak \ottnt{T}   \models  \ottnt{S'_{{\mathrm{2}}}}  \ottsym{:}  \ottsym{\{}  \Upsilon'  \mid  \varphi'  \ottsym{\}} \hyp{12}.
    \end{gather}
    Therefore, \(\sigma  \ottsym{:}  \Gamma  \models  \ottnt{V}  \triangleright  \ottnt{S'_{{\mathrm{2}}}}  \ottsym{:}  \ottsym{\{}   \ottmv{x} \mathord:\allowbreak \ottnt{T}   \triangleright  \Upsilon'  \mid  \varphi'  \ottsym{\}}\) follows from \ruleref{Sem-Push} and \hypref{12}.
  \item[\ruleref{RT-Drop}] We have
    \begin{gather}
       \ottnt{I} \ottsym{=} \ottkw{DROP}  \hyp{0}\\
       \Phi_{{\mathrm{1}}} \ottsym{=} \ottsym{\{}   \ottmv{x} \mathord:\allowbreak \ottnt{T}   \triangleright  \Upsilon  \mid  \varphi  \ottsym{\}}  \hyp{1}\\
       \Phi_{{\mathrm{2}}} \ottsym{=} \ottsym{\{}  \Upsilon  \mid  \exists \,  \ottmv{x} \mathord:\allowbreak \ottnt{T}   \ottsym{.}  \varphi  \ottsym{\}}  \hyp{2}
    \end{gather}
    for some \(\ottmv{x}\), \(\ottnt{T}\), \(\Upsilon\), and \(\varphi\).
    The last rule that derives \hypref{eval2} is \ruleref{E-Drop}.
    Therefore, we have
    % \propref{eval/inv(drop)}, \hypref{eval2}, and \hypref{0}, we have
    \begin{gather}
       \ottnt{S_{{\mathrm{1}}}} \ottsym{=} \ottnt{V}  \triangleright  \ottnt{S_{{\mathrm{2}}}}  \hyp{3}
    \end{gather}
    for some \(\ottnt{V}\).
    %
    % By \propref{semty/logic}, \hypref{semty2}, \hypref{1}, and \hypref{3}, we have
    From \hypref{semty2}, \hypref{3}, and \propref{subty/exists=>subst'}, we have \(\sigma  \ottsym{:}  \Gamma  \models  \ottnt{S_{{\mathrm{2}}}}  \ottsym{:}  \ottsym{\{}  \Upsilon  \mid  \exists \,  \ottmv{x} \mathord:\allowbreak \ottnt{T}   \ottsym{.}  \varphi  \wedge  \ottmv{x}  \ottsym{=}  \ottnt{V}  \ottsym{\}}\) and hence \(\sigma  \ottsym{:}  \Gamma  \models  \ottnt{S_{{\mathrm{2}}}}  \ottsym{:}  \ottsym{\{}  \Upsilon  \mid  \exists \,  \ottmv{x} \mathord:\allowbreak \ottnt{T}   \ottsym{.}  \varphi  \ottsym{\}}\) as required.

      % \hyp{5}.
      % \end{gather}
      % Then the goal follows by \propref{semty/logic/imply}, \propref{semty/wf},
      % \propref{fol/thm/misc(drop)}, and \hypref{5}.

    \item[\ruleref{RT-Dup}]
    We have
    \begin{gather}
       \ottnt{I} \ottsym{=} \ottkw{DUP}  \hyp{0}\\
       \Phi_{{\mathrm{1}}} \ottsym{=} \ottsym{\{}   \ottmv{x} \mathord:\allowbreak \ottnt{T}   \triangleright  \Upsilon  \mid  \varphi  \ottsym{\}}  \hyp{1}\\
       \Phi_{{\mathrm{2}}} \ottsym{=} \ottsym{\{}   \ottmv{x'} \mathord:\allowbreak \ottnt{T}   \triangleright   \ottmv{x} \mathord:\allowbreak \ottnt{T}   \triangleright  \Upsilon  \mid  \varphi  \wedge  \ottmv{x'}  \ottsym{=}  \ottmv{x}  \ottsym{\}}  \hyp{2}\\
       \ottmv{x'}  \notin   \text{dom}( \Gamma  \ottsym{,}   \widehat{  \ottmv{x} \mathord:\allowbreak \ottnt{T}   \triangleright  \Upsilon }  )   \hyp{2.1}
    \end{gather}
    for some \(\ottmv{x}\), \(\ottmv{x'}\), \(\ottnt{T}\), \(\Upsilon\), and \(\varphi\).
    The last rule that derives \hypref{eval2} is \ruleref{E-Dup}.
    Therefore, we have
    % By \propref{eval/inv(dup)} and \hypref{eval2}, and \hypref{0}, we have
    \begin{gather}
       \ottnt{S_{{\mathrm{1}}}} \ottsym{=} \ottnt{V}  \triangleright  \ottnt{S}  \hyp{3}\\
       \ottnt{S_{{\mathrm{2}}}} \ottsym{=} \ottnt{V}  \triangleright  \ottnt{V}  \triangleright  \ottnt{S}  \hyp{4}
    \end{gather}
    for some \(\ottnt{V}\) and \(\ottnt{S}\).
    %
    % By \propref{semty/logic} and \proref{subty/exists=>subst'}, \hypref{semty2}, \hypref{1}, and \hypref{3}, we have
    By \hypref{semty2}, we have \( \sigma  [  \ottmv{x}  \mapsto  \ottnt{V}  ]   \ottsym{:}  \Gamma  \ottsym{,}   \ottmv{x} \mathord:\allowbreak \ottnt{T}   \models  \ottnt{S}  \ottsym{:}  \ottsym{\{}  \Upsilon  \mid  \varphi  \ottsym{\}}\) and hence
    \(  \sigma  [  \ottmv{x}  \mapsto  \ottnt{V}  ]   [  \ottmv{x'}  \mapsto  \ottnt{V}  ]   \ottsym{:}  \Gamma  \ottsym{,}   \ottmv{x} \mathord:\allowbreak \ottnt{T}   \ottsym{,}   \ottmv{x'} \mathord:\allowbreak \ottnt{T}   \models  \ottnt{S}  \ottsym{:}  \ottsym{\{}  \Upsilon  \mid  \varphi  \ottsym{\}}\).
    % By \propref{semty/logic/imply}, \propref{semty/wf}, \propref{fol/thm/misc(dup)}, \hypref{2.1}, and \hypref{6}, we have
    Since \(  \sigma  [  \ottmv{x}  \mapsto  \ottnt{V}  ]   [  \ottmv{x'}  \mapsto  \ottnt{V}  ]   \ottsym{:}  \Gamma  \ottsym{,}   \ottmv{x} \mathord:\allowbreak \ottnt{T}   \ottsym{,}   \ottmv{x'} \mathord:\allowbreak \ottnt{T}   \models  \varphi  \implies  \ottsym{(}  \varphi  \wedge  \ottmv{x}  \ottsym{=}  \ottmv{x'}  \ottsym{)}\),
    we have \(  \sigma  [  \ottmv{x}  \mapsto  \ottnt{V}  ]   [  \ottmv{x'}  \mapsto  \ottnt{V}  ]   \ottsym{:}  \Gamma  \ottsym{,}   \ottmv{x} \mathord:\allowbreak \ottnt{T}   \ottsym{,}   \ottmv{x'} \mathord:\allowbreak \ottnt{T}   \models  \ottnt{S}  \ottsym{:}  \ottsym{\{}  \Upsilon  \mid  \varphi  \wedge  \ottmv{x}  \ottsym{=}  \ottmv{x'}  \ottsym{\}}\).
    % \KS{Not yet done.}
    % By \propref{semty/logic/imply}, \propref{semty/wf}, \propref{fol/thm/misc(dup)}, \hypref{2.1}, and \hypref{6}, we have
    % \begin{gather}
    %   \sigma  \ottsym{:}  \Gamma  \models  \ottnt{S}  \ottsym{:}  \ottsym{\{}  \Upsilon  \mid  \exists \,  \ottmv{x} \mathord:\allowbreak \ottnt{T}   \ottsym{.}  \ottsym{(}  \exists \,  \ottmv{x'} \mathord:\allowbreak \ottnt{T}   \ottsym{.}  \ottsym{(}  \varphi  \wedge  \ottmv{x'}  \ottsym{=}  \ottmv{x}  \ottsym{)}  \wedge  \ottmv{x'}  \ottsym{=}  \ottnt{V}  \ottsym{)}  \wedge  \ottmv{x}  \ottsym{=}  \ottnt{V}  \ottsym{\}} \hyp{7}.
    % \end{gather}
    % Hence, the goal follows by \propref{semty/logic} and \hypref{7}.
    Therefore, we have \(\sigma  \ottsym{:}  \Gamma  \models  \ottnt{S_{{\mathrm{2}}}}  \ottsym{:}  \ottsym{\{}  \Upsilon  \mid  \varphi  \wedge  \ottmv{x}  \ottsym{=}  \ottmv{x'}  \ottsym{\}}\) as required.

  \item[\ruleref{RT-Swap}] We have
    \begin{gather}
       \ottnt{I} \ottsym{=} \ottkw{SWAP}  \hyp{1}\\
       \Phi_{{\mathrm{1}}} \ottsym{=} \ottsym{\{}   \ottmv{x} \mathord:\allowbreak \ottnt{T}   \triangleright   \ottmv{x'} \mathord:\allowbreak \ottnt{T}   \triangleright  \Upsilon  \mid  \varphi  \ottsym{\}}  \hyp{2}\\
       \Phi_{{\mathrm{2}}} \ottsym{=} \ottsym{\{}   \ottmv{x'} \mathord:\allowbreak \ottnt{T}   \triangleright   \ottmv{x} \mathord:\allowbreak \ottnt{T}   \triangleright  \Upsilon  \mid  \varphi  \ottsym{\}}  \hyp{3}
    \end{gather}
    for some \(\ottmv{x}\), \(\ottmv{x'}\), \(\ottnt{T}\), \(\Upsilon\), and \(\varphi\).
    The last rule that derives \hypref{eval2} is \ruleref{E-Swap}.
    Therefore, we have
    \begin{gather}
       \ottnt{S_{{\mathrm{1}}}} \ottsym{=} \ottnt{V}  \triangleright  \ottnt{V'}  \triangleright  \ottnt{S}  \hyp{4}\\
       \ottnt{S_{{\mathrm{2}}}} \ottsym{=} \ottnt{V'}  \triangleright  \ottnt{V}  \triangleright  \ottnt{S}  \hyp{5}
    \end{gather}
    for some \(\ottnt{V}\), \(\ottnt{V'}\), and \(\ottnt{S}\).
    % 
    % By \propref{semty/logic}, \hypref{semty2}, \hypref{2}, and \hypref{4}, we have
    By \hypref{semty2}, we have \(  \sigma  [  \ottmv{x}  \mapsto  \ottnt{V}  ]   [  \ottmv{x'}  \mapsto  \ottnt{V'}  ]   \ottsym{:}  \Gamma  \ottsym{,}   \ottmv{x} \mathord:\allowbreak \ottnt{T}   \ottsym{,}   \ottmv{x'} \mathord:\allowbreak \ottnt{T}   \models  \ottnt{S}  \ottsym{:}  \ottsym{\{}  \Upsilon  \mid  \varphi  \ottsym{\}}\).
    % By \propref{semty/logic/imply}, \propref{semty/wf}, \propref{fol/thm/misc(swap)}, and \hypref{8}, we have
    Since $x \ne x'$, we have \(  \sigma  [  \ottmv{x'}  \mapsto  \ottnt{V'}  ]   [  \ottmv{x}  \mapsto  \ottnt{V}  ]   \ottsym{:}  \Gamma  \ottsym{,}   \ottmv{x'} \mathord:\allowbreak \ottnt{T}   \ottsym{,}   \ottmv{x} \mathord:\allowbreak \ottnt{T}   \models  \ottnt{S}  \ottsym{:}  \ottsym{\{}  \Upsilon  \mid  \varphi  \ottsym{\}}\).
    Therefore, we have \(\sigma  \ottsym{:}  \Gamma  \models  \ottnt{S_{{\mathrm{2}}}}  \ottsym{:}  \ottsym{\{}  \Upsilon  \mid  \varphi  \ottsym{\}}\).
    % \begin{gather}
    %   \sigma  \ottsym{:}  \Gamma  \models  \ottnt{S}  \ottsym{:}  \ottsym{\{}  \Upsilon  \mid  \exists \,  \ottmv{x} \mathord:\allowbreak \ottnt{T}   \ottsym{.}  \ottsym{(}  \exists \,  \ottmv{x'} \mathord:\allowbreak \ottnt{T'}   \ottsym{.}  \varphi  \wedge  \ottmv{x'}  \ottsym{=}  \ottnt{V'}  \ottsym{)}  \wedge  \ottmv{x}  \ottsym{=}  \ottnt{V}  \ottsym{\}}
    %   \hyp{9}.
    % \end{gather}
    % So, the goal follows by \propref{semty/logic} and \hypref{9}.

  \item[\ruleref{RT-Push}] We have
    \begin{gather}
       \ottnt{I} \ottsym{=} \ottkw{PUSH} \, \ottnt{T} \, \ottnt{V}  \hyp{1}\\
       \Phi_{{\mathrm{1}}} \ottsym{=} \ottsym{\{}  \Upsilon  \mid  \varphi  \ottsym{\}}  \hyp{2}\\
       \Phi_{{\mathrm{2}}} \ottsym{=} \ottsym{\{}   \ottmv{x} \mathord:\allowbreak \ottnt{T}   \triangleright  \Upsilon  \mid  \varphi  \wedge  \ottmv{x}  \ottsym{=}  \ottnt{V}  \ottsym{\}}  \hyp{3}\\
       \ottnt{V}  :  \ottnt{T}  \hyp{4}\\
       \ottmv{x}  \notin   \text{dom}( \Gamma  \ottsym{,}   \widehat{ \Upsilon }  )   \hyp{4.1}
    \end{gather}
    for some \(\ottmv{x}\), \(\ottnt{T}\), \(\Upsilon\), \(\varphi\), and \(\ottnt{V}\).
    The last rule that derives \hypref{semty2} is \ruleref{E-Push}.
    Therefore, we have
    \begin{gather}
       \ottnt{S_{{\mathrm{2}}}} \ottsym{=} \ottnt{V}  \triangleright  \ottnt{S_{{\mathrm{1}}}}  \hyp{5}.
    \end{gather}
    % By \propref{semty/logic/imply}, \propref{semty/wf},
    % \propref{fol/thm/misc(push)}, \hypref{semty2}, \hypref{2}, \hypref{4}, and
    % \hypref{4.1}, we have
    To show \(\sigma  \ottsym{:}  \Gamma  \models  \ottnt{S_{{\mathrm{2}}}}  \ottsym{:}  \ottsym{\{}   \ottmv{x} \mathord:\allowbreak \ottnt{T}   \triangleright  \Upsilon  \mid  \varphi  \wedge  \ottmv{x}  \ottsym{=}  \ottnt{V}  \ottsym{\}}\), it suffices to show
    \( \sigma  [  \ottmv{x}  \mapsto  \ottnt{V}  ]   \ottsym{:}  \Gamma  \ottsym{,}   \ottmv{x} \mathord:\allowbreak \ottnt{T}   \models  \ottnt{S_{{\mathrm{1}}}}  \ottsym{:}  \ottsym{\{}  \Upsilon  \mid  \varphi  \wedge  \ottmv{x}  \ottsym{=}  \ottnt{V}  \ottsym{\}}\) from \ruleref{Sem-Push},
    which follows from \(\sigma  \ottsym{:}  \Gamma  \models  \ottnt{S_{{\mathrm{1}}}}  \ottsym{:}  \ottsym{\{}  \Upsilon  \mid  \varphi  \ottsym{\}}\), \hypref{4.1}, and \( \sigma  [  \ottmv{x}  \mapsto  \ottnt{V}  ]   \ottsym{:}  \Gamma  \ottsym{,}   \ottmv{x} \mathord:\allowbreak \ottnt{T}   \models  \varphi  \implies  \ottsym{(}  \varphi  \wedge  \ottmv{x}  \ottsym{=}  \ottnt{V}  \ottsym{)}\).
    
  \item[\ruleref{RT-Not}]
    We have
    \begin{gather}
       \ottnt{I} \ottsym{=} \ottkw{NOT}  \hyp{1}\\
       \Phi_{{\mathrm{1}}} \ottsym{=} \ottsym{\{}   \ottmv{x} \mathord:\allowbreak \ottkw{int}   \triangleright  \Upsilon  \mid  \varphi  \ottsym{\}}  \hyp{2}\\
       \Phi_{{\mathrm{2}}} \ottsym{=} \ottsym{\{}   \ottmv{x'} \mathord:\allowbreak \ottkw{int}   \triangleright  \Upsilon  \mid  \exists \,  \ottmv{x} \mathord:\allowbreak \ottkw{int}   \ottsym{.}  \varphi  \wedge  \ottsym{(}  \ottmv{x}  \neq  0  \wedge  \ottmv{x'}  \ottsym{=}  0  \vee  \ottmv{x}  \ottsym{=}  0  \wedge  \ottmv{x'}  \ottsym{=}  1  \ottsym{)}  \ottsym{\}}  \hyp{3}\\
       \ottmv{x'}  \notin   \text{dom}( \Gamma  \ottsym{,}   \widehat{  \ottmv{x} \mathord:\allowbreak \ottkw{int}   \triangleright  \Upsilon }  )   \hyp{3.1}
    \end{gather}
    for some \(\ottmv{x}\), \(\Upsilon\), and \(\varphi\).
    The last rule that derives \hypref{eval2} is either \ruleref{E-NotT} or \ruleref{E-NotF}.
    We only show the case of \ruleref{E-NotT}; \ruleref{E-NotF} is similar.
    In this case, we have
    \begin{gather}
       \ottnt{S_{{\mathrm{1}}}} \ottsym{=} \ottnt{i}  \triangleright  \ottnt{S}  \hyp{4}\\
      \ottnt{i}  \neq  0 \hyp{5}\\
       \ottnt{S_{{\mathrm{2}}}} \ottsym{=} 0  \triangleright  \ottnt{S}  \hyp{6}.
    \end{gather}
    for some \(\ottnt{i}\) and \(\ottnt{S}\).
    By \hypref{semty2}, we have
    \begin{gather}
       \sigma  [  \ottmv{x}  \mapsto  \ottnt{i}  ]   \ottsym{:}  \Gamma  \ottsym{,}   \ottmv{x} \mathord:\allowbreak \ottnt{T}   \models  \ottnt{S}  \ottsym{:}  \ottsym{\{}  \Upsilon  \mid  \varphi  \ottsym{\}} \hyp{8}.
    \end{gather}
    From \propref{subty/exists=>subst'}, we have
    \begin{gather}
      \sigma  \ottsym{:}  \Gamma  \models  \ottnt{S}  \ottsym{:}  \ottsym{\{}  \Upsilon  \mid  \exists \,  \ottmv{x} \mathord:\allowbreak \ottnt{T}   \ottsym{.}  \varphi  \wedge  \ottmv{x}  \ottsym{=}  \ottnt{i}  \ottsym{\}} \hyp{8}.
    \end{gather}
    From \(\ottnt{i}  \neq  0\), we have \(\sigma  \ottsym{:}  \Gamma  \models  \ottnt{S}  \ottsym{:}  \ottsym{\{}  \Upsilon  \mid  \exists \,  \ottmv{x} \mathord:\allowbreak \ottnt{T}   \ottsym{.}  \varphi  \wedge  \ottmv{x}  \neq  0  \ottsym{\}}\).
    From \hypref{3.1}, we have
    \( \sigma  [  \ottmv{x'}  \mapsto  0  ]   \ottsym{:}  \Gamma  \ottsym{,}   \ottmv{x'} \mathord:\allowbreak \ottnt{T}   \models  \ottnt{S}  \ottsym{:}  \ottsym{\{}  \Upsilon  \mid  \exists \,  \ottmv{x} \mathord:\allowbreak \ottnt{T}   \ottsym{.}  \varphi  \wedge  \ottmv{x}  \neq  0  \ottsym{\}}\).
    From \( \sigma  [  \ottmv{x'}  \mapsto  0  ]   \ottsym{:}  \Gamma  \ottsym{,}   \ottmv{x'} \mathord:\allowbreak \ottnt{T}   \models  \ottsym{(}  \exists \,  \ottmv{x} \mathord:\allowbreak \ottnt{T}   \ottsym{.}  \varphi  \wedge  \ottmv{x}  \neq  0  \ottsym{)}  \implies  \ottsym{(}  \exists \,  \ottmv{x} \mathord:\allowbreak \ottnt{T}   \ottsym{.}  \varphi  \wedge  \ottmv{x}  \neq  0  \wedge  \ottmv{x'}  \ottsym{=}  0  \ottsym{)}\), we have
    \( \sigma  [  \ottmv{x'}  \mapsto  0  ]   \ottsym{:}  \Gamma  \ottsym{,}   \ottmv{x'} \mathord:\allowbreak \ottnt{T}   \models  \ottnt{S}  \ottsym{:}  \ottsym{\{}  \Upsilon  \mid  \exists \,  \ottmv{x} \mathord:\allowbreak \ottnt{T}   \ottsym{.}  \varphi  \wedge  \ottmv{x}  \neq  0  \wedge  \ottmv{x'}  \ottsym{=}  0  \ottsym{\}}\).
    Therefore, we have
    \(\sigma  \ottsym{:}  \Gamma  \models  0  \triangleright  \ottnt{S}  \ottsym{:}  \ottsym{\{}   \ottmv{x'} \mathord:\allowbreak \ottnt{T}   \triangleright  \Upsilon  \mid  \exists \,  \ottmv{x} \mathord:\allowbreak \ottnt{T}   \ottsym{.}  \varphi  \wedge  \ottmv{x}  \neq  0  \wedge  \ottmv{x'}  \ottsym{=}  0  \ottsym{\}}\),
    which is followed by
    \(\sigma  \ottsym{:}  \Gamma  \models  0  \triangleright  \ottnt{S}  \ottsym{:}  \ottsym{\{}   \ottmv{x'} \mathord:\allowbreak \ottnt{T}   \triangleright  \Upsilon  \mid  \exists \,  \ottmv{x} \mathord:\allowbreak \ottnt{T}   \ottsym{.}  \varphi  \wedge  \ottsym{(}  \ottsym{(}  \ottmv{x}  \neq  0  \wedge  \ottmv{x'}  \ottsym{=}  0  \ottsym{)}  \vee  \ottsym{(}  \ottmv{x}  \ottsym{=}  0  \wedge  \ottmv{x'}  \neq  1  \ottsym{)}  \ottsym{)}  \ottsym{\}}\) as required.

  \item[\ruleref{RT-Add}] We have
    We have 
    \begin{gather}
       \ottnt{I} \ottsym{=} \ottkw{ADD}  \hyp{1} \\
       \Phi_{{\mathrm{1}}} \ottsym{=} \ottsym{\{}   \ottmv{x_{{\mathrm{1}}}} \mathord:\allowbreak \ottkw{int}   \triangleright   \ottmv{x_{{\mathrm{2}}}} \mathord:\allowbreak \ottkw{int}   \triangleright  \Upsilon  \mid  \varphi  \ottsym{\}}  \hyp{2} \\
       \Phi_{{\mathrm{2}}} \ottsym{=} \ottsym{\{}   \ottmv{x_{{\mathrm{3}}}} \mathord:\allowbreak \ottkw{int}   \triangleright  \Upsilon  \mid  \exists \,  \ottmv{x_{{\mathrm{1}}}} \mathord:\allowbreak \ottkw{int}   \ottsym{,}   \ottmv{x_{{\mathrm{2}}}} \mathord:\allowbreak \ottkw{int}   \ottsym{.}  \varphi  \wedge  \ottmv{x_{{\mathrm{1}}}}  \ottsym{+}  \ottmv{x_{{\mathrm{2}}}}  \ottsym{=}  \ottmv{x_{{\mathrm{3}}}}  \ottsym{\}} 
      \hyp{3}\\
       \ottmv{x_{{\mathrm{3}}}}  \notin   \text{dom}( \Gamma  \ottsym{,}   \widehat{  \ottmv{x_{{\mathrm{1}}}} \mathord:\allowbreak \ottkw{int}   \triangleright   \ottmv{x_{{\mathrm{2}}}} \mathord:\allowbreak \ottkw{int}   \triangleright  \Upsilon }  )   \hyp{3.1}
    \end{gather}
    for some \(\ottmv{x_{{\mathrm{1}}}}\), \(\ottmv{x_{{\mathrm{2}}}}\), \(\ottmv{x_{{\mathrm{3}}}}\), \(\Upsilon\), and \(\varphi\).
    The last rule that derives \hypref{eval2} is \ruleref{E-Add}.  Therefore,
    \begin{gather}
       \ottnt{S_{{\mathrm{1}}}} \ottsym{=} \ottnt{i_{{\mathrm{1}}}}  \triangleright  \ottnt{i_{{\mathrm{2}}}}  \triangleright  \ottnt{S}  \hyp{4}\\
       \ottnt{S_{{\mathrm{2}}}} \ottsym{=} \ottnt{i_{{\mathrm{3}}}}  \triangleright  \ottnt{S}  \hyp{5}\\
       \ottnt{i_{{\mathrm{1}}}}  \ottsym{+}  \ottnt{i_{{\mathrm{2}}}} \ottsym{=} \ottnt{i_{{\mathrm{3}}}}  \hyp{6}
    \end{gather}
    for some \(\ottnt{i_{{\mathrm{1}}}}\), \(\ottnt{i_{{\mathrm{2}}}}\), \(\ottnt{i_{{\mathrm{3}}}}\), and \(\ottnt{S}\).
    By \hypref{semty2}, we have
    \begin{gather}
        \sigma  [  \ottmv{x_{{\mathrm{1}}}}  \mapsto  \ottnt{i_{{\mathrm{1}}}}  ]   [  \ottmv{x_{{\mathrm{2}}}}  \mapsto  \ottnt{i_{{\mathrm{2}}}}  ]   \ottsym{:}  \Gamma  \ottsym{,}   \ottmv{x_{{\mathrm{1}}}} \mathord:\allowbreak \ottkw{int}   \ottsym{,}   \ottmv{x_{{\mathrm{2}}}} \mathord:\allowbreak \ottkw{int}   \models  \ottnt{S}  \ottsym{:}  \ottsym{\{}  \Upsilon  \mid  \varphi  \ottsym{\}} \hyp{9}.
    \end{gather}
    From \hypref{3.1}, we have
    \begin{gather}
         \sigma  [  \ottmv{x_{{\mathrm{1}}}}  \mapsto  \ottnt{i_{{\mathrm{1}}}}  ]   [  \ottmv{x_{{\mathrm{2}}}}  \mapsto  \ottnt{i_{{\mathrm{2}}}}  ]   [  \ottmv{x_{{\mathrm{3}}}}  \mapsto  \ottnt{i_{{\mathrm{3}}}}  ]   \ottsym{:}  \Gamma  \ottsym{,}   \ottmv{x_{{\mathrm{1}}}} \mathord:\allowbreak \ottkw{int}   \ottsym{,}   \ottmv{x_{{\mathrm{2}}}} \mathord:\allowbreak \ottkw{int}   \ottsym{,}   \ottmv{x_{{\mathrm{3}}}} \mathord:\allowbreak \ottkw{int}   \models  \ottnt{S}  \ottsym{:}  \ottsym{\{}  \Upsilon  \mid  \varphi  \ottsym{\}} \hyp{9}.
    \end{gather}
    From \(   \sigma  [  \ottmv{x_{{\mathrm{1}}}}  \mapsto  \ottnt{i_{{\mathrm{1}}}}  ]   [  \ottmv{x_{{\mathrm{2}}}}  \mapsto  \ottnt{i_{{\mathrm{2}}}}  ]   [  \ottmv{x_{{\mathrm{3}}}}  \mapsto  \ottnt{i_{{\mathrm{3}}}}  ]   \ottsym{:}  \Gamma  \ottsym{,}   \ottmv{x_{{\mathrm{1}}}} \mathord:\allowbreak \ottkw{int}   \ottsym{,}   \ottmv{x_{{\mathrm{2}}}} \mathord:\allowbreak \ottkw{int}   \ottsym{,}   \ottmv{x_{{\mathrm{3}}}} \mathord:\allowbreak \ottkw{int}   \models  \varphi  \implies  \varphi  \wedge  \ottmv{x_{{\mathrm{1}}}}  \ottsym{+}  \ottmv{x_{{\mathrm{2}}}}  \ottsym{=}  \ottmv{x_{{\mathrm{3}}}}\), we have
    \begin{gather}
         \sigma  [  \ottmv{x_{{\mathrm{1}}}}  \mapsto  \ottnt{i_{{\mathrm{1}}}}  ]   [  \ottmv{x_{{\mathrm{2}}}}  \mapsto  \ottnt{i_{{\mathrm{2}}}}  ]   [  \ottmv{x_{{\mathrm{3}}}}  \mapsto  \ottnt{i_{{\mathrm{3}}}}  ]   \ottsym{:}  \Gamma  \ottsym{,}   \ottmv{x_{{\mathrm{1}}}} \mathord:\allowbreak \ottkw{int}   \ottsym{,}   \ottmv{x_{{\mathrm{2}}}} \mathord:\allowbreak \ottkw{int}   \ottsym{,}   \ottmv{x_{{\mathrm{3}}}} \mathord:\allowbreak \ottkw{int}   \models  \ottnt{S}  \ottsym{:}  \ottsym{\{}  \Upsilon  \mid  \varphi  \wedge  \ottmv{x_{{\mathrm{1}}}}  \ottsym{+}  \ottmv{x_{{\mathrm{2}}}}  \ottsym{=}  \ottmv{x_{{\mathrm{3}}}}  \ottsym{\}} \hyp{9}.
    \end{gather}
    and therefore
    \(  \sigma  [  \ottmv{x_{{\mathrm{1}}}}  \mapsto  \ottnt{i_{{\mathrm{1}}}}  ]   [  \ottmv{x_{{\mathrm{2}}}}  \mapsto  \ottnt{i_{{\mathrm{2}}}}  ]   \ottsym{:}  \Gamma  \ottsym{,}   \ottmv{x_{{\mathrm{1}}}} \mathord:\allowbreak \ottkw{int}   \ottsym{,}   \ottmv{x_{{\mathrm{2}}}} \mathord:\allowbreak \ottkw{int}   \models  \ottnt{i_{{\mathrm{3}}}}  \triangleright  \ottnt{S}  \ottsym{:}  \ottsym{\{}   \ottmv{x_{{\mathrm{3}}}} \mathord:\allowbreak \ottkw{int}   \triangleright  \Upsilon  \mid  \varphi  \wedge  \ottmv{x_{{\mathrm{1}}}}  \ottsym{+}  \ottmv{x_{{\mathrm{2}}}}  \ottsym{=}  \ottmv{x_{{\mathrm{3}}}}  \ottsym{\}}\).
    By \propref{subty/exists=>subst'}, we have
    \(\sigma  \ottsym{:}  \Gamma  \models  \ottnt{i_{{\mathrm{3}}}}  \triangleright  \ottnt{S}  \ottsym{:}  \ottsym{\{}   \ottmv{x_{{\mathrm{3}}}} \mathord:\allowbreak \ottkw{int}   \triangleright  \Upsilon  \mid  \exists \,  \ottmv{x_{{\mathrm{1}}}} \mathord:\allowbreak \ottkw{int}   \ottsym{,}   \ottmv{x_{{\mathrm{2}}}} \mathord:\allowbreak \ottkw{int}   \ottsym{.}  \varphi  \wedge  \ottmv{x_{{\mathrm{1}}}}  \ottsym{+}  \ottmv{x_{{\mathrm{2}}}}  \ottsym{=}  \ottmv{x_{{\mathrm{3}}}}  \ottsym{\}}\) as required.

  \item[\ruleref{RT-Pair}] Similar to the case for \ruleref{RT-Add}.

  \item[\ruleref{RT-CAR}]
    We have
    \begin{gather}
       \ottnt{I} \ottsym{=} \ottkw{CAR}  \hyp{1}\\
       \Phi_{{\mathrm{1}}} \ottsym{=} \ottsym{\{}   \ottmv{x} \mathord:\allowbreak \ottnt{T_{{\mathrm{1}}}}  \times  \ottnt{T_{{\mathrm{2}}}}   \triangleright  \Upsilon  \mid  \varphi  \ottsym{\}}  \hyp{2}\\
       \Phi_{{\mathrm{2}}} \ottsym{=} \ottsym{\{}   \ottmv{x_{{\mathrm{1}}}} \mathord:\allowbreak \ottnt{T_{{\mathrm{1}}}}   \triangleright  \Upsilon  \mid  \exists \,  \ottmv{x} \mathord:\allowbreak \ottnt{T_{{\mathrm{1}}}}  \times  \ottnt{T_{{\mathrm{2}}}}   \ottsym{,}   \ottmv{x_{{\mathrm{2}}}} \mathord:\allowbreak \ottnt{T_{{\mathrm{2}}}}   \ottsym{.}  \varphi  \wedge  \ottmv{x}  \ottsym{=}  \ottsym{(}  \ottmv{x_{{\mathrm{1}}}}  \ottsym{,}  \ottmv{x_{{\mathrm{2}}}}  \ottsym{)}  \ottsym{\}}  \hyp{3}\\
      \ottmv{x_{{\mathrm{1}}}}  \neq  \ottmv{x_{{\mathrm{2}}}} \hyp{3.1}\\
       \ottmv{x_{{\mathrm{1}}}}  \notin   \text{dom}( \Gamma  \ottsym{,}   \widehat{  \ottmv{x} \mathord:\allowbreak \ottnt{T_{{\mathrm{1}}}}  \times  \ottnt{T_{{\mathrm{2}}}}   \triangleright  \Upsilon }  )   \hyp{3.2}\\
       \ottmv{x_{{\mathrm{2}}}}  \notin   \text{dom}( \Gamma  \ottsym{,}   \widehat{  \ottmv{x} \mathord:\allowbreak \ottnt{T_{{\mathrm{1}}}}  \times  \ottnt{T_{{\mathrm{2}}}}   \triangleright  \Upsilon }  )   \hyp{3.3}
    \end{gather}
    for some \(\ottmv{x}\), \(\ottmv{x_{{\mathrm{1}}}}\), \(\ottmv{x_{{\mathrm{2}}}}\), \(\ottnt{T_{{\mathrm{1}}}}\), \(\ottnt{T_{{\mathrm{2}}}}\), \(\Upsilon\), and \(\varphi\).
    The last rule that derives \hypref{eval2} is \ruleref{E-CAR}.
    Therefore,
    \begin{gather}
       \ottnt{S_{{\mathrm{1}}}} \ottsym{=} \ottsym{(}  \ottnt{V_{{\mathrm{1}}}}  \ottsym{,}  \ottnt{V_{{\mathrm{2}}}}  \ottsym{)}  \triangleright  \ottnt{S}  \hyp{4}\\
       \ottnt{S_{{\mathrm{2}}}} \ottsym{=} \ottnt{V_{{\mathrm{1}}}}  \triangleright  \ottnt{S}  \hyp{5}
    \end{gather}
    for some \(\ottnt{V_{{\mathrm{1}}}}\), \(\ottnt{V_{{\mathrm{2}}}}\), and \(\ottnt{S}\).
    By \hypref{semty2}, we have
    \begin{gather}
       \sigma  [  \ottmv{x}  \mapsto  \ottsym{(}  \ottnt{V_{{\mathrm{1}}}}  \ottsym{,}  \ottnt{V_{{\mathrm{2}}}}  \ottsym{)}  ]   \ottsym{:}  \Gamma  \ottsym{,}   \ottmv{x} \mathord:\allowbreak \ottnt{T_{{\mathrm{1}}}}  \times  \ottnt{T_{{\mathrm{2}}}}   \models  \ottnt{S}  \ottsym{:}  \ottsym{\{}  \Upsilon  \mid  \varphi  \ottsym{\}} \hyp{7}
    \end{gather}
    and hence
    \begin{gather}
      \sigma  \ottsym{:}  \Gamma  \models  \ottnt{S}  \ottsym{:}  \ottsym{\{}  \Upsilon  \mid  \exists \,  \ottmv{x} \mathord:\allowbreak \ottnt{T_{{\mathrm{1}}}}  \times  \ottnt{T_{{\mathrm{2}}}}   \ottsym{.}  \varphi  \wedge  \ottmv{x}  \ottsym{=}  \ottsym{(}  \ottnt{V_{{\mathrm{1}}}}  \ottsym{,}  \ottnt{V_{{\mathrm{2}}}}  \ottsym{)}  \ottsym{\}} \hyp{7}.
    \end{gather}
    $\sigma  \ottsym{:}  \Gamma  \models  \ottnt{S}  \ottsym{:}  \ottsym{\{}  \Upsilon  \mid  \exists \,  \ottmv{x} \mathord:\allowbreak \ottnt{T_{{\mathrm{1}}}}  \times  \ottnt{T_{{\mathrm{2}}}}   \ottsym{.}  \varphi  \wedge  \ottmv{x}  \ottsym{=}  \ottsym{(}  \ottnt{V_{{\mathrm{1}}}}  \ottsym{,}  \ottnt{V_{{\mathrm{2}}}}  \ottsym{)}  \ottsym{\}}$ implies
    $\sigma  \ottsym{:}  \Gamma  \models  \ottnt{S}  \ottsym{:}  \ottsym{\{}  \Upsilon  \mid  \exists \,  \ottmv{x} \mathord:\allowbreak \ottnt{T_{{\mathrm{1}}}}  \times  \ottnt{T_{{\mathrm{2}}}}   \ottsym{,}   \ottmv{x_{{\mathrm{1}}}} \mathord:\allowbreak \ottnt{T_{{\mathrm{1}}}}   \ottsym{,}   \ottmv{x_{{\mathrm{2}}}} \mathord:\allowbreak \ottnt{T_{{\mathrm{2}}}}   \ottsym{.}  \varphi  \wedge  \ottmv{x}  \ottsym{=}  \ottsym{(}  \ottmv{x_{{\mathrm{1}}}}  \ottsym{,}  \ottmv{x_{{\mathrm{2}}}}  \ottsym{)}  \wedge  \ottmv{x_{{\mathrm{1}}}}  \ottsym{=}  \ottnt{V_{{\mathrm{1}}}}  \wedge  \ottmv{x_{{\mathrm{2}}}}  \ottsym{=}  \ottnt{V_{{\mathrm{2}}}}  \ottsym{\}}$.
    Therefore, $\sigma  \ottsym{:}  \Gamma  \models  \ottnt{S}  \ottsym{:}  \ottsym{\{}  \Upsilon  \mid  \exists \,  \ottmv{x_{{\mathrm{1}}}} \mathord:\allowbreak \ottnt{T_{{\mathrm{1}}}}   \ottsym{.}  \ottsym{(}  \exists \,  \ottmv{x} \mathord:\allowbreak \ottnt{T_{{\mathrm{1}}}}  \times  \ottnt{T_{{\mathrm{2}}}}   \ottsym{,}   \ottmv{x_{{\mathrm{2}}}} \mathord:\allowbreak \ottnt{T_{{\mathrm{2}}}}   \ottsym{.}  \varphi  \ottsym{)}  \wedge  \ottmv{x_{{\mathrm{1}}}}  \ottsym{=}  \ottnt{V_{{\mathrm{1}}}}  \ottsym{\}}$,
    which implies
    $\sigma  \ottsym{:}  \Gamma  \models  \ottnt{V_{{\mathrm{1}}}}  \triangleright  \ottnt{S}  \ottsym{:}  \ottsym{\{}  \Upsilon  \mid  \exists \,  \ottmv{x} \mathord:\allowbreak \ottnt{T_{{\mathrm{1}}}}  \times  \ottnt{T_{{\mathrm{2}}}}   \ottsym{,}   \ottmv{x_{{\mathrm{2}}}} \mathord:\allowbreak \ottnt{T_{{\mathrm{2}}}}   \ottsym{.}  \varphi  \ottsym{\}}$ as required.
    
    % By the definition of $\sigma  \ottsym{:}  \Gamma  \models  \ottnt{S}  \ottsym{:}  \Phi$, we have
    % $\sigma  \ottsym{:}  \Gamma  \models  \ottnt{S}  \ottsym{:}  \ottsym{\{}  \Upsilon  \mid  \exists \,  \ottmv{x} \mathord:\allowbreak \ottnt{T_{{\mathrm{1}}}}  \times  \ottnt{T_{{\mathrm{2}}}}   \ottsym{.}  \varphi  \wedge  \ottmv{x}  \ottsym{=}  \ottsym{(}  \ottnt{V_{{\mathrm{1}}}}  \ottsym{,}  \ottnt{V_{{\mathrm{2}}}}  \ottsym{)}  \ottsym{\}}$

    % By \propref{semty/logic/imply}, \propref{semty/wf},
    % \propref{fol/thm/misc(car)}, \hypref{3.1}, \hypref{3.2}, \hypref{3.3}, and
    % \hypref{7}, we have
    % \begin{gather}
    %   \sigma  \ottsym{:}  \Gamma  \models  \ottnt{S}  \ottsym{:}  \ottsym{\{}  \Upsilon  \mid  \exists \,  \ottmv{x_{{\mathrm{1}}}} \mathord:\allowbreak \ottnt{T_{{\mathrm{1}}}}   \ottsym{.}  \ottsym{(}  \exists \,  \ottmv{x} \mathord:\allowbreak \ottnt{T_{{\mathrm{1}}}}  \times  \ottnt{T_{{\mathrm{2}}}}   \ottsym{,}   \ottmv{x_{{\mathrm{2}}}} \mathord:\allowbreak \ottnt{T_{{\mathrm{2}}}}   \ottsym{.}  \varphi  \wedge  \ottmv{x}  \ottsym{=}  \ottsym{(}  \ottmv{x_{{\mathrm{1}}}}  \ottsym{,}  \ottmv{x_{{\mathrm{2}}}}  \ottsym{)}  \ottsym{)}  \wedge  \ottmv{x_{{\mathrm{1}}}}  \ottsym{=}  \ottnt{V_{{\mathrm{1}}}}  \ottsym{\}} \hyp{10}.
    % \end{gather}
    % So the goal follows by \propref{semty/logic} and \hypref{10}.

  \item[\ruleref{RT-CDR}] Similar to the case for \ruleref{RT-CAR}.

  \item[\ruleref{RT-Nil}]
    Similar to the case for \ruleref{RT-Push}.
    % \AI{We can just say ``Similar to the case for
    %   \ruleref{RT-PUSH}''.}  We have
    % \begin{gather}
    %    \ottnt{I} \ottsym{=} \ottkw{NIL} \, \ottnt{T}  \hyp{1}\\
    %    \Phi_{{\mathrm{1}}} \ottsym{=} \ottsym{\{}  \Upsilon  \mid  \varphi  \ottsym{\}}  \hyp{2}\\
    %    \Phi_{{\mathrm{2}}} \ottsym{=} \ottsym{\{}   \ottmv{x} \mathord:\allowbreak \ottnt{T} \, \ottkw{list}   \triangleright  \Upsilon  \mid  \varphi  \wedge  \ottmv{x}  \ottsym{=}  \ottsym{[}  \ottsym{]}  \ottsym{\}}  \hyp{3}\\
    %    \ottmv{x}  \notin   \text{dom}( \Gamma  \ottsym{,}   \widehat{ \Upsilon }  )   \hyp{3.1}
    % \end{gather}
    % for some \(\ottmv{x}\), \(\ottnt{T}\), and \(\varphi\).  By \propref{eval/inv(nil)},
    % \hypref{eval2}, and \hypref{1}, we have
    % \begin{gather}
    %    \ottnt{S_{{\mathrm{2}}}} \ottsym{=} \ottsym{[}  \ottsym{]}  \triangleright  \ottnt{S_{{\mathrm{1}}}}  \hyp{4}.
    % \end{gather}
    % We can derive
    % \begin{gather}
    %    \ottsym{[}  \ottsym{]}  :  \ottnt{T} \, \ottkw{list}  \hyp{5}
    % \end{gather}
    % by \ruleref{RTV-Nil}.  By \propref{semty/logic/imply}, \propref{semty/wf},
    % \propref{fol/thm/misc(push)}, \hypref{semty2}, \hypref{3.1}, and \hypref{5},
    % we have
    % \begin{gather}
    %   \sigma  \ottsym{:}  \Gamma  \models  \ottnt{S_{{\mathrm{1}}}}  \ottsym{:}  \ottsym{\{}  \Upsilon  \mid  \exists \,  \ottmv{x} \mathord:\allowbreak \ottnt{T} \, \ottkw{list}   \ottsym{.}  \ottsym{(}  \varphi  \wedge  \ottmv{x}  \ottsym{=}  \ottsym{[}  \ottsym{]}  \ottsym{)}  \wedge  \ottmv{x}  \ottsym{=}  \ottsym{[}  \ottsym{]}  \ottsym{\}} \hyp{6}.
    % \end{gather}
    % So the goal follows by \propref{semty/logic} and \hypref{6}.

  \item[\ruleref{RT-Cons}]
    Similar to the case for \ruleref{RT-Pair}.

  \item[\ruleref{RT-If}]
    
    We have
    \begin{gather}
       \ottnt{I} \ottsym{=} \ottkw{IF} \, \mathit{IS}_{{\mathrm{1}}} \, \mathit{IS}_{{\mathrm{2}}}  \hyp{1}\\
       \Phi_{{\mathrm{1}}} \ottsym{=} \ottsym{\{}   \ottmv{x} \mathord:\allowbreak \ottkw{int}   \triangleright  \Upsilon  \mid  \varphi  \ottsym{\}}  \hyp{2}\\
       \Gamma \vdash \ottsym{\{}  \Upsilon  \mid  \exists \,  \ottmv{x} \mathord:\allowbreak \ottkw{int}   \ottsym{.}  \varphi  \wedge  \ottmv{x}  \neq  0  \ottsym{\}} \; \mathit{IS}_{{\mathrm{1}}} \; \Phi_{{\mathrm{2}}}  \hyp{3}\\
       \Gamma \vdash \ottsym{\{}  \Upsilon  \mid  \exists \,  \ottmv{x} \mathord:\allowbreak \ottkw{int}   \ottsym{.}  \varphi  \wedge  \ottmv{x}  \ottsym{=}  0  \ottsym{\}} \; \mathit{IS}_{{\mathrm{2}}} \; \Phi_{{\mathrm{2}}}  \hyp{4}
    \end{gather}
    for some \(\mathit{IS}_{{\mathrm{1}}}\), \(\mathit{IS}_{{\mathrm{2}}}\), \(\ottmv{x}\), \(\Upsilon\), and \(\varphi\).
    The last rule that derives \hypref{eval2} is \ruleref{E-IfT} or \ruleref{E-IfF}.
    We conduct case analysis.
    
    % By \propref{eval/inv(if)}, \hypref{eval2}, and \hypref{1}, we have the
    % following two cases.
    \begin{match}
    \item[\ruleref{E-IfT}] For some \(\ottnt{i}\) and \(\ottnt{S}\),
      \begin{gather}
        \ottnt{i}  \neq  0 \hyp{5}\\
         \ottnt{S_{{\mathrm{1}}}} \ottsym{=} \ottnt{i}  \triangleright  \ottnt{S}  \hyp{6}\\
        \ottnt{S}  \vdash  \mathit{IS}_{{\mathrm{1}}}  \Downarrow  \ottnt{S_{{\mathrm{2}}}} \hyp{7}.
      \end{gather}
      From \hypref{semty2}, \hypref{2}, \hypref{6}, and \propref{subty/exists=>subst'}, we have
      \begin{gather}
        \sigma  \ottsym{:}  \Gamma  \models  \ottnt{S}  \ottsym{:}  \ottsym{\{}  \Upsilon  \mid  \exists \,  \ottmv{x} \mathord:\allowbreak \ottkw{int}   \ottsym{.}  \varphi  \wedge  \ottmv{x}  \ottsym{=}  \ottnt{i}  \ottsym{\}} \hyp{9}.
      \end{gather}
      From \hypref{5}, we have
      \begin{gather}
        \sigma  \ottsym{:}  \Gamma  \models  \ottnt{S}  \ottsym{:}  \ottsym{\{}  \Upsilon  \mid  \exists \,  \ottmv{x} \mathord:\allowbreak \ottkw{int}   \ottsym{.}  \varphi  \wedge  \ottmv{x}  \neq  0  \ottsym{\}} \hyp{10}.
      \end{gather}
      Then, the goal follows From IH and \hypref{3}.

    \item[\ruleref{E-IfF}]
      Similar to the case for \ruleref{E-IfT}.
      % For some \(\ottnt{S}\),
      % \begin{gather}
      %    \ottnt{S_{{\mathrm{1}}}} \ottsym{=} 0  \triangleright  \ottnt{S}  \hyp{11}\\
      %   \ottnt{S}  \vdash  \mathit{IS}_{{\mathrm{2}}}  \Downarrow  \ottnt{S_{{\mathrm{2}}}} \hyp{12}.
      % \end{gather}
      % By \propref{semty/logic}, \hypref{semty2}, \hypref{2}, and \hypref{11}, we
      % have
      % \begin{gather}
      %   \sigma  \ottsym{:}  \Gamma  \models  \ottnt{S}  \ottsym{:}  \ottsym{\{}  \Upsilon  \mid  \exists \,  \ottmv{x} \mathord:\allowbreak \ottkw{int}   \ottsym{.}  \varphi  \wedge  \ottmv{x}  \ottsym{=}  0  \ottsym{\}} \hyp{13}.
      % \end{gather}
      % Now the goal follows by applying IH of \hypref{4} to \hypref{12} and
      % \hypref{13}.
    \end{match}

  \item[\ruleref{RT-Loop}] We have
    \begin{gather}
       \ottnt{I} \ottsym{=} \ottkw{LOOP} \, \mathit{IS}  \hyp{1}\\
       \Phi_{{\mathrm{1}}} \ottsym{=} \ottsym{\{}   \ottmv{x} \mathord:\allowbreak \ottkw{int}   \triangleright  \Upsilon  \mid  \varphi  \ottsym{\}}  \hyp{2}\\
       \Phi_{{\mathrm{2}}} \ottsym{=} \ottsym{\{}  \Upsilon  \mid  \exists \,  \ottmv{x} \mathord:\allowbreak \ottkw{int}   \ottsym{.}  \varphi  \wedge  \ottmv{x}  \ottsym{=}  0  \ottsym{\}}  \hyp{3}\\
       \Gamma \vdash \ottsym{\{}  \Upsilon  \mid  \exists \,  \ottmv{x} \mathord:\allowbreak \ottkw{int}   \ottsym{.}  \varphi  \wedge  \ottmv{x}  \neq  0  \ottsym{\}} \; \mathit{IS} \; \ottsym{\{}   \ottmv{x} \mathord:\allowbreak \ottkw{int}   \triangleright  \Upsilon  \mid  \varphi  \ottsym{\}}  \hyp{4}\\
       \ottnt{S_{{\mathrm{1}}}} \ottsym{=} \ottnt{i}  \triangleright  \ottnt{S}  \hyp{5}
    \end{gather}
    for some \(\mathit{IS}\), \(\ottmv{x}\), \(\Upsilon\), \(\ottnt{S}\), and
    \(\varphi\).  By IH and \hypref{4}, we have
    \begin{multline}
      \text{For any } \ottnt{S'_{{\mathrm{1}}}}, \ottnt{S'_{{\mathrm{2}}}}, \text{ if } \ottnt{S'_{{\mathrm{1}}}}  \vdash  \mathit{IS}  \Downarrow  \ottnt{S'_{{\mathrm{2}}}} \text{
        and } \sigma  \ottsym{:}  \Gamma  \models  \ottnt{S'_{{\mathrm{1}}}}  \ottsym{:}  \ottsym{\{}  \Upsilon  \mid  \exists \,  \ottmv{x} \mathord:\allowbreak \ottkw{int}   \ottsym{.}  \varphi  \wedge  \ottmv{x}  \neq  0  \ottsym{\}},\\ \text{ then
      } \sigma  \ottsym{:}  \Gamma  \models  \ottnt{S'_{{\mathrm{2}}}}  \ottsym{:}  \ottsym{\{}   \ottmv{x} \mathord:\allowbreak \ottkw{int}   \triangleright  \Upsilon  \mid  \varphi  \ottsym{\}} \hyp{5}.
    \end{multline}
    Then, the goal follows from \propref{soundness/loop}, \hypref{eval2},
    \hypref{semty2}, \hypref{1}, \hypref{2} and \hypref{5}.

  \item[\ruleref{RT-IfCons}]
    Similar to the case for \ruleref{RT-If}.

  \item[\ruleref{RT-Iter}] We have
    \begin{gather}
       \ottnt{I} \ottsym{=} \ottkw{ITER} \, \mathit{IS}  \hyp{1}\\
       \Phi_{{\mathrm{1}}} \ottsym{=} \ottsym{\{}   \ottmv{x} \mathord:\allowbreak \ottnt{T} \, \ottkw{list}   \triangleright  \Upsilon  \mid  \varphi  \ottsym{\}}  \hyp{2}\\
       \Phi_{{\mathrm{2}}} \ottsym{=} \ottsym{\{}  \Upsilon  \mid  \exists \,  \ottmv{x} \mathord:\allowbreak \ottnt{T} \, \ottkw{list}   \ottsym{.}  \varphi  \wedge  \ottmv{x}  \ottsym{=}  \ottsym{[}  \ottsym{]}  \ottsym{\}}  \hyp{3}\\
       \Gamma  \ottsym{,}   \ottmv{x_{{\mathrm{2}}}} \mathord:\allowbreak \ottnt{T} \, \ottkw{list}  \vdash \ottsym{\{}   \ottmv{x_{{\mathrm{1}}}} \mathord:\allowbreak \ottnt{T}   \triangleright  \Upsilon  \mid  \exists \,  \ottmv{x} \mathord:\allowbreak \ottnt{T} \, \ottkw{list}   \ottsym{.}  \varphi  \wedge  \ottmv{x_{{\mathrm{1}}}}  ::  \ottmv{x_{{\mathrm{2}}}}  \ottsym{=}  \ottmv{x}  \ottsym{\}} \; \mathit{IS} \; \ottsym{\{}  \Upsilon  \mid  \exists \,  \ottmv{x} \mathord:\allowbreak \ottnt{T} \, \ottkw{list}   \ottsym{.}  \varphi  \wedge  \ottmv{x_{{\mathrm{2}}}}  \ottsym{=}  \ottmv{x}  \ottsym{\}}  \hyp{5}\\
      \ottmv{x_{{\mathrm{1}}}}  \neq  \ottmv{x_{{\mathrm{2}}}} \hyp{5.1}\\
       \ottmv{x_{{\mathrm{1}}}}  \notin   \text{dom}( \Gamma  \ottsym{,}   \widehat{  \ottmv{x} \mathord:\allowbreak \ottnt{T} \, \ottkw{list}   \triangleright  \Upsilon }  )   \hyp{5.2}\\
       \ottmv{x_{{\mathrm{2}}}}  \notin   \text{dom}( \Gamma  \ottsym{,}   \widehat{  \ottmv{x} \mathord:\allowbreak \ottnt{T} \, \ottkw{list}   \triangleright  \Upsilon }  )   \hyp{5.3}
    \end{gather}
    for some \(\mathit{IS}\), \(\ottmv{x}\), \(\ottmv{x_{{\mathrm{1}}}}\), \(\ottmv{x_{{\mathrm{2}}}}\), \(\ottnt{T}\),
    \(\Upsilon\), and \(\varphi\).
    From IH and \hypref{5}, we have
    \begin{multline}
      \text{For any } \ottnt{S'_{{\mathrm{1}}}}, \ottnt{S'_{{\mathrm{2}}}}, \sigma',
      \text{ if } \ottnt{S'_{{\mathrm{1}}}}  \vdash  \mathit{IS}  \Downarrow  \ottnt{S'_{{\mathrm{2}}}} \text{ and } \sigma'  \ottsym{:}  \Gamma  \ottsym{,}   \ottmv{x_{{\mathrm{2}}}} \mathord:\allowbreak \ottnt{T} \, \ottkw{list}   \models  \ottnt{S'_{{\mathrm{1}}}}  \ottsym{:}  \ottsym{\{}   \ottmv{x_{{\mathrm{1}}}} \mathord:\allowbreak \ottnt{T}   \triangleright  \Upsilon  \mid  \exists \,  \ottmv{x} \mathord:\allowbreak \ottnt{T} \, \ottkw{list}   \ottsym{.}  \varphi  \wedge  \ottmv{x_{{\mathrm{1}}}}  ::  \ottmv{x_{{\mathrm{2}}}}  \ottsym{=}  \ottmv{x}  \ottsym{\}},\\
      \text{ then } \sigma'  \ottsym{:}  \Gamma  \ottsym{,}   \ottmv{x_{{\mathrm{2}}}} \mathord:\allowbreak \ottnt{T} \, \ottkw{list}   \models  \ottnt{S'_{{\mathrm{2}}}}  \ottsym{:}  \ottsym{\{}  \Upsilon  \mid  \exists \,  \ottmv{x} \mathord:\allowbreak \ottnt{T} \, \ottkw{list}   \ottsym{.}  \varphi  \wedge  \ottmv{x_{{\mathrm{2}}}}  \ottsym{=}  \ottmv{x}  \ottsym{\}} \hyp{6}.
    \end{multline}
    Then, the goal follows from \propref{soundness/iter}. %, \hypref{eval2}, \hypref{semty2}, \hypref{1}, \hypref{2}, \hypref{5.1}, \hypref{5.2},
    \hypref{5.3}, and \hypref{6}.

  \item[\ruleref{RT-Lambda}] Since the last rule that derive \hypref{eval2} is \ruleref{E-Lambda}, we have
    \begin{gather}
       \ottnt{I} \ottsym{=} \ottkw{LAMBDA} \, \ottnt{T_{{\mathrm{1}}}} \, \ottnt{T_{{\mathrm{2}}}} \, \mathit{IS}  \hyp{1}\\
       \Phi_{{\mathrm{1}}} \ottsym{=} \ottsym{\{}  \Upsilon  \mid  \varphi  \ottsym{\}}  \hyp{2}\\
       \Phi_{{\mathrm{2}}} \ottsym{=} \ottsym{\{}   \ottmv{x} \mathord:\allowbreak \ottnt{T_{{\mathrm{1}}}}  \to  \ottnt{T_{{\mathrm{2}}}}   \triangleright  \Upsilon  \mid  \varphi  \wedge  \forall \,  \ottmv{y'_{{\mathrm{1}}}} \mathord:\allowbreak \ottnt{T_{{\mathrm{1}}}}   \ottsym{,}   \ottmv{y_{{\mathrm{1}}}} \mathord:\allowbreak \ottnt{T_{{\mathrm{1}}}}   \ottsym{,}   \ottmv{y_{{\mathrm{2}}}} \mathord:\allowbreak \ottnt{T_{{\mathrm{2}}}}   \ottsym{.}  \ottmv{y'_{{\mathrm{1}}}}  \ottsym{=}  \ottmv{y_{{\mathrm{1}}}}  \wedge  \varphi_{{\mathrm{1}}}  \wedge   \ottkw{call} ( \ottmv{x} ,  \ottmv{y'_{{\mathrm{1}}}} ) =  \ottmv{y_{{\mathrm{2}}}}   \implies  \varphi_{{\mathrm{2}}}  \ottsym{\}}  \hyp{3}\\
        \ottmv{y'_{{\mathrm{1}}}} \mathord:\allowbreak \ottnt{T_{{\mathrm{1}}}}  \vdash \ottsym{\{}   \ottmv{y_{{\mathrm{1}}}} \mathord:\allowbreak \ottnt{T_{{\mathrm{1}}}}   \triangleright  \ddagger  \mid  \ottmv{y'_{{\mathrm{1}}}}  \ottsym{=}  \ottmv{y_{{\mathrm{1}}}}  \wedge  \varphi_{{\mathrm{1}}}  \ottsym{\}} \; \mathit{IS} \; \ottsym{\{}   \ottmv{y_{{\mathrm{2}}}} \mathord:\allowbreak \ottnt{T_{{\mathrm{2}}}}   \triangleright  \ddagger  \mid  \varphi_{{\mathrm{2}}}  \ottsym{\}}  \hyp{4}\\
       \ottmv{x}  \notin    \text{dom}( \Gamma  \ottsym{,}   \widehat{ \Upsilon }  )   \cup  \ottsym{\{}  \ottmv{y_{{\mathrm{1}}}}  \ottsym{,}  \ottmv{y'_{{\mathrm{1}}}}  \ottsym{,}  \ottmv{y_{{\mathrm{2}}}}  \ottsym{\}}   \hyp{4.1}\\
      \ottmv{y_{{\mathrm{1}}}}  \neq  \ottmv{y'_{{\mathrm{1}}}} \hyp{4.2}\\
        \ottmv{y'_{{\mathrm{1}}}} \mathord:\allowbreak \ottnt{T_{{\mathrm{1}}}}   \ottsym{,}   \ottmv{y_{{\mathrm{1}}}} \mathord:\allowbreak \ottnt{T_{{\mathrm{1}}}}    \vdash   \varphi_{{\mathrm{1}}}  : \mathord{*}  \hyp{4.3}\\
       \ottnt{S_{{\mathrm{2}}}} \ottsym{=}  \langle  \mathit{IS}  \rangle   \triangleright  \ottnt{S_{{\mathrm{1}}}} \hyp{5}
    \end{gather}
    for some \(\mathit{IS}\), \(\ottmv{x}\), \(\ottmv{y_{{\mathrm{1}}}}\), \(\ottmv{y'_{{\mathrm{1}}}}\), \(\ottmv{y_{{\mathrm{2}}}}\),
    \(\ottnt{T_{{\mathrm{1}}}}\), \(\ottnt{T_{{\mathrm{2}}}}\), \(\Upsilon\), \(\varphi\), \(\varphi_{{\mathrm{1}}}\), and \(\varphi_{{\mathrm{2}}}\).
    From IH and \hypref{4}, we have
    \begin{multline}
      \text{For any } \ottnt{V_{{\mathrm{1}}}}, \ottnt{V_{{\mathrm{2}}}}, \sigma, \text{ if } \ottnt{V_{{\mathrm{1}}}}  \triangleright  \ddagger  \vdash  \mathit{IS}  \Downarrow  \ottnt{V_{{\mathrm{2}}}}  \triangleright  \ddagger \text{ and } \sigma  \ottsym{:}   \ottmv{y'_{{\mathrm{1}}}} \mathord:\allowbreak \ottnt{T_{{\mathrm{1}}}}   \models  \ottnt{V_{{\mathrm{1}}}}  \triangleright  \ddagger  \ottsym{:}  \ottsym{\{}   \ottmv{y_{{\mathrm{1}}}} \mathord:\allowbreak \ottnt{T_{{\mathrm{1}}}}   \triangleright  \ddagger  \mid  \ottmv{y'_{{\mathrm{1}}}}  \ottsym{=}  \ottmv{y_{{\mathrm{1}}}}  \wedge  \varphi_{{\mathrm{1}}}  \ottsym{\}},\\
      \text{ then } \sigma  \ottsym{:}   \ottmv{y'_{{\mathrm{1}}}} \mathord:\allowbreak \ottnt{T_{{\mathrm{1}}}}   \models  \ottnt{V_{{\mathrm{2}}}}  \triangleright  \ddagger  \ottsym{:}  \ottsym{\{}   \ottmv{y_{{\mathrm{2}}}} \mathord:\allowbreak \ottnt{T_{{\mathrm{2}}}}   \triangleright  \ddagger  \mid  \varphi_{{\mathrm{2}}}  \ottsym{\}} \hyp{5}.
    \end{multline}
    % \YN{Actually, this is a consequence fact of IH...}  We can
    % derive \(  \ottmv{y'_{{\mathrm{1}}}} \mathord:\allowbreak \ottnt{T_{{\mathrm{1}}}}   \ottsym{,}   \ottmv{y_{{\mathrm{1}}}} \mathord:\allowbreak \ottnt{T_{{\mathrm{1}}}}    \vdash   \ottmv{y'_{{\mathrm{1}}}}  \ottsym{=}  \ottmv{y_{{\mathrm{1}}}}  \wedge  \varphi_{{\mathrm{1}}}  : \mathord{*} \) by \hypref{4.3}, and so
    % by \propref{typing/wf} and \hypref{4}, we have
    % \begin{gather}
    %     \ottmv{y'_{{\mathrm{1}}}} \mathord:\allowbreak \ottnt{T_{{\mathrm{1}}}}   \ottsym{,}   \ottmv{y_{{\mathrm{1}}}} \mathord:\allowbreak \ottnt{T_{{\mathrm{2}}}}    \vdash   \varphi_{{\mathrm{2}}}  : \mathord{*}  \hyp{5.1}.
    % \end{gather}
    % By \propref{typing/conservative} and \hypref{4}, we have
    % \(\ottnt{T_{{\mathrm{1}}}}  \triangleright  \ddagger  \vdash  \mathit{IS}  \Rightarrow  \ottnt{T_{{\mathrm{2}}}}  \triangleright  \ddagger\), and so by \ruleref{RTV-Fun}, we have
    % \begin{gather}
    %     \langle  \mathit{IS}  \rangle   :  \ottnt{T_{{\mathrm{1}}}}  \to  \ottnt{T_{{\mathrm{2}}}}  \hyp{5.2}.
    % \end{gather}
    By \propref{soundness/lambda}, \hypref{4.2}, \hypref{4.3}, \hypref{5},
    \hypref{5.1}, and \hypref{5.2}, we have
    \begin{gather}
      \Gamma  \ottsym{,}   \widehat{ \Upsilon }   \models  \forall \,  \ottmv{y'_{{\mathrm{1}}}} \mathord:\allowbreak \ottnt{T_{{\mathrm{1}}}}   \ottsym{,}   \ottmv{y_{{\mathrm{1}}}} \mathord:\allowbreak \ottnt{T_{{\mathrm{1}}}}   \ottsym{,}   \ottmv{y_{{\mathrm{2}}}} \mathord:\allowbreak \ottnt{T_{{\mathrm{2}}}}   \ottsym{.}  \ottmv{y'_{{\mathrm{1}}}}  \ottsym{=}  \ottmv{y_{{\mathrm{1}}}}  \wedge  \varphi_{{\mathrm{1}}}  \wedge   \ottkw{call} (  \langle  \mathit{IS}  \rangle  ,  \ottmv{y'_{{\mathrm{1}}}} ) =  \ottmv{y_{{\mathrm{2}}}}   \implies  \varphi_{{\mathrm{2}}} \hyp{6}
    \end{gather}
    and hence, from \hypref{semty2}, \hypref{2}, and \hypref{6}, we have
    \begin{gather}
      \sigma  \ottsym{:}  \Gamma  \models  \ottnt{S_{{\mathrm{1}}}}  \ottsym{:}  \ottsym{\{}  \Upsilon  \mid  \varphi  \wedge  \forall \,  \ottmv{y'_{{\mathrm{1}}}} \mathord:\allowbreak \ottnt{T_{{\mathrm{1}}}}   \ottsym{,}   \ottmv{y_{{\mathrm{1}}}} \mathord:\allowbreak \ottnt{T_{{\mathrm{1}}}}   \ottsym{,}   \ottmv{y_{{\mathrm{2}}}} \mathord:\allowbreak \ottnt{T_{{\mathrm{2}}}}   \ottsym{.}  \ottmv{y'_{{\mathrm{1}}}}  \ottsym{=}  \ottmv{y_{{\mathrm{1}}}}  \wedge  \varphi_{{\mathrm{1}}}  \wedge   \ottkw{call} (  \langle  \mathit{IS}  \rangle  ,  \ottmv{y'_{{\mathrm{1}}}} ) =  \ottmv{y_{{\mathrm{2}}}}   \implies  \varphi_{{\mathrm{2}}}  \ottsym{\}} \hyp{7}.
    \end{gather}
    By \hypref{4.1}, and \hypref{7}, we have
    \begin{multline}
      \sigma  \ottsym{:}  \Gamma  \models  \ottnt{S_{{\mathrm{1}}}}  \ottsym{:}  \ottsym{\{}  \Upsilon  \mid  \exists \,  \ottmv{x} \mathord:\allowbreak \ottnt{T_{{\mathrm{1}}}}  \to  \ottnt{T_{{\mathrm{2}}}}   \ottsym{.}  \ottsym{(}   \varphi  \wedge {} \\  \forall \,  \ottmv{y'_{{\mathrm{1}}}} \mathord:\allowbreak \ottnt{T_{{\mathrm{1}}}}   \ottsym{,}   \ottmv{y_{{\mathrm{1}}}} \mathord:\allowbreak \ottnt{T_{{\mathrm{1}}}}   \ottsym{,}   \ottmv{y_{{\mathrm{2}}}} \mathord:\allowbreak \ottnt{T_{{\mathrm{2}}}}   \ottsym{.}  \ottmv{y'_{{\mathrm{1}}}}  \ottsym{=}  \ottmv{y_{{\mathrm{1}}}}   \wedge  \varphi_{{\mathrm{1}}}  \wedge   \ottkw{call} ( \ottmv{x} ,  \ottmv{y'_{{\mathrm{1}}}} ) =  \ottmv{y_{{\mathrm{2}}}}   \implies  \varphi_{{\mathrm{2}}}  \ottsym{)}  \wedge  \ottmv{x}  \ottsym{=}   \langle  \mathit{IS}  \rangle   \ottsym{\}}
      \hyp{9}.
    \end{multline}
    Therefore, from \propref{subty/exists=>subst'} and \hypref{9}, we have
    \begin{multline*}
      \sigma  \ottsym{:}  \Gamma  \models   \langle  \mathit{IS}  \rangle   \triangleright  \ottnt{S_{{\mathrm{1}}}}  \ottsym{:}  \ottsym{\{}   \ottmv{x} \mathord:\allowbreak \ottnt{T_{{\mathrm{1}}}}  \to  \ottnt{T_{{\mathrm{2}}}}   \triangleright  \Upsilon  \mid   \varphi  \wedge {} \\  \forall \,  \ottmv{y'_{{\mathrm{1}}}} \mathord:\allowbreak \ottnt{T_{{\mathrm{1}}}}   \ottsym{,}   \ottmv{y_{{\mathrm{1}}}} \mathord:\allowbreak \ottnt{T_{{\mathrm{1}}}}   \ottsym{,}   \ottmv{y_{{\mathrm{2}}}} \mathord:\allowbreak \ottnt{T_{{\mathrm{2}}}}   \ottsym{.}  \ottmv{y'_{{\mathrm{1}}}}  \ottsym{=}  \ottmv{y_{{\mathrm{1}}}}   \wedge  \varphi_{{\mathrm{1}}}  \wedge   \ottkw{call} ( \ottmv{x} ,  \ottmv{y'_{{\mathrm{1}}}} ) =  \ottmv{y_{{\mathrm{2}}}}   \implies  \varphi_{{\mathrm{2}}}  \ottsym{\}}
    \end{multline*}
    as required.

  \item[\ruleref{RT-Exec}]
    We have
    \begin{gather}
       \ottnt{I} \ottsym{=} \ottkw{EXEC}  \hyp{1}\\
       \Phi_{{\mathrm{1}}} \ottsym{=} \ottsym{\{}   \ottmv{x_{{\mathrm{1}}}} \mathord:\allowbreak \ottnt{T_{{\mathrm{1}}}}   \triangleright   \ottmv{x_{{\mathrm{2}}}} \mathord:\allowbreak \ottnt{T_{{\mathrm{1}}}}  \to  \ottnt{T_{{\mathrm{2}}}}   \triangleright  \Upsilon  \mid  \varphi  \ottsym{\}}  \hyp{2}\\
       \Phi_{{\mathrm{2}}} \ottsym{=} \ottsym{\{}   \ottmv{x_{{\mathrm{3}}}} \mathord:\allowbreak \ottnt{T_{{\mathrm{2}}}}   \triangleright  \Upsilon  \mid  \exists \,  \ottmv{x_{{\mathrm{1}}}} \mathord:\allowbreak \ottnt{T_{{\mathrm{1}}}}   \ottsym{,}   \ottmv{x_{{\mathrm{2}}}} \mathord:\allowbreak \ottnt{T_{{\mathrm{1}}}}  \to  \ottnt{T_{{\mathrm{2}}}}   \ottsym{.}  \varphi  \wedge   \ottkw{call} ( \ottmv{x_{{\mathrm{2}}}} ,  \ottmv{x_{{\mathrm{1}}}} ) =  \ottmv{x_{{\mathrm{3}}}}   \ottsym{\}} 
      \hyp{3}\\
       \ottmv{x_{{\mathrm{3}}}}  \notin   \text{dom}( \Gamma  \ottsym{,}   \widehat{  \ottmv{x_{{\mathrm{1}}}} \mathord:\allowbreak \ottnt{T_{{\mathrm{1}}}}   \triangleright   \ottmv{x_{{\mathrm{2}}}} \mathord:\allowbreak \ottnt{T_{{\mathrm{1}}}}  \to  \ottnt{T_{{\mathrm{2}}}}   \triangleright  \Upsilon }  )   \hyp{3.1}
    \end{gather}
    for some \(\ottmv{x_{{\mathrm{1}}}}\), \(\ottmv{x_{{\mathrm{2}}}}\), \(\ottmv{x_{{\mathrm{3}}}}\), \(\ottnt{T_{{\mathrm{1}}}}\),
    \(\ottnt{T_{{\mathrm{2}}}}\), \(\Upsilon\), and \(\varphi\).  Since the last rule that
    derives \hypref{eval2} is \ruleref{E-Exec}, we have
    \begin{gather}
       \ottnt{S_{{\mathrm{1}}}} \ottsym{=} \ottnt{V_{{\mathrm{1}}}}  \triangleright   \langle  \mathit{IS}  \rangle   \triangleright  \ottnt{S}  \hyp{4}\\
       \ottnt{S_{{\mathrm{2}}}} \ottsym{=} \ottnt{V_{{\mathrm{2}}}}  \triangleright  \ottnt{S}  \hyp{5}\\
      \ottnt{V_{{\mathrm{1}}}}  \triangleright  \ddagger  \vdash  \mathit{IS}  \Downarrow  \ottnt{V_{{\mathrm{2}}}}  \triangleright  \ddagger \hyp{6}
    \end{gather}
    for some \(\ottnt{V_{{\mathrm{1}}}}\), \(\ottnt{V_{{\mathrm{2}}}}\), \(\mathit{IS}\), and \(\ottnt{S}\).  By
    \hypref{semty2}, \hypref{2}, and \hypref{4}, we have
    \begin{gather}
      \sigma  \ottsym{:}  \Gamma  \models  \ottnt{S}  \ottsym{:}  \ottsym{\{}  \Upsilon  \mid  \exists \,  \ottmv{x_{{\mathrm{2}}}} \mathord:\allowbreak \ottnt{T_{{\mathrm{1}}}}  \to  \ottnt{T_{{\mathrm{2}}}}   \ottsym{.}  \ottsym{(}  \exists \,  \ottmv{x_{{\mathrm{1}}}} \mathord:\allowbreak \ottnt{T_{{\mathrm{1}}}}   \ottsym{.}  \varphi  \wedge  \ottmv{x_{{\mathrm{1}}}}  \ottsym{=}  \ottnt{V_{{\mathrm{1}}}}  \ottsym{)}  \wedge  \ottmv{x_{{\mathrm{2}}}}  \ottsym{=}   \langle  \mathit{IS}  \rangle   \ottsym{\}} \hyp{9}.
    \end{gather}
    % By \propref{semty/conservative}, \hypref{semty2}, \hypref{2}, and
    % \hypref{4}, we have
    % \( \ottnt{V_{{\mathrm{1}}}}  \triangleright   \langle  \mathit{IS}  \rangle   \triangleright  \ottnt{S}  :  \ottnt{T_{{\mathrm{1}}}}  \triangleright  \ottnt{T_{{\mathrm{1}}}}  \to  \ottnt{T_{{\mathrm{2}}}}  \triangleright   \lfloor  \ottsym{\{}  \Upsilon  \mid  \varphi  \ottsym{\}}  \rfloor  \), and so by
    % \propref{st/inv(push)}, we have
    % \begin{gather}
    %    \ottnt{V_{{\mathrm{1}}}}  :  \ottnt{T_{{\mathrm{1}}}}  \hyp{7}\\
    %     \langle  \mathit{IS}  \rangle   :  \ottnt{T_{{\mathrm{1}}}}  \to  \ottnt{T_{{\mathrm{2}}}}  \hyp{8}.
    % \end{gather}
    % By \propref{vtyping/inv(fun)} and \hypref{8}, we have
    % \begin{gather}
    %   \ottnt{T_{{\mathrm{1}}}}  \triangleright  \ddagger  \vdash  \mathit{IS}  \Rightarrow  \ottnt{T_{{\mathrm{2}}}}  \triangleright  \ddagger \hyp{9.5}.
    % \end{gather}
    % We can derive \( \ottnt{V_{{\mathrm{1}}}}  \triangleright  \ddagger  :  \ottnt{T_{{\mathrm{1}}}}  \triangleright  \ddagger \) by \ruleref{ST-Bottom},
    % \ruleref{ST-Push}, \hypref{7}, and so, we have \( \ottnt{V_{{\mathrm{2}}}}  \triangleright  \ddagger  :  \ottnt{T_{{\mathrm{2}}}}  \triangleright  \ddagger \)
    % by \propref{styping/sound}, \hypref{6}, and \hypref{9.5}.  Hence,
    % \begin{gather}
    %    \ottnt{V_{{\mathrm{2}}}}  :  \ottnt{T_{{\mathrm{2}}}}  \hyp{9.8}
    % \end{gather}
    % by \propref{st/inv(push)}.
    By \propref{soundness/exec}, \hypref{6}, \hypref{7}, and \hypref{8}, we have
    \begin{gather}
      \Gamma  \ottsym{,}   \widehat{ \Upsilon }   \models   \ottkw{call} (  \langle  \mathit{IS}  \rangle  ,  \ottnt{V_{{\mathrm{1}}}} ) =  \ottnt{V_{{\mathrm{2}}}}  \hyp{10}.
    \end{gather}
    Therefore, from \hypref{9}, and \hypref{10}, we have
    \begin{gather}
      \sigma  \ottsym{:}  \Gamma  \models  \ottnt{S}  \ottsym{:}  \ottsym{\{}  \Upsilon  \mid  \ottsym{(}  \exists \,  \ottmv{x_{{\mathrm{2}}}} \mathord:\allowbreak \ottnt{T_{{\mathrm{1}}}}  \to  \ottnt{T_{{\mathrm{2}}}}   \ottsym{.}  \ottsym{(}  \exists \,  \ottmv{x_{{\mathrm{1}}}} \mathord:\allowbreak \ottnt{T_{{\mathrm{1}}}}   \ottsym{.}  \varphi  \wedge  \ottmv{x_{{\mathrm{1}}}}  \ottsym{=}  \ottnt{V_{{\mathrm{1}}}}  \ottsym{)}  \wedge  \ottmv{x_{{\mathrm{2}}}}  \ottsym{=}   \langle  \mathit{IS}  \rangle   \ottsym{)}  \wedge   \ottkw{call} (  \langle  \mathit{IS}  \rangle  ,  \ottnt{V_{{\mathrm{1}}}} ) =  \ottnt{V_{{\mathrm{2}}}}   \ottsym{\}} \hyp{11}.
    \end{gather}
    From \hypref{3.1} and \hypref{11}, we have
    $\sigma  \ottsym{:}  \Gamma  \models  \ottnt{S}  \ottsym{:}  \ottsym{\{}  \Upsilon  \mid  \exists \,  \ottmv{x_{{\mathrm{3}}}} \mathord:\allowbreak \ottnt{T_{{\mathrm{2}}}}   \ottsym{.}  \ottsym{(}  \exists \,  \ottmv{x_{{\mathrm{1}}}} \mathord:\allowbreak \ottnt{T_{{\mathrm{1}}}}   \ottsym{,}   \ottmv{x_{{\mathrm{2}}}} \mathord:\allowbreak \ottnt{T_{{\mathrm{1}}}}  \to  \ottnt{T_{{\mathrm{2}}}}   \ottsym{.}  \varphi  \wedge   \ottkw{call} ( \ottmv{x_{{\mathrm{2}}}} ,  \ottmv{x_{{\mathrm{1}}}} ) =  \ottmv{x_{{\mathrm{3}}}}   \ottsym{)}  \wedge  \ottmv{x_{{\mathrm{3}}}}  \ottsym{=}  \ottnt{V_{{\mathrm{2}}}}  \ottsym{\}}$ and hence
    $\sigma  \ottsym{:}  \Gamma  \models  \ottnt{V_{{\mathrm{2}}}}  \triangleright  \ottnt{S}  \ottsym{:}  \ottsym{\{}   \ottmv{x_{{\mathrm{3}}}} \mathord:\allowbreak \ottnt{T_{{\mathrm{2}}}}   \triangleright  \Upsilon  \mid  \ottsym{(}  \exists \,  \ottmv{x_{{\mathrm{1}}}} \mathord:\allowbreak \ottnt{T_{{\mathrm{1}}}}   \ottsym{,}   \ottmv{x_{{\mathrm{2}}}} \mathord:\allowbreak \ottnt{T_{{\mathrm{1}}}}  \to  \ottnt{T_{{\mathrm{2}}}}   \ottsym{.}  \varphi  \wedge   \ottkw{call} ( \ottmv{x_{{\mathrm{2}}}} ,  \ottmv{x_{{\mathrm{1}}}} ) =  \ottmv{x_{{\mathrm{3}}}}   \ottsym{)}  \ottsym{\}}$ as required.

  \item[\ruleref{RT-TransferTokens}] We have
    \begin{gather}
       \ottnt{I} \ottsym{=} \texttt{TRANSFER\symbol{95}TOKENS} \, \ottnt{T}  \hyp{1}\\
       \Phi_{{\mathrm{1}}} \ottsym{=} \ottsym{\{}   \ottmv{x_{{\mathrm{1}}}} \mathord:\allowbreak \ottnt{T}   \triangleright   \ottmv{x_{{\mathrm{2}}}} \mathord:\allowbreak \ottkw{int}   \triangleright   \ottmv{x_{{\mathrm{3}}}} \mathord:\allowbreak \ottkw{address}   \triangleright  \Upsilon  \mid  \varphi  \ottsym{\}}  \hyp{2}\\
       \Phi_{{\mathrm{2}}} \ottsym{=} \ottsym{\{}   \ottmv{x_{{\mathrm{4}}}} \mathord:\allowbreak \ottkw{operation}   \triangleright  \Upsilon  \mid  \exists \,  \ottmv{x_{{\mathrm{1}}}} \mathord:\allowbreak \ottnt{T}   \ottsym{,}   \ottmv{x_{{\mathrm{2}}}} \mathord:\allowbreak \ottkw{int}   \ottsym{,}   \ottmv{x_{{\mathrm{3}}}} \mathord:\allowbreak \ottkw{address}   \ottsym{.}  \varphi  \wedge  \ottmv{x_{{\mathrm{4}}}}  \ottsym{=}   \texttt{Transfer}  (  \ottmv{x_{{\mathrm{1}}}} ,  \ottmv{x_{{\mathrm{2}}}} ,  \ottmv{x_{{\mathrm{3}}}}  )   \ottsym{\}}  \hyp{3}\\
       \ottmv{x_{{\mathrm{4}}}}  \notin   \text{dom}( \Gamma  \ottsym{,}   \widehat{  \ottmv{x_{{\mathrm{1}}}} \mathord:\allowbreak \ottnt{T}   \triangleright   \ottmv{x_{{\mathrm{2}}}} \mathord:\allowbreak \ottkw{int}   \triangleright   \ottmv{x_{{\mathrm{3}}}} \mathord:\allowbreak \ottkw{address}   \triangleright  \Upsilon }  )   \hyp{3.1}
    \end{gather}
    for some \(\ottmv{x_{{\mathrm{1}}}}\), \(\ottmv{x_{{\mathrm{2}}}}\), \(\ottmv{x_{{\mathrm{3}}}}\), \(\ottmv{x_{{\mathrm{4}}}}\), \(\ottnt{T}\),
    \(\Upsilon\), and \(\varphi\).  The last rule that derives \hypref{eval2} is \ruleref{E-TransferTokens}; therefore
    \begin{gather}
       \ottnt{S_{{\mathrm{1}}}} \ottsym{=} \ottnt{V}  \triangleright  \ottnt{i}  \triangleright  \ottnt{a}  \triangleright  \ottnt{S}  \hyp{4}\\
       \ottnt{S_{{\mathrm{2}}}} \ottsym{=}  \texttt{Transfer}  (  \ottnt{V} ,  \ottnt{i} ,  \ottnt{a}  )   \triangleright  \ottnt{S}  \hyp{5}
    \end{gather}
    for some \(\ottnt{V}\), \(\ottnt{i}\), \(\ottnt{a}\), and \(\ottnt{S}\).  By
    \hypref{semty2}, \hypref{2}, and \hypref{4}, we have
    \begin{gather}
      \sigma  \ottsym{:}  \Gamma  \models  \ottnt{S}  \ottsym{:}  \ottsym{\{}  \Upsilon  \mid  \exists \,  \ottmv{x_{{\mathrm{3}}}} \mathord:\allowbreak \ottkw{address}   \ottsym{.}  \ottsym{(}  \exists \,  \ottmv{x_{{\mathrm{2}}}} \mathord:\allowbreak \ottkw{int}   \ottsym{.}  \ottsym{(}  \exists \,  \ottmv{x_{{\mathrm{1}}}} \mathord:\allowbreak \ottnt{T}   \ottsym{.}  \varphi  \wedge  \ottmv{x_{{\mathrm{1}}}}  \ottsym{=}  \ottnt{V}  \ottsym{)}  \wedge  \ottmv{x_{{\mathrm{2}}}}  \ottsym{=}  \ottnt{i}  \ottsym{)}  \wedge  \ottmv{x_{{\mathrm{3}}}}  \ottsym{=}  \ottnt{a}  \ottsym{\}} \hyp{9}.
    \end{gather}
    By \hypref{3.1} and \hypref{9}, we have
    \begin{multline}
      \sigma  \ottsym{:}  \Gamma  \models  \ottnt{S}  \ottsym{:}  \ottsym{\{}  \Upsilon  \mid  \exists \,  \ottmv{x_{{\mathrm{4}}}} \mathord:\allowbreak \ottkw{operation}   \ottsym{.}  \ottsym{(}   \exists \,  \ottmv{x_{{\mathrm{1}}}} \mathord:\allowbreak \ottnt{T}   \ottsym{,}   \ottmv{x_{{\mathrm{2}}}} \mathord:\allowbreak \ottkw{int}   \ottsym{,}   \ottmv{x_{{\mathrm{3}}}} \mathord:\allowbreak \ottkw{address}   \ottsym{.}  \varphi  \wedge {} \\  \ottmv{x_{{\mathrm{4}}}}  \ottsym{=}   \texttt{Transfer}  (  \ottmv{x_{{\mathrm{1}}}} ,  \ottmv{x_{{\mathrm{2}}}} ,  \ottmv{x_{{\mathrm{3}}}}  )    \ottsym{)}  \wedge  \ottmv{x_{{\mathrm{4}}}}  \ottsym{=}   \texttt{Transfer}  (  \ottnt{V} ,  \ottnt{i} ,  \ottnt{a}  )   \ottsym{\}} \hyp{10}.
    \end{multline}
    Therefore, we have
    \begin{multline*}
      \sigma  \ottsym{:}  \Gamma  \models   \texttt{Transfer}  (  \ottnt{V} ,  \ottnt{i} ,  \ottnt{a}  )   \triangleright  \ottnt{S}  \ottsym{:}  \ottsym{\{}   \ottmv{x_{{\mathrm{4}}}} \mathord:\allowbreak \ottkw{operation}   \triangleright  \Upsilon  \mid  \ottsym{(}   \exists \,  \ottmv{x_{{\mathrm{1}}}} \mathord:\allowbreak \ottnt{T}   \ottsym{,}   \ottmv{x_{{\mathrm{2}}}} \mathord:\allowbreak \ottkw{int}   \ottsym{,}   \ottmv{x_{{\mathrm{3}}}} \mathord:\allowbreak \ottkw{address}   \ottsym{.}  \varphi  \wedge {} \\  \ottmv{x_{{\mathrm{4}}}}  \ottsym{=}   \texttt{Transfer}  (  \ottmv{x_{{\mathrm{1}}}} ,  \ottmv{x_{{\mathrm{2}}}} ,  \ottmv{x_{{\mathrm{3}}}}  )    \ottsym{)}  \ottsym{\}}
    \end{multline*}
    as required.

  \item[\ruleref{RT-Sub}] We have
    \begin{gather}
      \Gamma  \vdash  \Phi_{{\mathrm{1}}}  \ottsym{<:}  \Phi'_{{\mathrm{1}}} \hyp{1}\\
      \Gamma  \vdash  \Phi'_{{\mathrm{2}}}  \ottsym{<:}  \Phi_{{\mathrm{2}}} \hyp{2}\\
       \Gamma \vdash \Phi'_{{\mathrm{1}}} \; \ottnt{I} \; \Phi'_{{\mathrm{2}}}  \hyp{3}
    \end{gather}
    for some \(\Phi'_{{\mathrm{1}}}\) and \(\Phi'_{{\mathrm{2}}}\).
    By \propref{soundness/subtyping}, \hypref{semty2}, and \hypref{1}, we have
    \begin{gather}
      \sigma  \ottsym{:}  \Gamma  \models  \ottnt{S_{{\mathrm{1}}}}  \ottsym{:}  \Phi'_{{\mathrm{1}}} \hyp{4}.
    \end{gather}
    By IH, \hypref{3}, \hypref{eval2}, and \hypref{4}, we have
    \begin{gather}
      \sigma  \ottsym{:}  \Gamma  \models  \ottnt{S_{{\mathrm{2}}}}  \ottsym{:}  \Phi'_{{\mathrm{2}}} \hyp{5}.
    \end{gather}
    Then, the goal follows from \propref{soundness/subtyping}.
    \qedhere
  \end{match}
\end{prop}

%%% Local Variables:
%%% mode: latex
%%% TeX-master: "../paper.otex"
%%% End:

% \section*{Declarations}
% \subsection*{Funding}
% This work has been partially supported by a research grant from Tezos Foundations.
% \subsection*{Conflicts of intereset/Competing interests}
% No conflicts.
% \subsection*{Availability of data and material}
% \url{https://gitlab.com/aigarashi/ReFX}
% \subsection*{Code availability}
% \url{https://gitlab.com/aigarashi/ReFX}

\else
\vfill

{\small\medskip\noindent{\bf Open Access} This chapter is licensed under the terms of the Creative Commons\break Attribution 4.0 International License (\url{http://creativecommons.org/licenses/by/4.0/}), which permits use, sharing, adaptation, distribution and reproduction in any medium or format, as long as you give appropriate credit to the original author(s) and the source, provide a link to the Creative Commons license and indicate if changes were made.}

{\small \spaceskip .28em plus .1em minus .1em The images or other third party material in this chapter are included in the chapter's Creative Commons license, unless indicated otherwise in a credit line to the material.~If material is not included in the chapter's Creative Commons license and your intended\break use is not permitted by statutory regulation or exceeds the permitted use, you will need to obtain permission directly from the copyright holder.}

\medskip\noindent\includegraphics{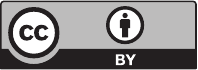}
\fi

\end{document}